\documentclass[structabstract]{aa}  
\pdfoutput=1
\usepackage{hyperref}
\usepackage{graphicx}
\usepackage{txfonts}

\usepackage[percent]{overpic}
\usepackage{color}
\usepackage{multirow}

\usepackage{natbib} 
\bibpunct{(}{)}{;}{a}{}{,} 

\newcommand{\mytextbf}[1]{{#1}}

\begin{document}
   \title{\mytextbf{Correlations} between sunspots and their moat flows}


   \author{J. L\"ohner-B\"ottcher \and R. Schlichenmaier}
    \institute{Kiepenheuer-Institut f\"ur Sonnenphysik, Sch\"oneckstr. 6, 79104 Freiburg\\
              \email{jlb, schliche @kis.uni-freiburg.de}}

   \date{Received Oct 11, 2012; accepted Nov 29, 2012}

 
  \abstract
{The presence of the moat flow around sunspots is intimately linked to the mere existence of sunspots.}
{We characterize the moat flow (MF) and Evershed flow (EF) in sunspots to enhance our knowledge of sunspot structures and photospheric flow properties.}
{We calibrated HMI synoptic Doppler maps and used them to \mytextbf{analyze}  \mytextbf{3h time averages} of 31 circular, stable, and fully developed sunspots at heliocentric angles of some $50^{\circ}$. Assuming axially \mytextbf{symmetrical} flow fields, we infer the azimuthally averaged horizontal velocity component of the MF and EF \mytextbf{from 51 velocity maps}. We studied the MF properties (velocity and extension) and elaborate on how these components depend on sunspot parameters (sunspot size and EF velocity). To explore the weekly and monthly evolution of MFs, we compare spots rotating from the eastern to western limbs and spots that \mytextbf{reappear} on the eastern limb.}
{Our  calibration procedure of HMI Doppler maps yields reliable and consistent results. \mytextbf{In 3h averages, w}e find the MF decreases on average from some $1000\pm200$\,m/s just outside the spot boundary to 500\,m/s after an additional 4\,Mm. The average MF extension lies at $9.2\pm5\,\rm{ Mm}$, where the velocity drops below some 180\,m/s. Neither the MF velocity nor its extension depend significantly on the sunspot size or EF velocity. But, the EF velocity does show a tendency to be enhanced with sunspot size. On a time scale of a week and a month, we find decreasing MF extensions and a tendency for the MF velocity to increase for strongly decaying sunspots, whereas the changing EF velocity has no impact on the MF.}
{\mytextbf{On 3h averages, t}he EF velocity scales with the size of sunspots, while the MF properties show no significant correlation with the EF or with the sunspot size. This we interpret as a hint that the physical origins of EF and MF are distinct.}
   \titlerunning{ \mytextbf{Correlations} between sunspots and their moat flows} 
   \authorrunning{L\"ohner-B\"ottcher \& Schlichenmaier}
   \maketitle
%
\section{Introduction}\label{intro}
As complex magnetic formations, sunspots strongly affects the photospheric dynamics of the granular convective pattern of the plasma in which they are embedded. They suppress the upward propagation of heat in the magnetic core, seen as the cooler and therefore darker umbra, and they harbor the well-known penumbral Evershed flow observed as \mytextbf{a} radially outwards-directed, horizontal flow in the photospheric layers. In photospheric observations this leads to a blueshift and  \mytextbf{blueward}  asymmetry of the spectral line for penumbral filaments facing the center of the solar disk and a redshift and  \mytextbf{redward} asymmetry for filaments facing the solar limb \citep[e.g.,][]{evershed1909,maltby1964,schliche2004}. In general, the line shift, as well as the asymmetry, increases from the inner to the outer penumbra. In photospheric layers the Evershed flow is close to horizontal with maximum inclinations of $5^\circ$ to $10^\circ$, exhibiting \mytextbf{downward flows} in the outer penumbra \citep[e.g.,][]{michard1951,balthasar+etal1996,schliche+schmidt2000,franz2009}. Average horizontal velocities are in the range of 3-4 km/s \citep{shine1994}, but can exceed 6 km/s or more on small scales \citep[e.g.,][]{rouppe2002,bellotrubio2003a}. The Evershed flow seems to stop abruptly at the white light boundary of the spot \citep{wiehr1992,schliche1999}, which is in line with \mytextbf{downward flows} in the outer penumbra. \mytextbf{Only a small fraction of the flow continues to flow into the magnetic canopy that surrounds the spot \citep{rezaei+al2006}.}

Whereas the Evershed flow \mytextbf{(EF)} is magnetized \citep[e.g.,][]{schliche+collados2002}, the annular region around fully-fledged sunspots, called the moat cell, is largely \mytextbf{nonmagnetic}. It harbors the moat flow \mytextbf{(MF)}, a radially oriented outflow of plasma adjacent to the outer penumbral boundary. Embedded in these motions, small, moving magnetic features (MMFs) migrate away from the sunspot \citep{sheeley1969,sheeley1972,vrabec1971,harvey+harvey1973}. The existence of both the MF and the EF depend on the presence of a penumbra \citep{vrabec1974}. Consequently, the moat flow can only be observed for sunspot sides with a well-developed penumbra and lacks pores \citep{sobotka1999}.

The moat flow develops immediately after the formation of a penumbra \mytextbf{\citep{pardon1979,schlichenmaier+etal2010} and was observed even after the decay of the penumbra \citep{verma+etal2012}. The MF is present as a stationary flow through the spot's lifetime, which varies between several weeks to several months \citep[see, e.g.,][]{solanki2003}: \citet{sobotka+roudier2007} do not find any significant variations in the MF within half a day.} During the initiation of the MF, the magnetic components in the vicinity of the sunspot are pushed to the periphery and leave the moat cell largely \mytextbf{nonmagnetic} with single MMFs. The extension of the moat field reaches between 10\,Mm to 20\,Mm from the penumbral boundary for small sunspots and can be roughly twice the spot radius for larger spots \citep{brickhouse+labonte1988}\mytextbf{, but \citet{sobotka+roudier2007} find no such correspondence.} 

The moat flow velocity ranges between 0.5\,km/s to 1\,km/s and can be seen by tracking the bright granular features \mytextbf{\citep[e.g.,][]{rimmele1997,dominguez+etal2007,balthasar+muglach} or \mytextbf{Doppler shift} measurements \citep{balthasar+etal1996}}. \mytextbf{That the MF consists of migrating granules, indicates that the MF is as nonmagnetic as granules, i.e., largely nonmagnetic.}

Also helioseismic measurements have revealed the existence of the MF as an outflow extending up to 30\,Mm with \mytextbf{a} maximum velocity of 1\,km/s just next to the penumbral boundary \citep{gizon+etal2000} in the first 2\,Mm of the solar surface. \mytextbf{ According to recent studies \citep{sun1997,gizon+etal2009,featherstone+etal2011}, the MF is also detectable in deeper layers, but} has slower speeds compared to surface measurements.

\mytextbf{\citet{cabrera2006} suggest that a link exists between EF and MMFs.} Magnetic velocity packages, called the Evershed clouds inside the penumbra \citep{shine1994,rimmele1994,cabrera2007} propagate outwards to the extension of penumbral filaments and the moat region where they are embedded as MMFs. \mytextbf{MMFs can travel from the penumbra into the vicinity of the sunspot \citep{dalda+pillet2005,dalda+rubio2008a,dalda+rubio2008b}. Also \citet{schliche2002} proposes a scenario that is consistent with observed MMFs. In this way, a magneto-convective overshoot instability in an Evershed flux tube leads to the migrating feature in the \mytextbf{MF} region. There is still discordance about a link between the EF and MF.}
\citet{dominguez+etal2007} observed irregular sunspots and found that the moat flow is only present \mytextbf{in radial extensions} of penumbral filaments, but not perpendicular to them. \mytextbf{This indicates that the MF is an extension of the EF. However, this seems to be impossible since the EF is magnetized, while the MF is intrinsically unmagnetic, as mentioned above.}

\mytextbf{In this paper, we analyze the EF and MF and elaborate on a possible link between the two. To that end, we utilize \mytextbf{Doppler shift} measurements of HMI. In Sec.~\ref{sec:red} we perform a thorough calibration of HMI Doppler maps. In Sec.~\ref{sec:obs} we describe the criteria for data selection and the method we used to analyze the flow properties. In Sec.~\ref{sec:res} we present our results for the flow properties and elaborate on correlations between spot and MF properties and on how they change during the evolution of a spot. The statistics are based on 51 maps constructed from 31 different spots. The data sets we used and the radial dependencies of the flow velocity of all spots, which form the basis of our analysis, are given in the Appendix. In Sec.~\ref{sec:concl}, we discuss the present understanding of MFs in the context of our results and summarize our results and conclusions.}

\section{Calibration of HMI Doppler maps}\label{sec:red}
\paragraph{Data:} Our work is based on synoptic \mytextbf{720s intensity} maps and Doppler maps\footnote{A velocity map of the Sun according to spectral \mytextbf{line shifts} due to the Doppler effect.} in the \ion{Fe}{I} spectral line at 6173.3\,\AA~from the JSOC webpage recorded by the Helioseismic Magnetic Imager (HMI) of the Solar Dynamics Observatory (SDO) up to maximum values of $\pm$\,6.5\,km/s and a spatial resolution of $\rm{ \approx 0.5\arcsec/px}$.

\paragraph{Calibration by subtracting systematic components:}
To yield undisturbed flow velocities relative to the solar surface, we need to define a rest frame ($v\!=\!0$). To this end, we construct a \mytextbf{time averaged} velocity map, $\langle v(x,y)\rangle_t$, in which the effects of granulation, oscillations, and supergranulation are removed. This average is composed of three systematic, large-scale components:

\begin{equation}
\langle v(x,y)\rangle_t = v_{\rm clv}(r) + v_{\rm rot}(d,l,B_0) + v_{\rm res}(x,y)\label{eq:components}
\end{equation}

The first term is the \mytextbf{c}enter-to-limb variation in the convective blueshift, $v_{\rm clv}$. It is radially \mytextbf{symmetrical} and is several hundred m/s, depending on the distance, r, (or heliocentric angle $\theta$) to disk center. The second term is the differential rotation, $v_{\rm rot}$. It is axially \mytextbf{symmetrical} with velocities up to $\pm2000\,{\rm m/s}$ with respect to the meridian. The values depend on the distance $d$ from the meridian, the latitude $l$ and the inclination $B_0$ of the rotation axis, which varies annually between $\pm7.27^\circ$. The third term is a \mytextbf{nonsymmetrical} residual, $v_{\rm res}$, with velocities up to $\pm150\,{\rm m/s}$, which contains instrumental effects and other systematic flow fields of smaller magnitude. The large-scale meridional flow with velocities of \mytextbf{less than} 20\,m/s can be neglected for our sunspot studies, and if existing, it is included in these three components. In the next four sections we determine the four terms of Eq.~\ref{eq:components}.

\subsection{Construction of \mytextbf{a time-averaged} velocity map: $\langle v(x,y)\rangle_{t}$} 
\begin{figure}[htbp]
\centering
\includegraphics*[height=3.73cm]{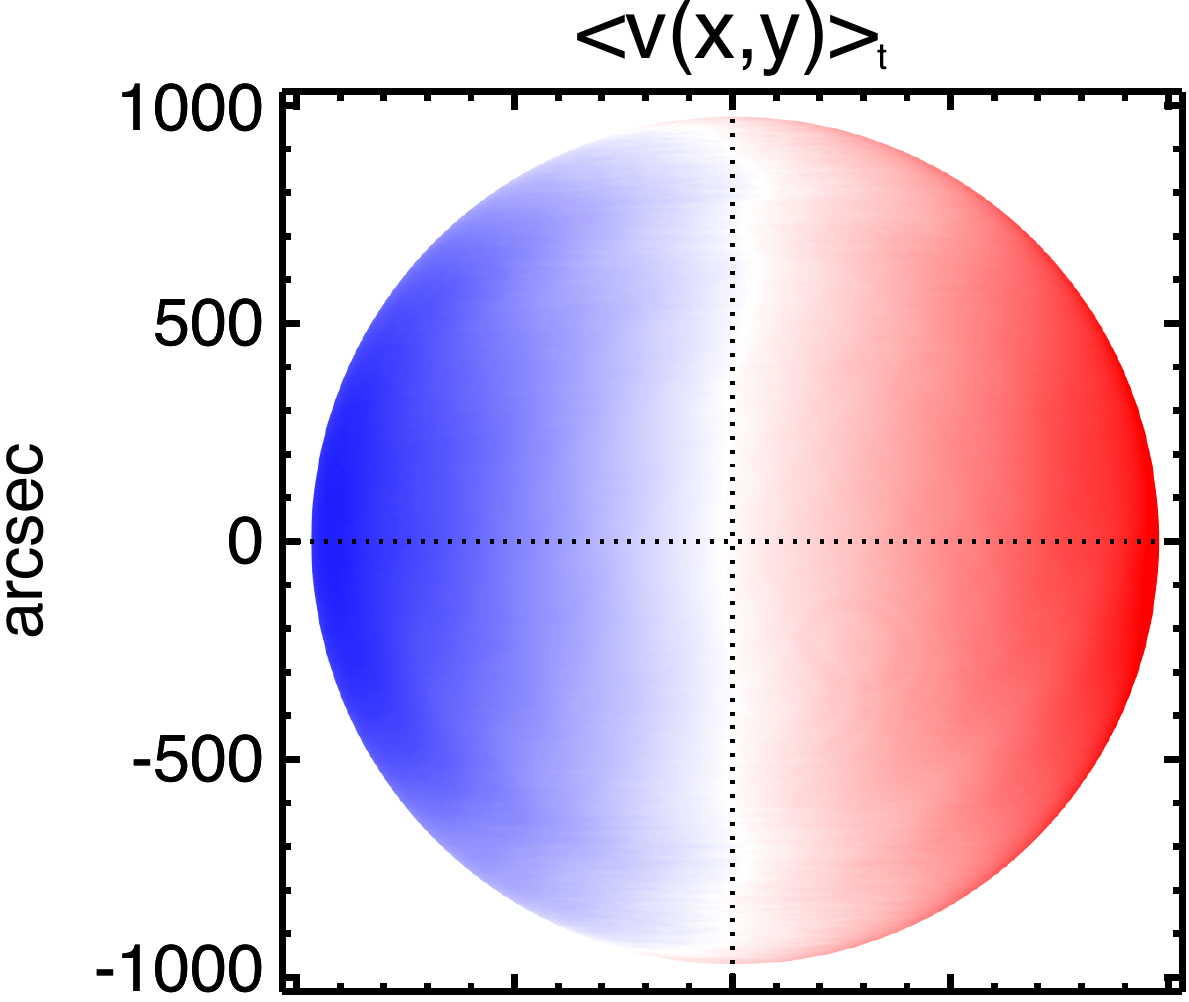}
\includegraphics*[height=3.73cm]{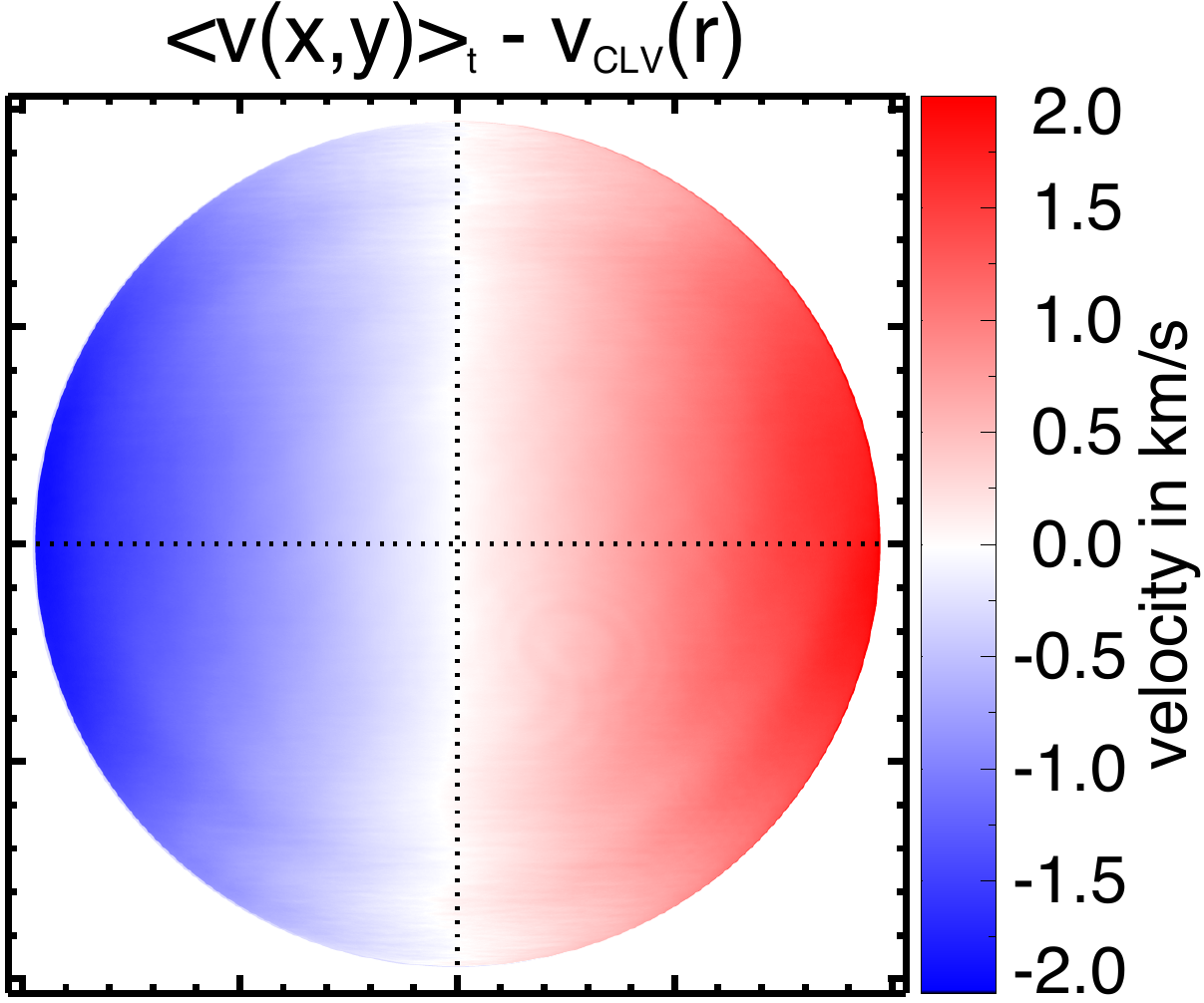}
\caption{\label{fig:reduction}--- Left panel: Averaged velocity map, $\langle v(x,y)\rangle_{t=61\,{\rm days}}$, with maximum \mytextbf{blueshifts} (eastern limb) and \mytextbf{redshifts} (western limb) with \mytextbf{some \mbox{2\,km/s}. The} axes \mytextbf{are} given in arcsec. --- Right panel: Reduced  velocity map, $\langle v(x,y)\rangle_{t}-v_{\rm clv}$.}
\end{figure}

In velocity maps, supergranules produce a \mytextbf{redshift} on the limb side and a \mytextbf{blueshift} on the opposite side when they are off disk center. Because a supergranule migrates across the disk its average velocity signal vanishes when it is next to the equator, since \mytextbf{blueshifts} are eliminated by the following \mytextbf{redshifts}. However, close to the poles, the \mytextbf{blueshifts} and \mytextbf{redshifts} are separated in latitude such that a single supergranule leaves noticeable traces in the Doppler map. To diminish these traces a long time series is necessary. In our case we have averaged \mytextbf{720s Doppler maps of} 61 entire days (a total of 7320 \mytextbf{maps }) between June and December 2010, shown in the left hand panel of Fig.~\ref{fig:reduction}.

In order not to spoil the time average by active regions, we have masked regions according to the darkening in the continuum intensity and replaced the content of the mask in the Doppler map by the velocities on the opposite side of the heliocentric equator. Thanks to the solar activity minimum, this method was appropriate. For consistency, we normalized the solar radius of all Doppler maps and eliminated the respective observer motion\footnote{Keywords for observer motion in the \mytextbf{FITS file header}: $OBS\_VR$ (radial), $OBS\_VW$ (westward), and $OBS\_VN$ (northward).} relative to the Sun. Since we favored time periods with inclination $B_0\approx0^\circ$, the obtained velocity map $\langle v(x,y)\rangle_{t=61\,{\rm days}}$ should serve as a good basis for modeling the systematic components named in Eq.~\ref{eq:components}. In $\langle v(x,y)\rangle_{t}$ the velocity at disk center is about 50 m/s.

\begin{figure}[!htdp]
\begin{center}
\begin{tabular}{|l|}
\hline
\large{a)}\\
\includegraphics*[trim = 1mm 0mm 0mm 0mm, clip, width=7.8cm]{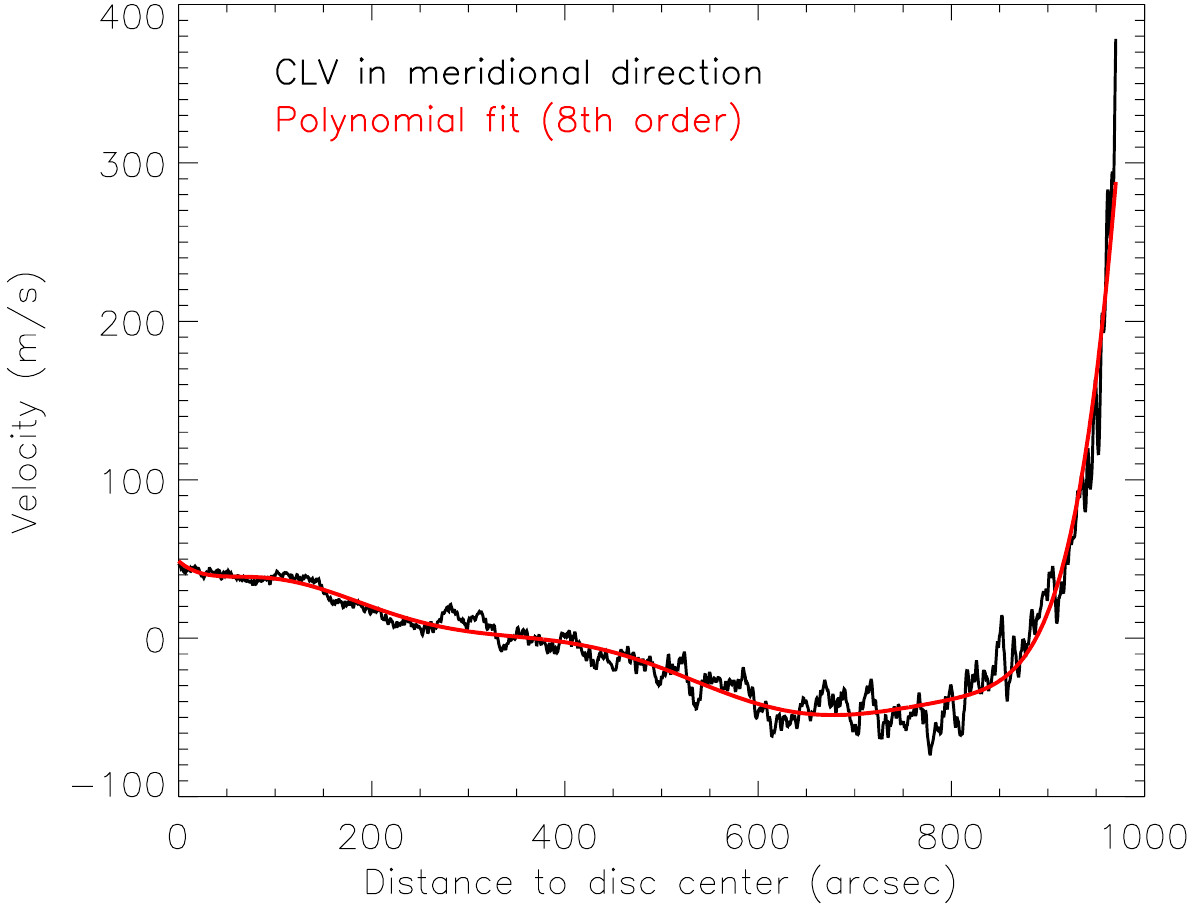}\\
\hline
\large{b)}\\
\includegraphics*[width=7.2cm]{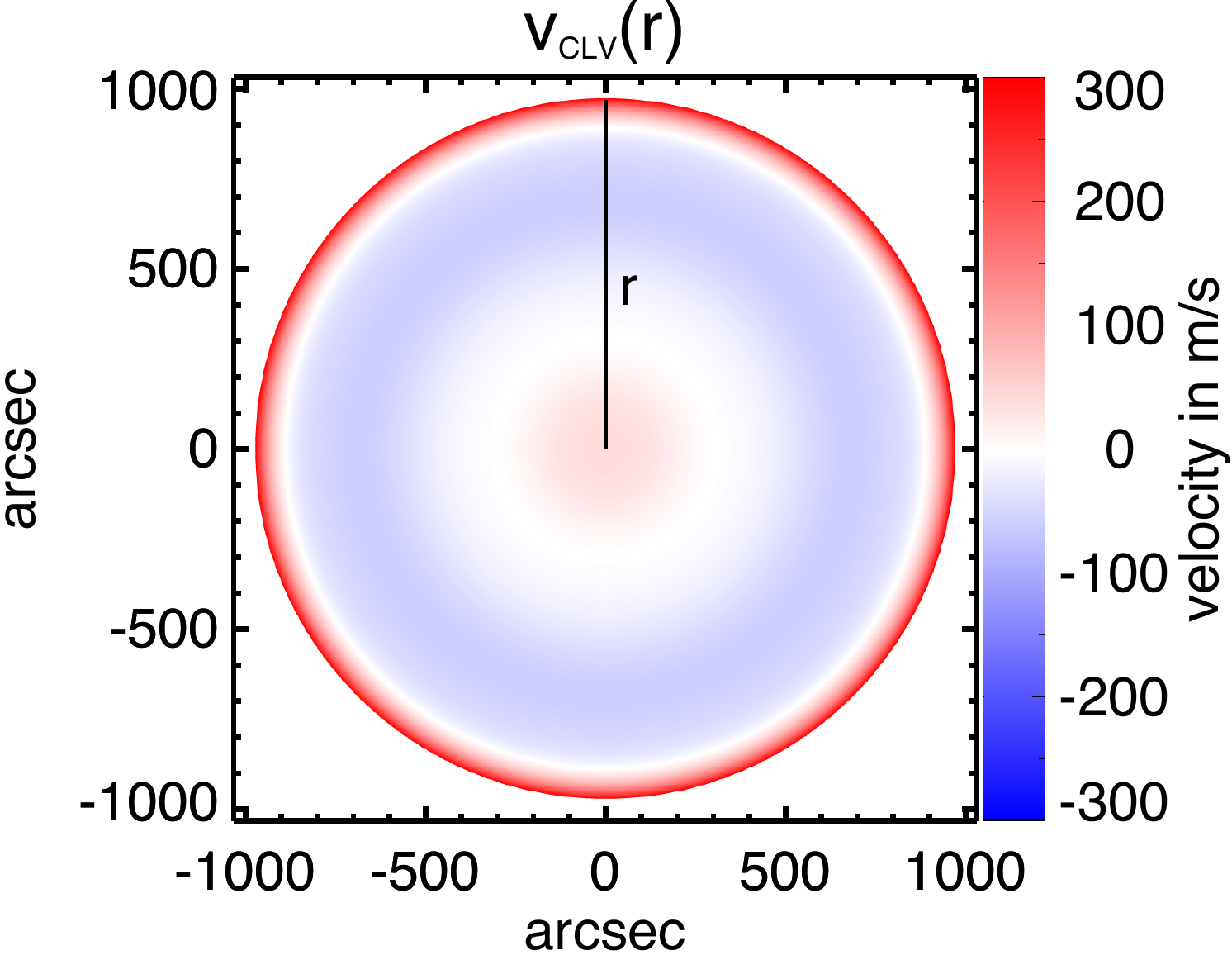}\\
\hline
\large{c)}\\
\includegraphics[width=7.2cm]{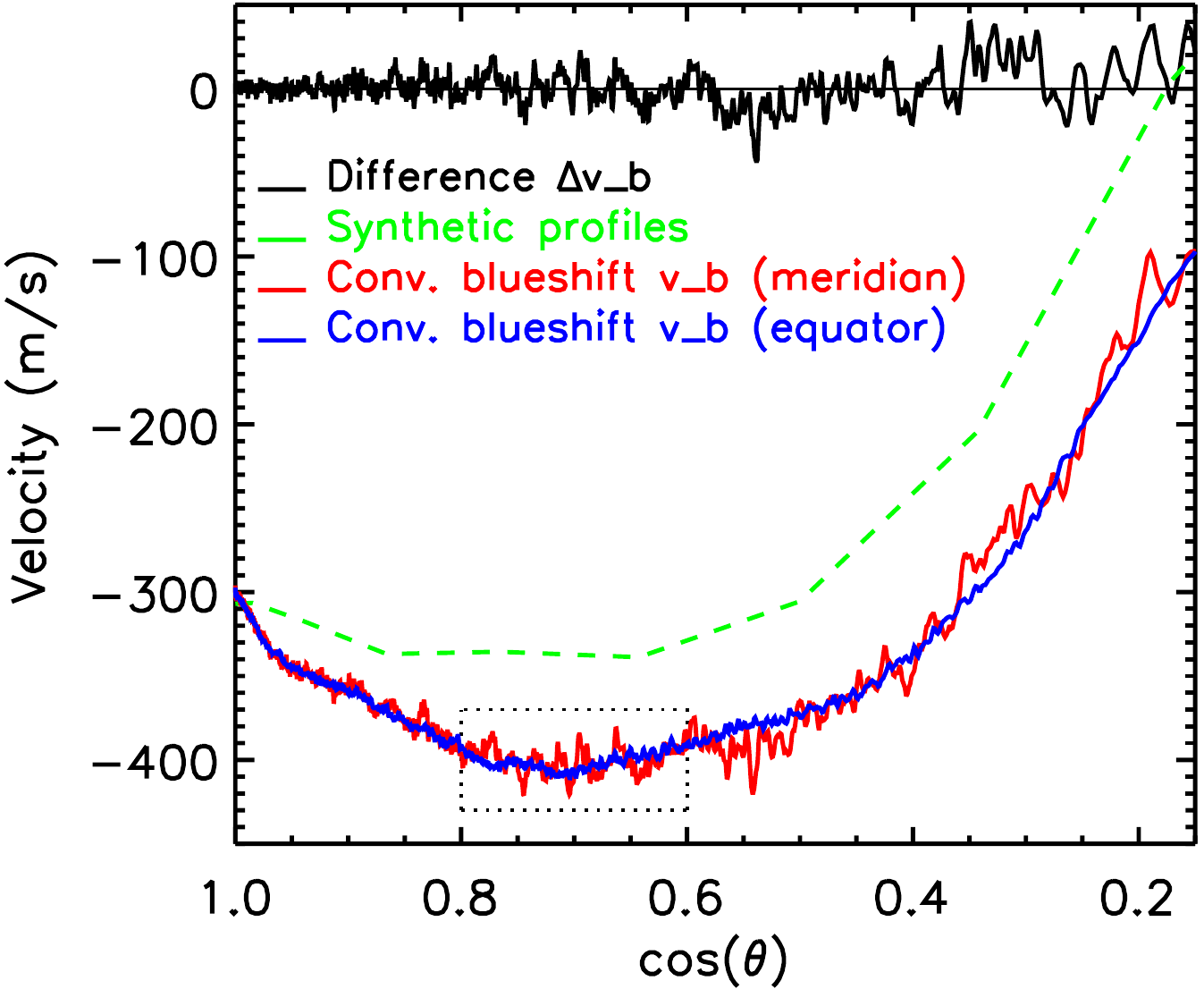}\\
\hline
\end{tabular}
\end{center}
\caption{\label{fig:clv}--- \textbf{a)} Average CLV, $\langle v(0, y)\rangle_{t}$, along the meridional centerline \mytextbf{with an 8th-order} polynomial fit, $v_{\rm clv}(y)$ (red curve) in m/s against the distance $y$ to the \mytextbf{disk} center in arcsec. -- \textbf{b)} Calibrated, radially symmetrical \mytextbf{CLV model}, $v_{\rm clv}(r)$, with axis given in arcsec. -- \textbf{c)} Convective blueshift, $v_{\rm b}(r)$, in $\tilde{v}_{\rm clv}$ along the meridian (red solid curve) and equator (blue solid curve) and their difference (black solid curve) displayed in m/s against $\cos\theta$ with heliocentric angle, $\theta$. A synthesis of the spectral line from COBOLD simulations (green dashed curve) is added. The dotted black box refers to the \mytextbf{preferred} sunspot location.}
\end{figure}

\subsection{Radially \mytextbf{symmetrical} velocity component: $v_{\rm clv}(r)$}
As seen in the left hand panel of Fig.~\ref{fig:reduction}, the \mytextbf{LOS velocity of the} solar rotation, which should be zero for the meridian and axially symmetrical with respect to the centerlines, is deformed. This deformation along the meridian contains the \mytextbf{center-to-limb} variation (CLV) of the convective \mytextbf{blueshift} and the meridional flow. To determine \mytextbf{the sum of both effects} {\em \mytextbf{a priori}} (without knowing the differential rotation), we decided to use the average, $\langle v(0\pm dx, y)\rangle_{t;\,dx=10px;\,\pm y}$, of \mytextbf{a} \mytextbf{20-pixel} wide stripe along the meridian of $\langle v(x,y)\rangle_t$ with respect to the distance, $r$, to the disk center as shown in Fig.~\ref{fig:clv}a. Assuming radial symmetry, the curves of the northern and southern hemispheres are averaged. The resulting $\langle v(0, y)\rangle_{t}$ is fitted by a \mytextbf{8th-order} polynomial, $v_{\rm clv}(y)$. The curve runs from $+50$\,m/s at the \mytextbf{disk} center to $-60$\,m/s at \mytextbf{some} $700{\rm \arcsec}$ and strong \mytextbf{redshifts} exceeding 300\,m/s at the solar limb. This CLV is in good accordance with recent studies of the CLV of the \ion{Fe}{i} line \citep[e.g.,][]{balthasar1985, cruzrodriguez2011}. As the up- and downturns, which are caused by the \mytextbf{supergranular} migration at higher latitudes, are on the order of some 30\,m/s, we neglect the impact of the meridional flow\footnote{A large-scale \mytextbf{axisymmetrical} flow directing from the equator to the poles on the solar surface with velocities up to 20 m/s at latitudes around 35$^\circ$ \citep{HathRight2010,komm2011}} on $v_{\rm clv}(y)$ and further calibrations. We use the \mytextbf{polynomial} $v_{\rm clv}(y)$ to create a radially \mytextbf{symmetrical CLV model}, $v_{\rm clv}(r)$, as displayed in Fig.~\ref{fig:clv}b.

Since the convective blueshift is the result of the bigger impact of hot upwards moving granules than the cool, downwards directed, \mytextbf{intergranular} fraction on a spatially averaged solar surface, the \mytextbf{line shift} depends on the observed spectral line and the \mytextbf{line of sight}. According to calculations based on the Li\`{e}ge atlas \citep{liege}, the convective blueshift of the \ion{Fe}{i} spectral line at 6173.3\,\AA~in the center is \mbox{$v_{\rm b}\!=\!-305\,{\rm m/s}$}. For observational studies in Sec.~\ref{sec:obs}, we have to offset $v_{\rm clv}(r)$ by \mbox{$v_{\rm off}\!\approx\!-350{\rm m/s}$} to \mytextbf{obtain} the convective \mytextbf{blueshift}, $v_{\rm b}(r)$, with its CLV.

After the calibration of $v_{\rm rot}(d,l,B_0)$ and $v_{\rm res}(x,y)$ in Sects.~\ref{sec:rota} and \ref{sec:resi} we can check the accuracy of the data calibration by computing the CLV velocity map {\it a posteriori}: 
\begin{eqnarray*}
\tilde{v}_{\rm clv}(x, y)=\langle v(x,y)\rangle_t - v_{\rm rot}(d,l,B_0) - v_{\rm res}(x,y)\qquad.
\end{eqnarray*}
Then, we compare the CLV in the meridional and equatorial directions, i.e. \mbox{$\langle \tilde{v}_{\rm clv}(0\!\pm\!dx, y)\rangle_{dx=10{\rm px}; \pm y}$} and \mbox{$\langle \tilde{v}_{\rm clv}(x, 0\!\pm\!dy)\rangle_{dy=10{\rm px}; \pm x}$}. The curves are displayed in \mytextbf{Fig.~\ref{fig:clv}c and  closely resemble} each other. The velocities were offset by \mbox{$v_{\rm off}$} to achieve the run from \mbox{$v_{\rm b}(\cos\theta\!=\!1)\!=\!-305\,{\rm m/s}$} to \mbox{$v_{\rm b}(\cos\theta=0.8\ldots0.6)\!\approx\!-400\,{\rm m/s}$} \mytextbf{(preferred spot location)} and lower velocities at the solar limb. \mytextbf{While the equatorial direction exhibits minor fluctuations, the meridional direction displays stronger fluctuations owing to the} residuals of supergranules. The difference \mytextbf{between both curves shown in the upper part of Fig.~\ref{fig:clv}c reveals an} rms smaller than 10 m/s in low latitudes. The \mytextbf{dashed green} curve in Fig.~\ref{fig:clv}c results from synthetic profiles for \ion{Fe}{i} 6173.3\,\AA~based on COBOLD simulations \citep{freytag,beeck}. The synthetic curve differs by less than 100\,m/s, so is in good qualitative agreement with our measurements.

\subsection{Differential rotation: $v_{\rm rot}(d,l,B_0)$}\label{sec:rota}
Since the rotation velocity is differential on the solar surface and depends on the \mytextbf{line of sight} (LOS), we have to generate a specific, axially symmetrical solar rotation model, $v_{\rm rot}(d,l,B_0)$, with respect to the distance, $d$, from the meridian, the latitude, $l$, and the inclination, $B_0$, of the rotation axis. 

\begin{figure}[htbp]
\centering
\includegraphics[trim = 0mm 0mm 0mm 5mm, clip,width=7.5cm]{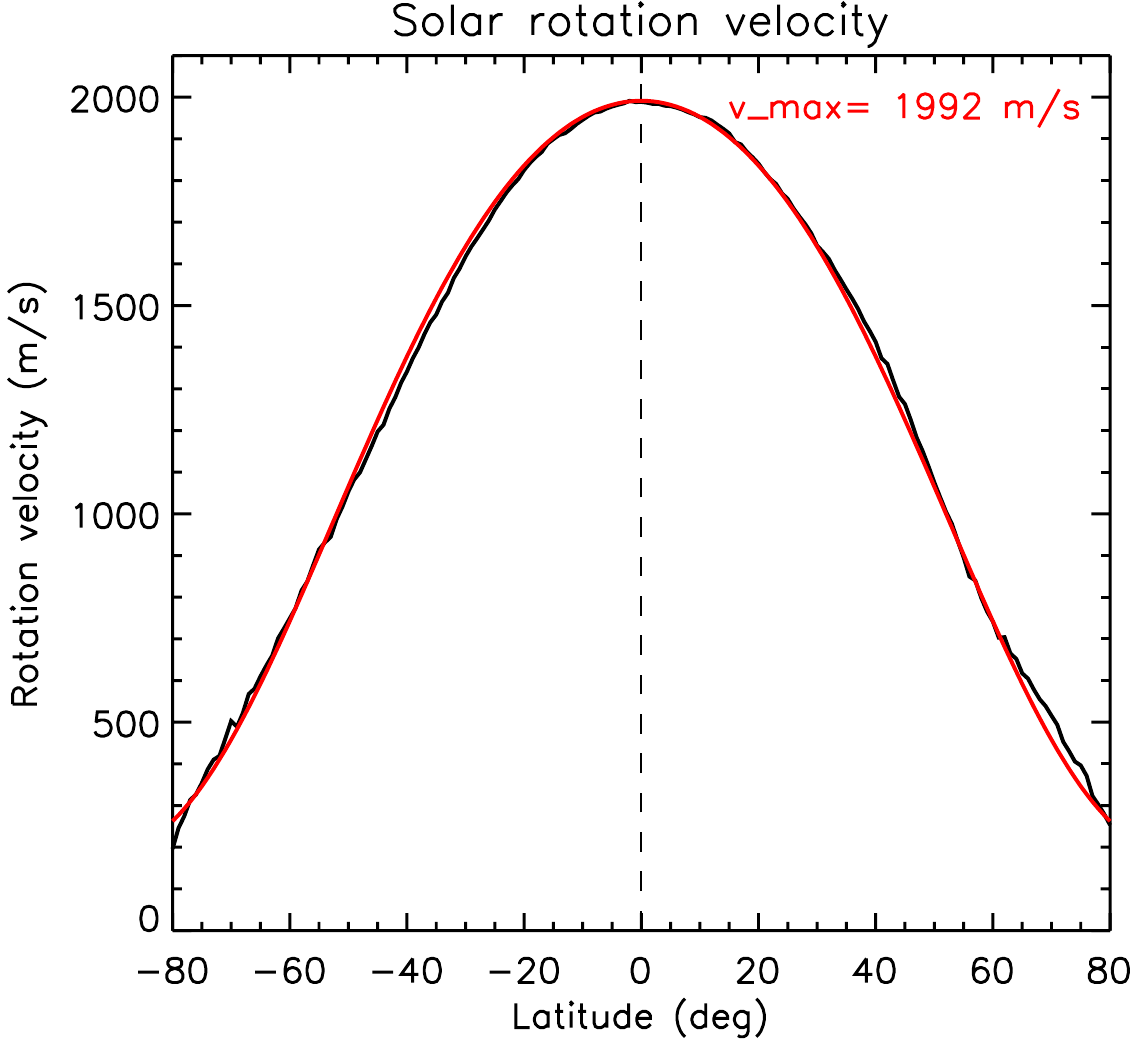}
\caption{\label{fig:sunrot}--- Absolute rotation velocities, $v_{\rm rot}(l,B_0\!=\!0)$, in m/s (black curve) on the solar surface for latitudes up to $l\!=\!\pm80^\circ$ fitted by $\Omega(l)$ (red curve) \mytextbf{and} an equatorial maximum of \mbox{$\Omega(0)\!=\!1992\,{\rm m/s}$}.}
\end{figure}
We start out with the \mytextbf{time-averaged} velocity map reduced by the radially \mytextbf{symmetrical} component, i.~e.~ $\langle v(x,y)\rangle_t - v_{\rm clv}(r)$,  displayed in the right hand panel of Fig.~\ref{fig:reduction}. We determine the absolute latitudinal velocity, $v_{\rm rot}(l)$ for $B_0\!=\!0^\circ$, by performing linear regressions for the \mytextbf{LOS velocities} along latitudinal cuts, $v(d/R_l)$, with $R_l$ being the distance of the solar limb from the meridian: $v_{\rm rot}(d, l)\!=\!v_{\rm rot}(l)\cdot (d/R_l)$. \mytextbf{Figure \ref{fig:sunrot} shows the latitudinal dependence, $v_{\rm rot}(l)$.} 

To compare our rotation model with literature values, we follow \citet{stix2002} and \mytextbf{approximate $v_{\rm rot}(l)$} by
\begin{equation} 
\Omega(l) = {\rm A} + {\rm B}\cdot\sin^2 l +{\rm C}\cdot\sin^4 l \label{omega}\qquad.
\end{equation}
The \mytextbf{approximation, $\Omega(l)$, is plotted} in Fig.~\ref{fig:sunrot}. We obtained \mytextbf{a} differential decrease to higher latitudes\mytextbf{. The} equatorial maximum, \mbox{${\rm A\!=\!14.10\pm0.03^\circ{\rm/day}}$},  \mytextbf{corresponds} to \mbox{$\Omega(0)\!=\!1992\,{\rm m/s}$}. \mytextbf{The coefficients} \mbox{$B\!=\!-9.0^\circ{\rm/day}$} and \mbox{$C\!=\!-2.5^\circ{\rm/day}$} have standard deviations on the order of ${\rm 0.1^\circ/day}$. Since the result depends on the method of measurement \citep[see:][]{stix2002}, we compared our rotational velocity with the \mytextbf{Doppler shift} measurements of \citet{snodgrass1984} \mytextbf{who obtained an} equatorial rate of \mbox{$A\!=\!14.05\pm0.01^\circ{\rm/day}$} or \mbox{$\Omega(0)\!=\!1975\,{\rm m/s}$}\mytextbf{. This is in good accordance with our findings}. \mytextbf{However, compared to Snodgras} coefficients, \mbox{$B\!=\!-1.49^\circ{\rm/day}$} and \mbox{$C\!=\!-2.61^\circ{\rm/day}$} with deviations on the order of some $0.1\,^\circ{\rm /day}$, \mytextbf{we obtained} a stronger decrease in the rotational velocity to higher latitudes. 

\begin{figure}[htbp]
\includegraphics*[height=3.73cm]{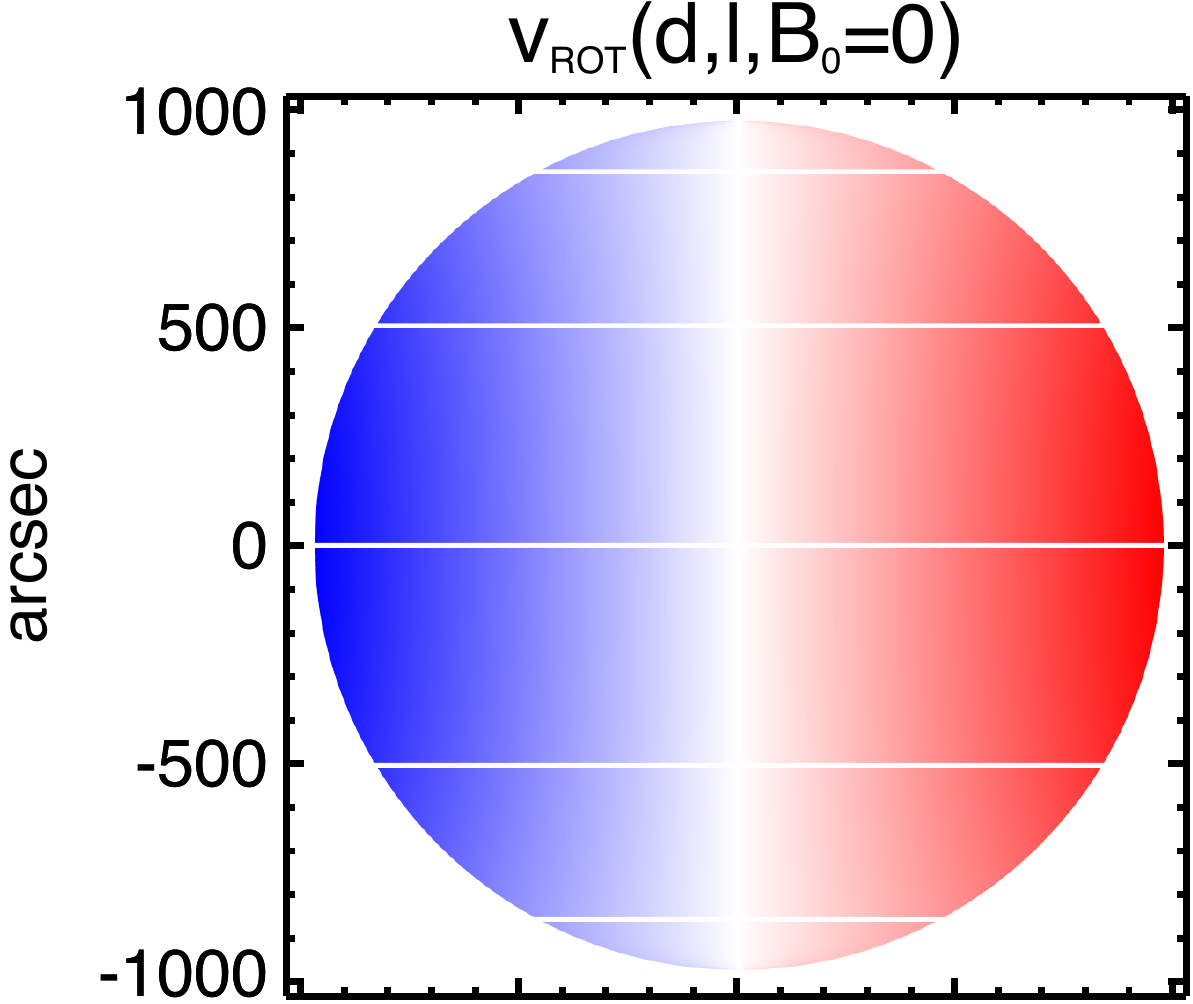}
\includegraphics*[height=3.73cm]{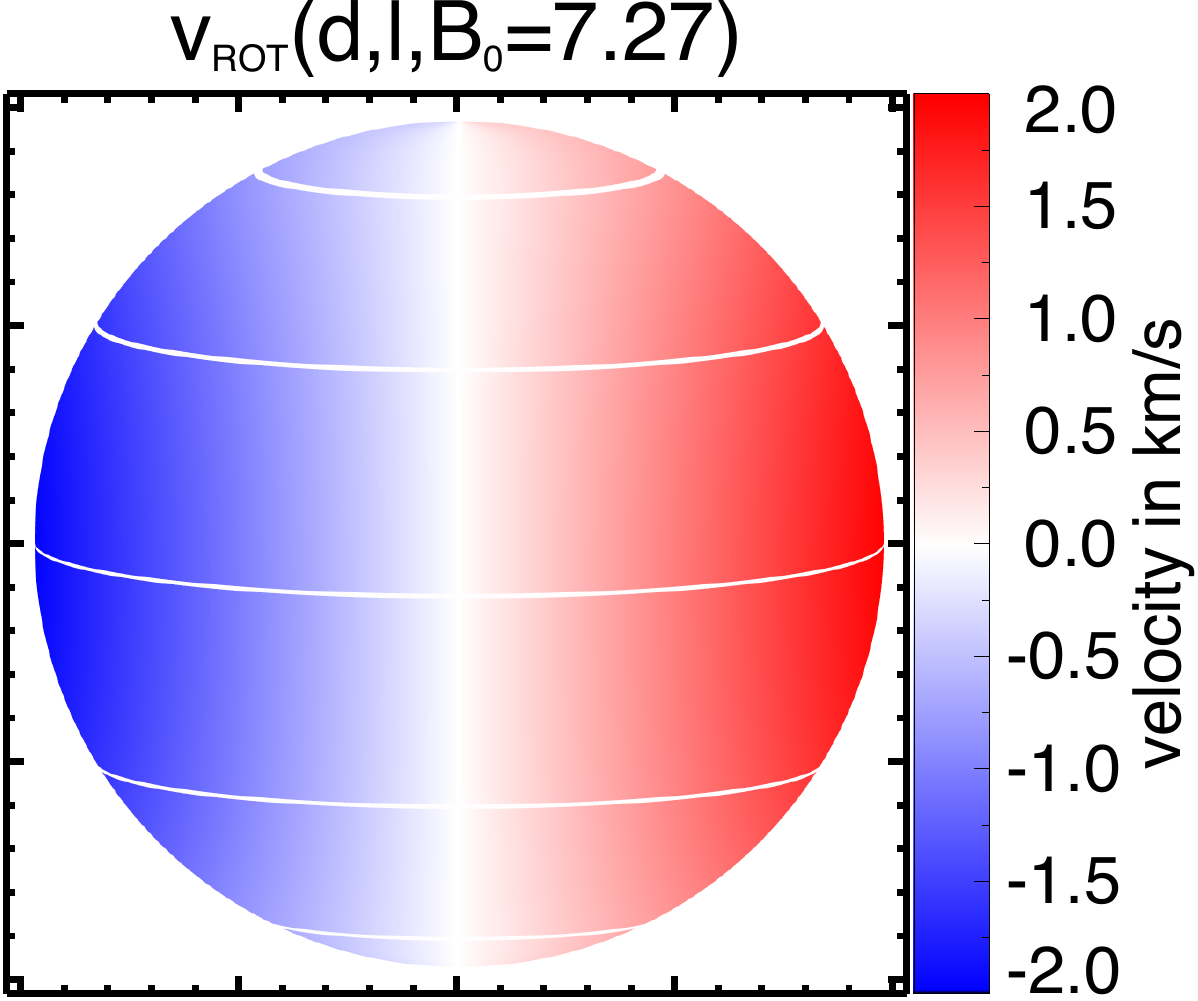}
\caption{\label{fig:modell}--- Computed differential rotation maps, \mbox{$v_{\rm rot}(d,l,B_0\!=\!0^\circ)$} (left panel) and \mbox{$v_{\rm rot}(d,l,B_0\!=\!7.27^\circ)$} (right panel), with maximum velocities \mbox{$v_{\rm rot}(l\!=\!0,B_0\!=\!0^\circ)\!=\!1992\,{\rm m/s}$} and \mbox{$v_{\rm rot}(l\!=\!0,B_0\!=\!7.27^\circ)\!=\!1976\,{\rm m/s}$}, spatially displayed in arcsec.}
\end{figure}

The differential rotation model, $v_{\rm rot}(d,l,B_0)$, displayed as an example in Fig.~\ref{fig:modell} for $B_0\!=\!0^\circ$ (left panel) and $B_0\!=\!7.27^\circ$ (right panel), generates the \mytextbf{LOS velocities} for every position on the solar \mytextbf{disk} according to the annually varying $B_0\!=\!\pm7.27^\circ$.

\begin{figure}[htbp]
\centering
\includegraphics*[width=7.5cm]{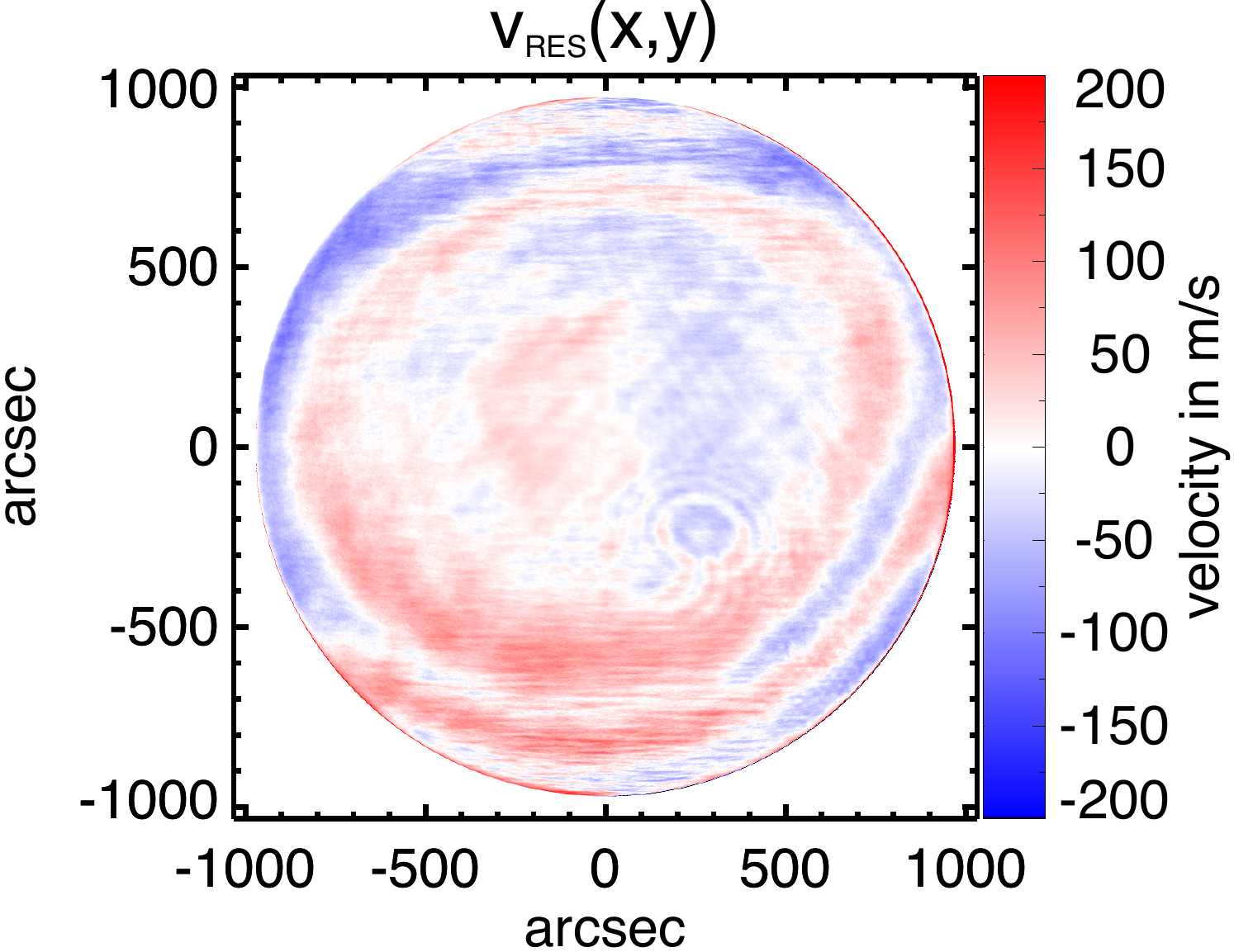}
\caption{\label{fig:fringes}--- Residual velocity map, $v_{\rm res}(x,y)$, with axis given in arcsec.}
\end{figure}
\subsection{Residual velocity map: $v_{\rm res}(x,y)$}\label{sec:resi}
Augural annular patterns \mytextbf{(see Fig.~\ref{fig:reduction})} indicate a systematical, \mytextbf{nonsymmetrical} residual, $v_{\rm res}(x,y)$, that contains instrumental effects and other systematic flow fields of smaller magnitude. Subtracting $v_{\rm clv}$ and $v_{\rm rot}$ from the \mytextbf{61 day } average, $\langle v\rangle_t$, we obtain the residual \mytextbf{nonsymmetrical} velocity map: \begin{equation} v_{\rm res}(x,y)=\langle v(x,y)\rangle_t - v_{\rm clv}(r) - v_{\rm rot}(d,l,B_0\!=\!0^\circ)\qquad,\end{equation} which is displayed in Fig.~\ref{fig:fringes}. Our result agrees well with the HMI calibration studies of \citet{howe+etal2011} and \citet{centeno+etal2011}. Several large circular patterns and smaller interference rings were checked as spatially fixed with velocities varying between \mbox{$\pm150$\,{\rm m/s}}. The residual map is subtracted {\it \mytextbf{a priori}} for all Doppler maps in Sec.~\ref{sec:obs}. \mytextbf{C}hanges in time are expected to have small amplitude \mytextbf{ so} will not \mytextbf{affect} our  analysis of horizontal flow velocities.
\section{Data selection and analysis}\label{sec:obs}
\paragraph{Overview:}
In this section, we \mytextbf{analyze} the sunspot moat flow and Evershed flow in the solar photosphere by the calibrated \mytextbf{720s Doppler} maps recorded by HMI. We have selected circular sunspots and constructed 3h time averages (out of 15 such Doppler maps) of the field of view (FOV)\mytextbf{. By, doing this, we} reduce the amplitudes of granulation and oscillation (see Sec.~\ref{sec:selection}). We \mytextbf{analyze}  the flow field in the penumbra and in the surrounding moat cell.

\begin{figure*}[htbp]
\begin{overpic}[trim = 8.5mm 6.3mm 1mm 3.4mm, clip, height=4.3cm]{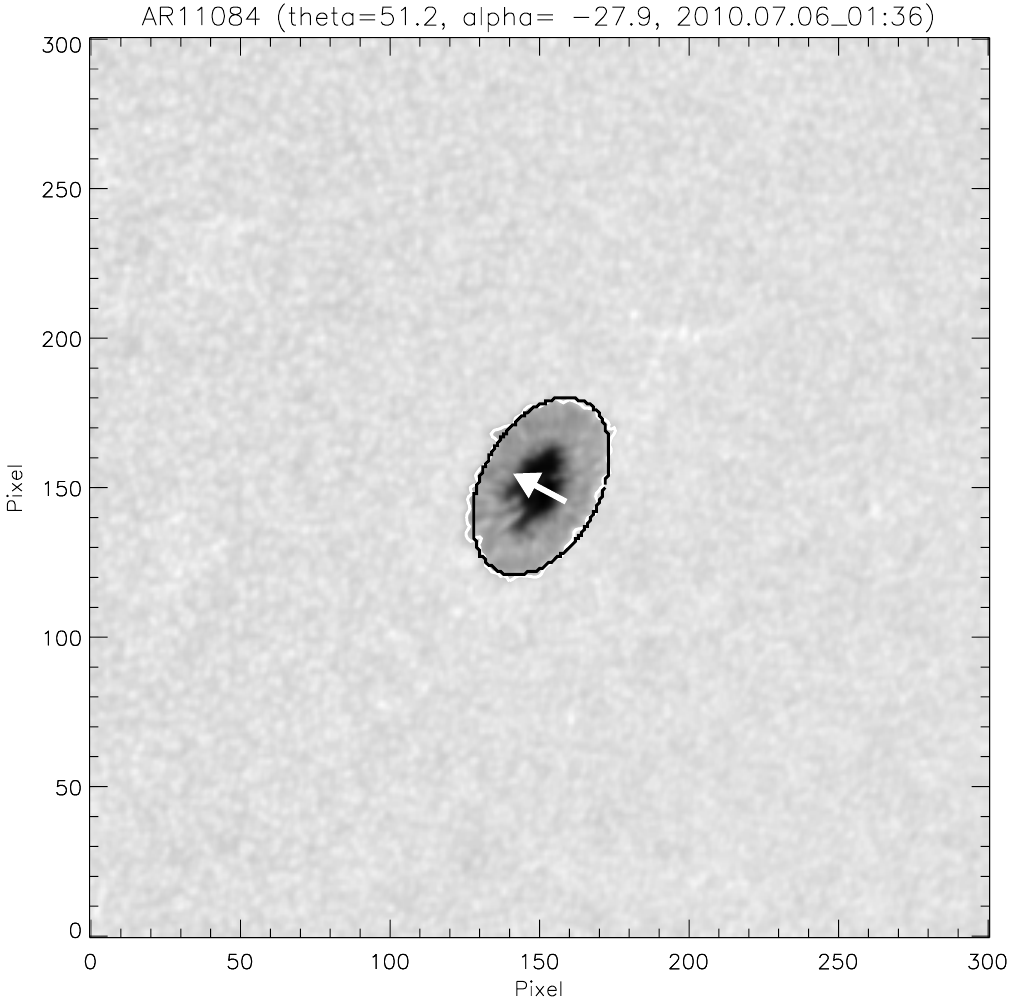}
\put(3,89){\textcolor{black}{\large\textbf{a)}}}
\end{overpic}
\begin{overpic}[trim = 8.5mm 6.3mm 19.5mm 3.4mm, clip, height=4.29cm]{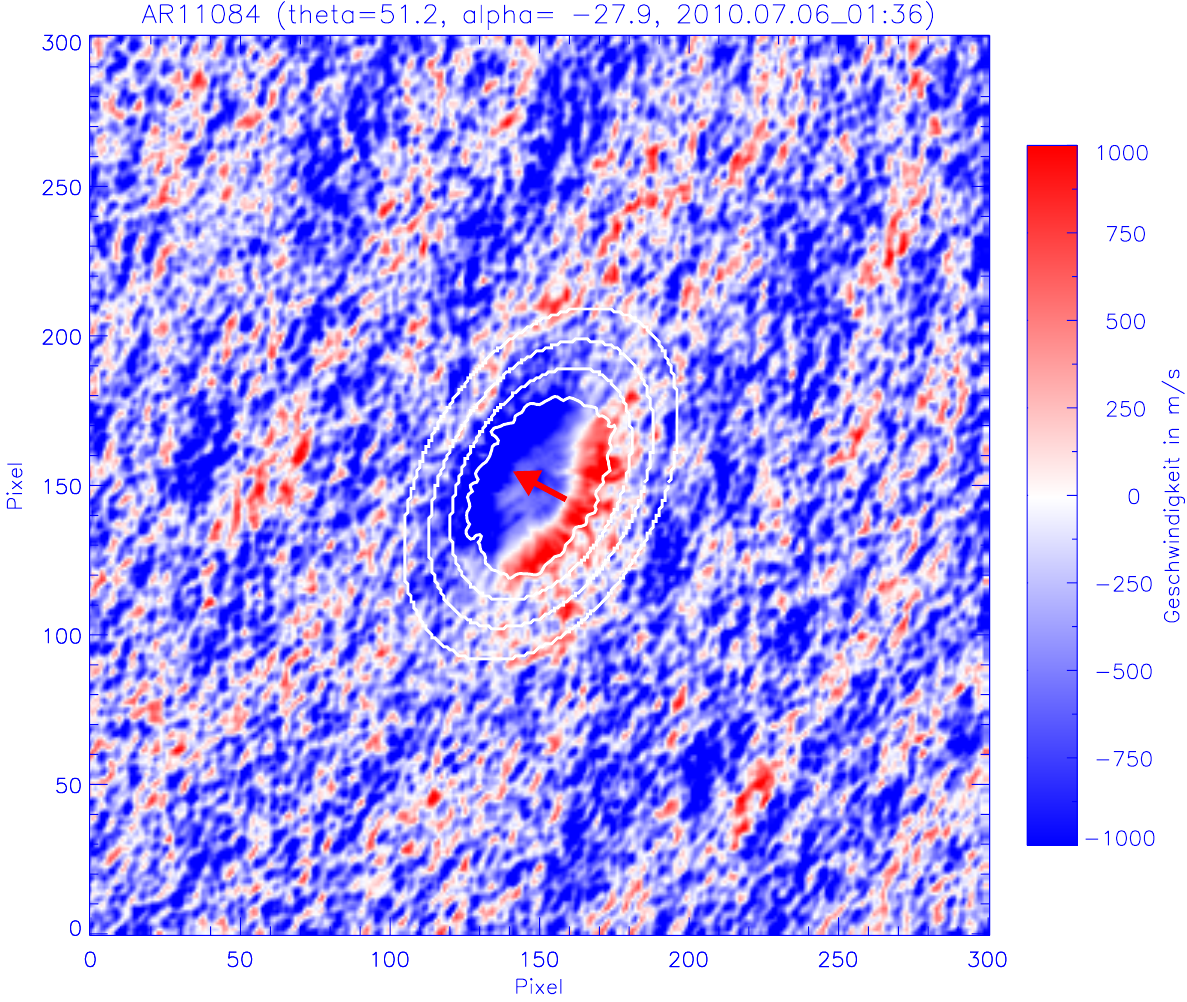}
\put(3,89){\textcolor{black}{\large\textbf{b)}}}
\end{overpic}
\begin{overpic}[trim = 16.7mm 10mm 26.2mm 5mm, clip, height=4.305cm]{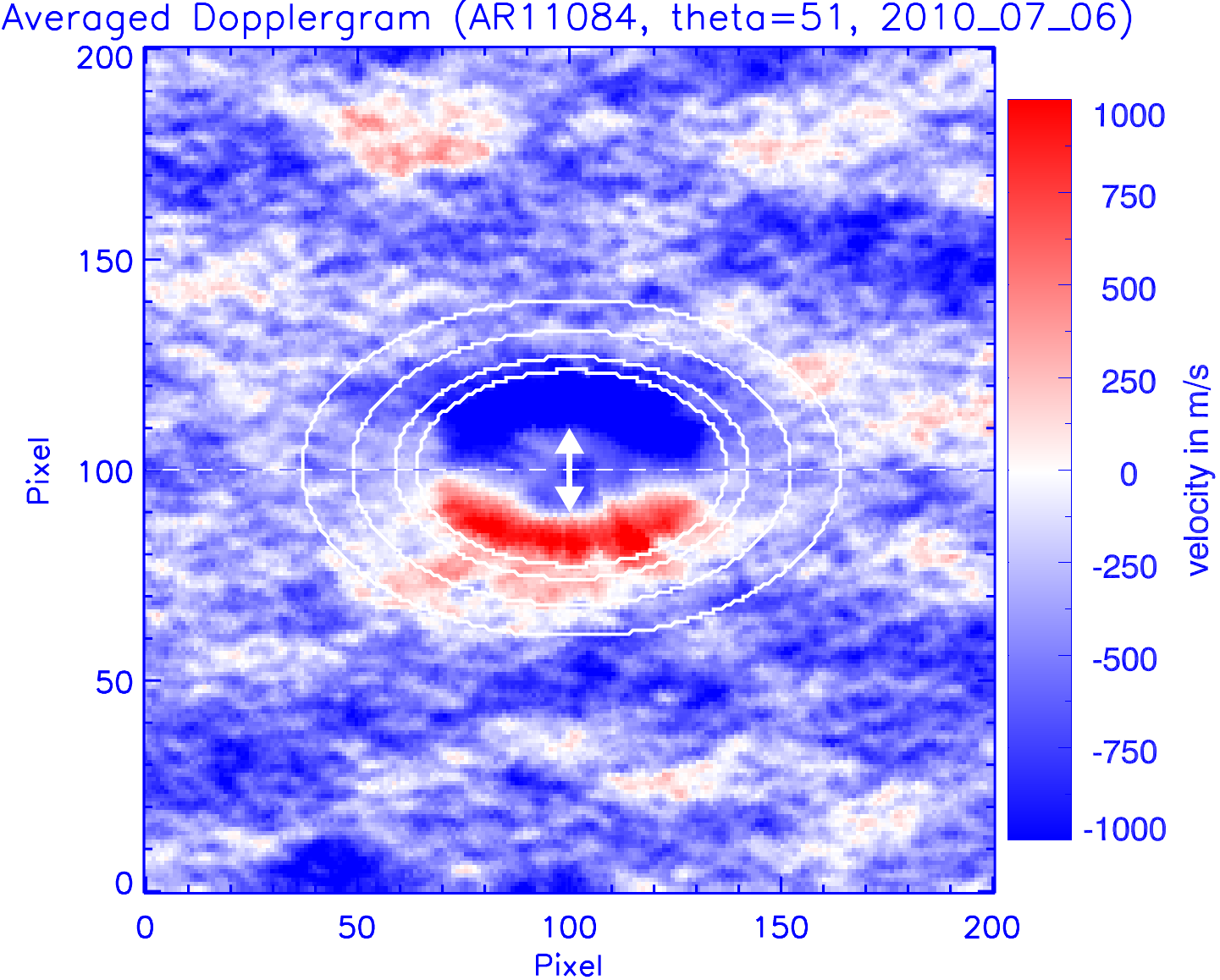}
\put(3,89){\textcolor{black}{\large\textbf{c)}}}      
\end{overpic}
\begin{overpic}[trim = 16.6mm 10mm 0mm 5.5mm, clip, height=4.3cm]{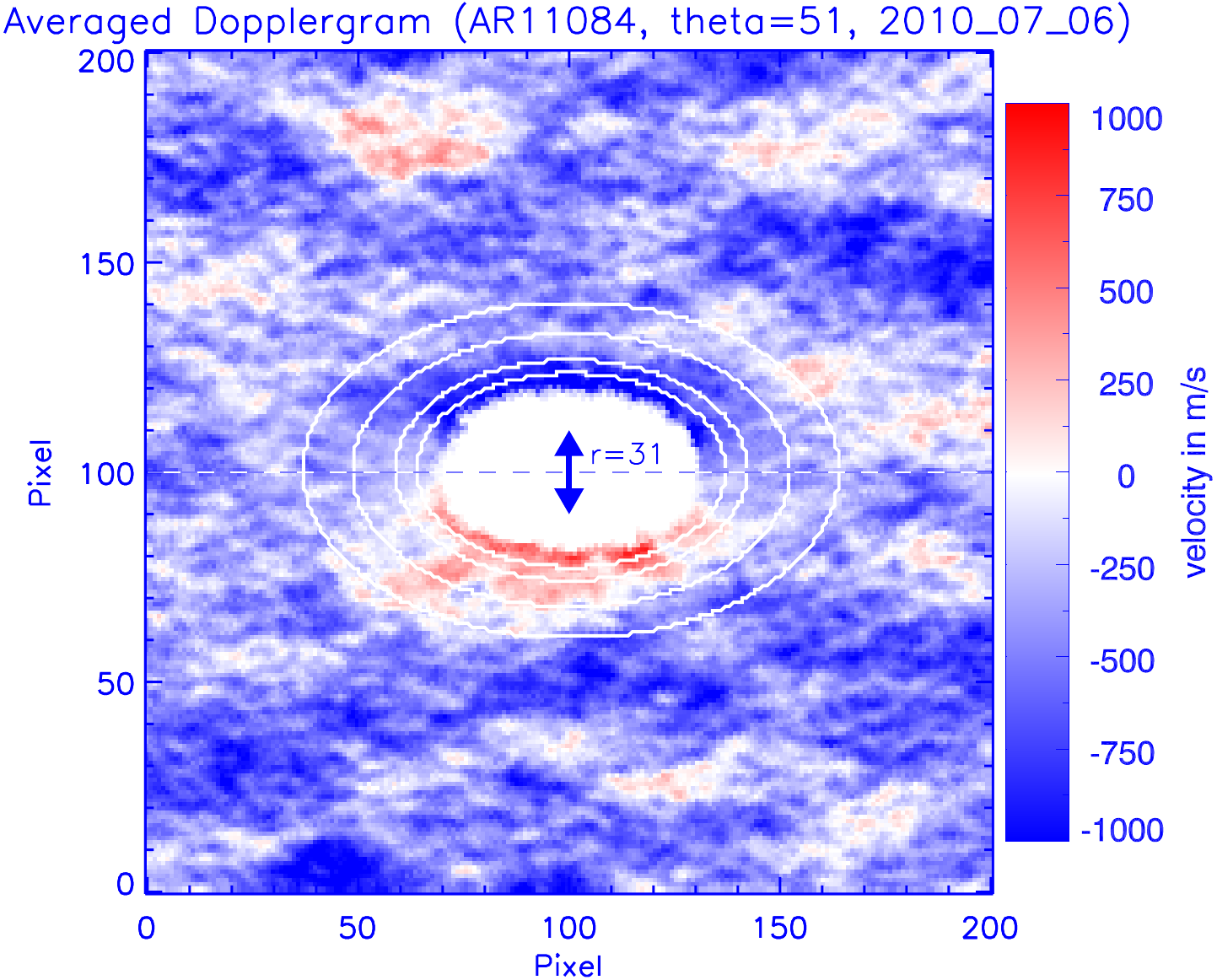}
\put(3,71){\textcolor{black}{\large\textbf{d)}}}      
\end{overpic}
\begin{overpic}[trim = 16.5mm 10mm 26mm 5mm, clip,height=4.3cm]{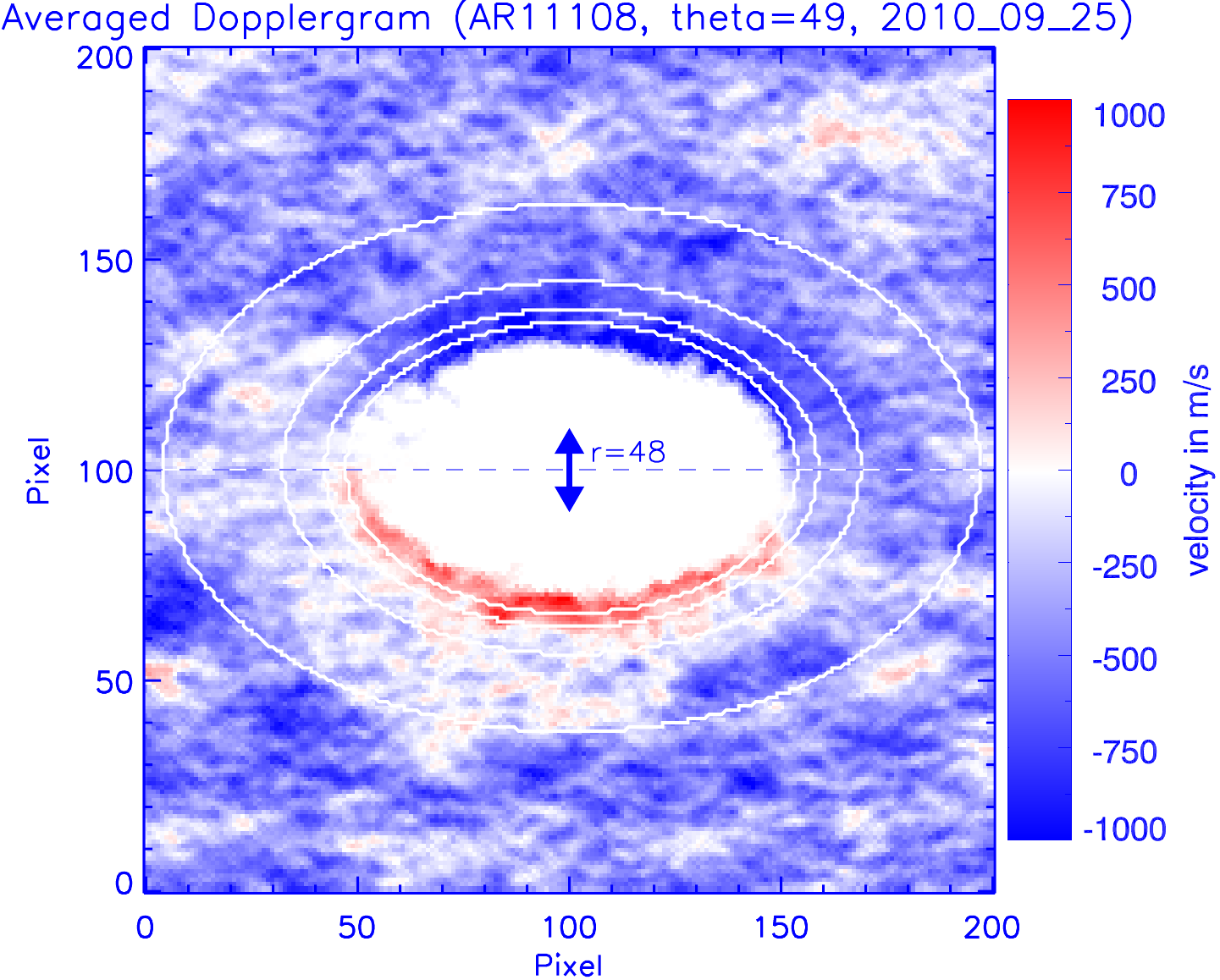}
\put(3,89){\textcolor{black}{\large\textbf{e)}}}      
\end{overpic}
\begin{overpic}[trim = 16.5mm 10mm 26mmmm 5mm, clip,height=4.3cm]{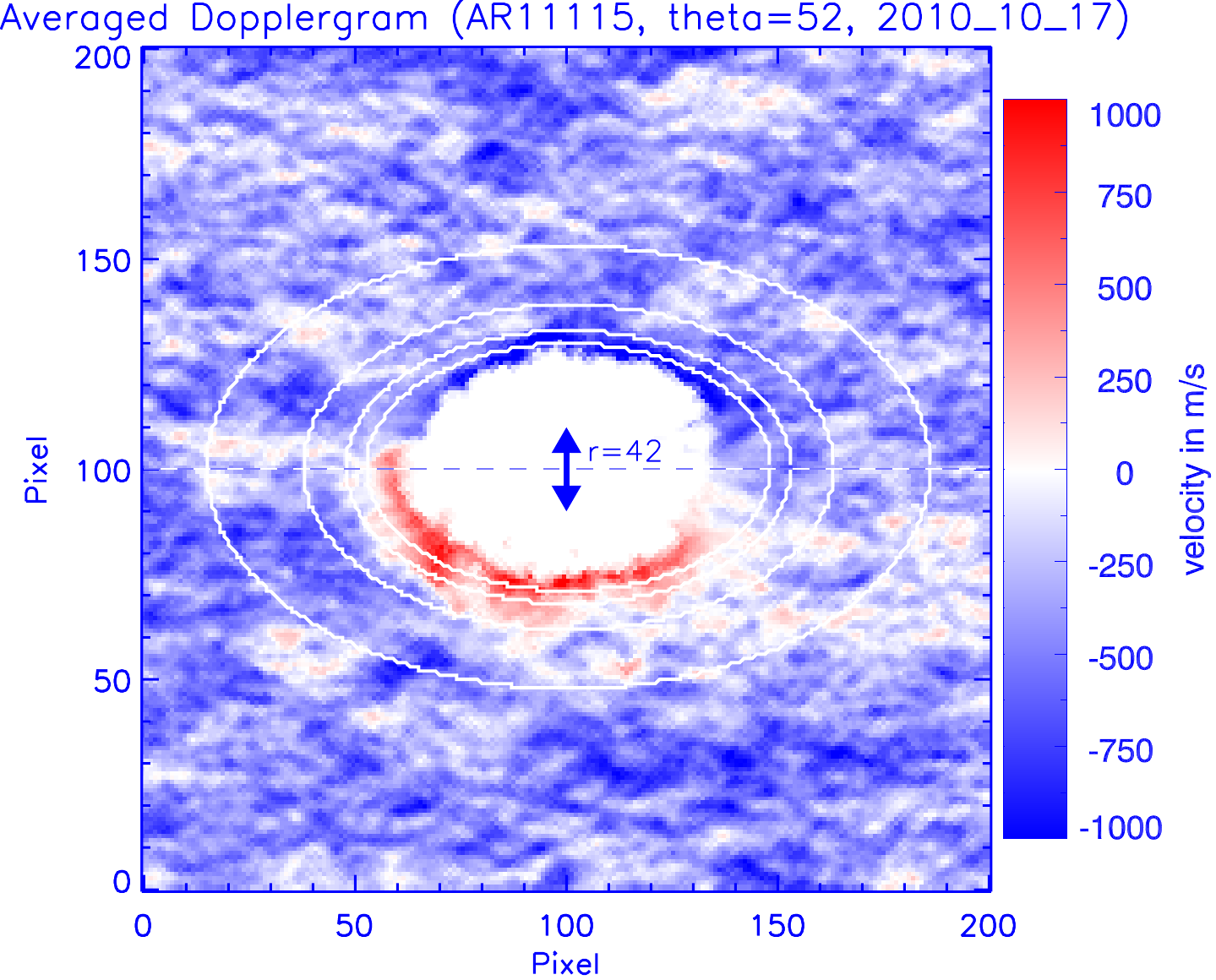}
\put(3,89){\textcolor{black}{\large\textbf{f)}}}      
\end{overpic}
\begin{overpic}[trim = 16.5mm 10mm 26mm 5mm, clip,height=4.3cm]{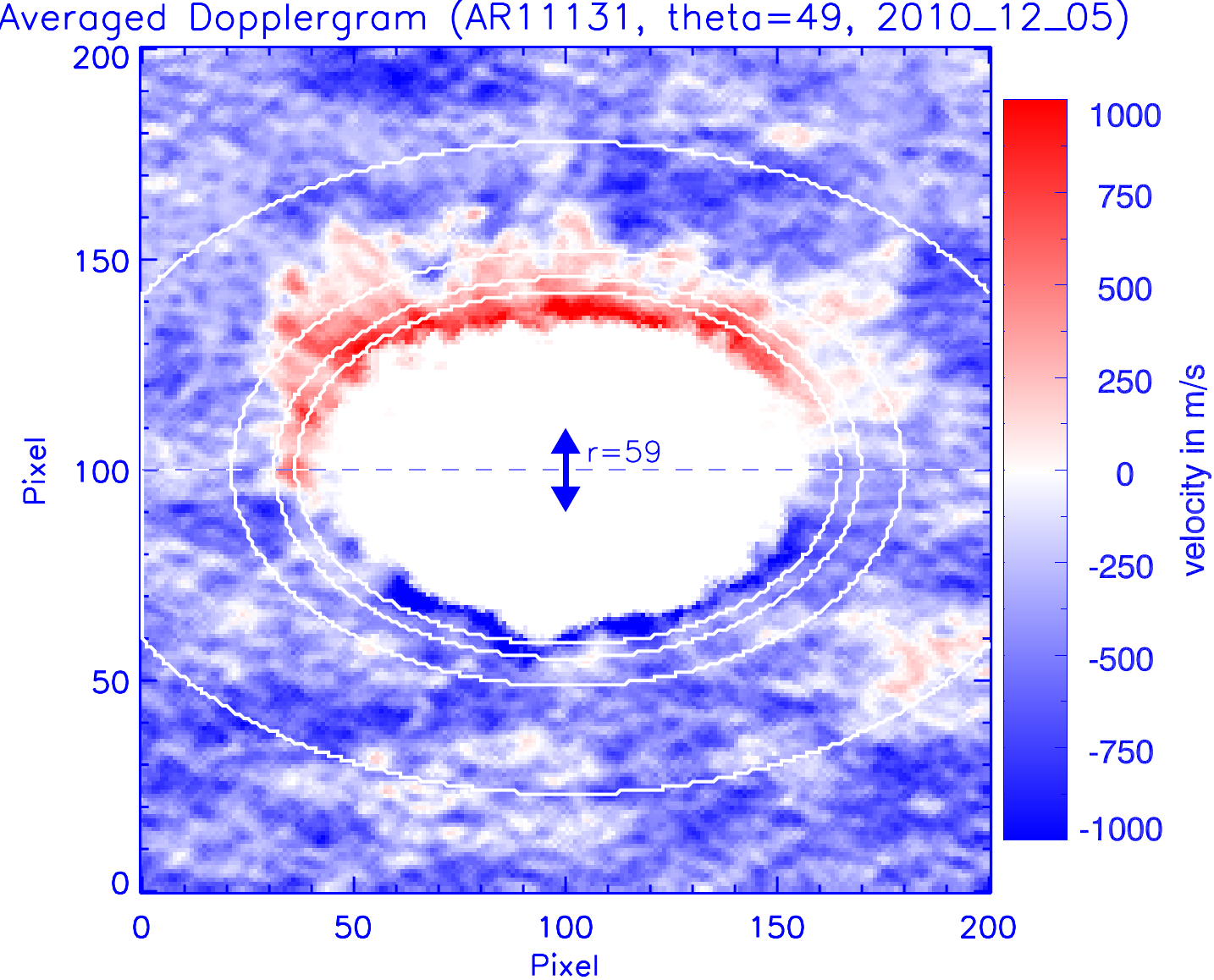}
\put(3,89){\textcolor{black}{\large\textbf{g)}}}      
\end{overpic}
\begin{overpic}[trim = 16.5mm 10mm 0mm 5mm, clip,height=4.3cm]{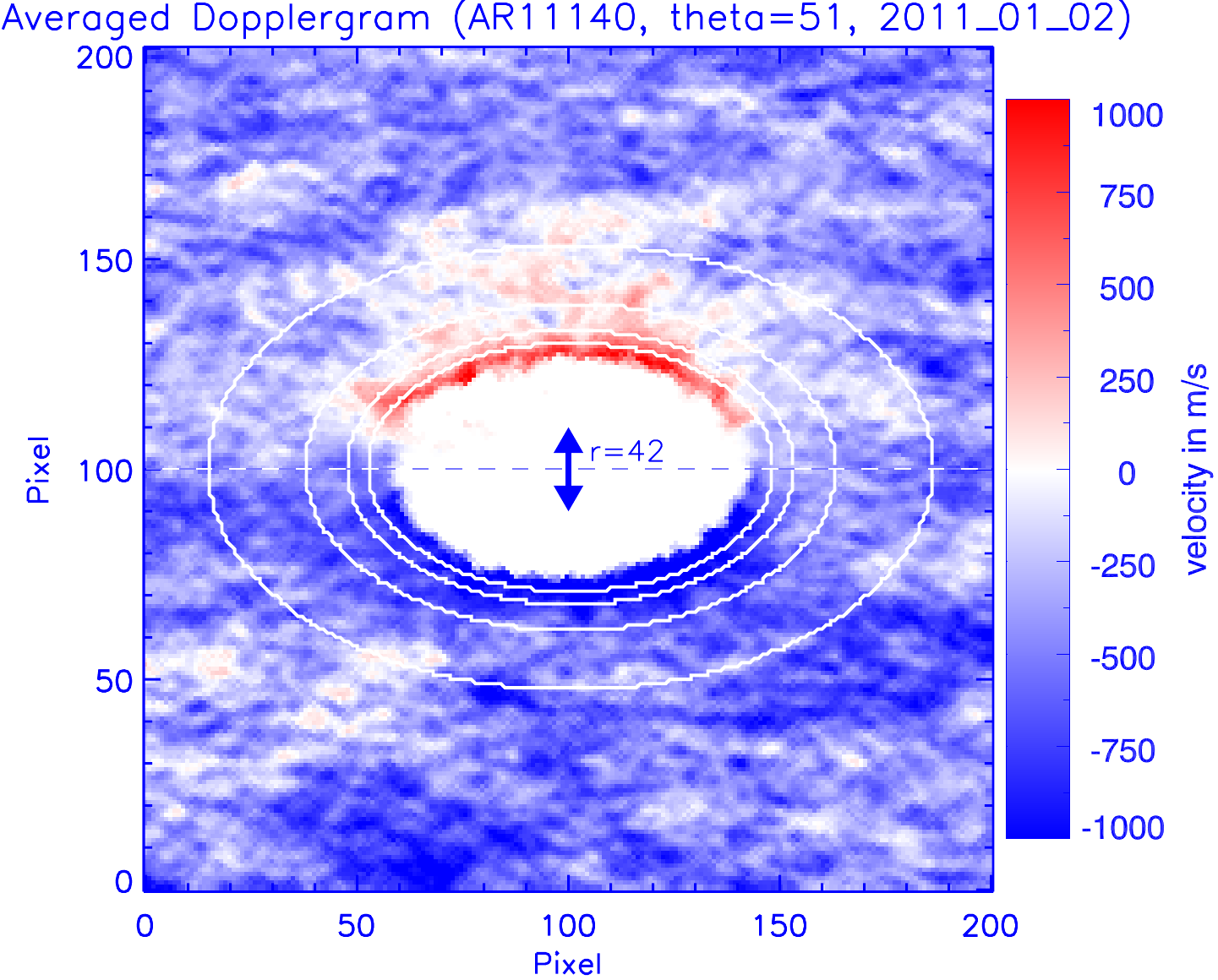}
\put(3,71){\textcolor{black}{\large\textbf{h)}}}      
\end{overpic}
\caption{\label{fig:dopcont}--- \textbf{a)} 720s intensity map and \textbf{b)} corresponding 720s velocity map, $\tilde{v}_{\rm dop}$, of AR11084 located at \mbox{$19^\circ {\rm S}, 48^\circ {\rm W}$}, respectively $\theta=51^\circ$, recorded 2010 July 6 at 01:36 UT (see Table~\ref{apptab}: No. 1b). An arrow is pointing to the \mytextbf{disk} center. The penumbral contour (white line), \mytextbf{the theoretical,} circular spot shape (black ellipse), and foreshortened circles with rising radii are displayed as well. The FOV size is 301\,px ($\approx150\arcsec$). -- \textbf{c)} \mytextbf{Aligned 3h average map, \mbox{$v_{\rm 3h}$, of AR11084}. -- \textbf{d\,--\,h)} Masked versions of spots No. 1b, 4a, 6a, 7a, 9a (Table~\ref{apptab}) at a FOV size of 201\,px ($\approx100\arcsec$) and white ellipses with $R_{\rm S}+x\,|_{x=5,10,20,R_{\rm S}}$. The velocities are displayed up to \mbox{$\pm1000\,{\rm m/s}$}.}}
\end{figure*}

\subsection{Data selection and preprocessing:}\label{sec:selection}
The moat flow is coupled to the presence of a penumbra and follows the direction of the penumbral filaments . Thus, we searched for sunspots with a fully-fledged, circular penumbra and found 31 applicable, single sunspots in a slight decay\footnote{Z\"urich Classification: Group H} in the time between June 2010 and January 2012. The selected sunspots vary in size between 9\,Mm and 22\,Mm and are listed in Table~\ref{apptab} of the appendix. A contour of the outer penumbra for each spot is determined by means of an intensity threshold for a spatially averaged image. The center of a sunspot is determined as the center-of-gravity of all points inside its penumbral contour.

To \mytextbf{analyze} predominantly horizontal flows with \mytextbf{LOS Doppler} maps, we selected the observing position of the sunspot center at heliocentric angles \mytextbf{of some} $\theta\!\approx\!50^\circ$. There, the LOS component of the horizontal flow is significant, while the calibration uncertainties that increase towards the solar limb are still small.  

Each sunspot FOV from the size of 301\,x\,301\,px is tracked by its averaged center, $(x_{\rm c},y_{\rm c})$, for 15 successive \mytextbf{720s intensity} maps \mytextbf{(Fig.~\ref{fig:dopcont}a)}, while the respective Doppler maps, $v_{\rm dop}(x,y)$, are calibrated by 
\begin{equation}
\tilde{v}_{\rm dop}(x,y)\!=\!v_{\rm dop}(x,y)-v_{\rm rot}(d,l,B_0)-v_{\rm res}(x,y)+v_{\rm off}(x,y)
\end{equation}
to \mytextbf{obtain} velocity maps, $\tilde{v}_{\rm dop}(x,y)$, \mytextbf{(Fig.~\ref{fig:dopcont}b)} free of all systematical velocity components (described in Sec.~\ref{sec:red})\mytextbf{, but} the surface convection\mytextbf{. } $v_{\rm off}$ describes the offset of the convective blueshift and its CLV (Fig.~\ref{fig:clv}c) to values around \mbox{$-400$\,m/s} at the chosen observing position. 
 
 The sunspot and flow properties are stable within three hours . To diminish 5-min oscillations and granulation, we averaged the tracked and aligned sunspot FOVs (201\,x\,201\,px) of all the 15 successive velocity maps and refer to this \mytextbf{3h average} as \mbox{$v_{\rm 3h}$} \mytextbf{(Figs.\ref{fig:dopcont}c--h)}. Likewise, the \mytextbf{received LOS velocities} are adapted for an average heliocentric angle, $\theta$. For better illustrating the MF, the \mytextbf{3h average} is shown \mytextbf{here} with a masked umbra and penumbra.

We studied the weekly evolution of the observed components by tracking 20 sunspots near the eastern limb to a second proper position ($\theta\!\approx\!50^\circ$) after some 6 to 8 days and compared the results of both \mbox{$v_{\rm 3h}$} analysis. As this was not possible for 11 \mytextbf{of 31 spots}, we achieve a sample of 51 \mytextbf{velocity maps}. Tracking three long-lasting sunspots across the far side of the Sun, they reappear at the eastern \mytextbf{front side} \mytextbf{(Table~\ref{apptab}: No. 4$\,\rightarrow$\,6, 7\,$\rightarrow$\,9, 11\,$\rightarrow$\,13; Figs.~\ref{fig:dopcont}e--h)} which allows us to study the monthly evolution.

\subsection{Method of flow analysis}
We \mytextbf{analyze}  \mytextbf{flow fields of} fully calibrated velocity maps, \mbox{$v_{\rm 3h}$}, by applying a method that determines the azimuthally averaged flow properties \citep{schliche+schmidt2000}. In this way, we assume axially \mytextbf{symmetrical} flow fields for circular sunspots with a fully developed penumbra. The foreshortening effects of a circular sunspot at a certain heliocentric angle, $\theta$, cause the elliptical shape as shown for AR11084 \mytextbf{(Figs.~\ref{fig:dopcont}a--d)}. To observe the spatial velocity dependency of the flow fields, we use an automatic procedure that generates ellipses as foreshortened circles (with radii, $r$, in px) based on \mytextbf{the} heliocentric angle and rotated according to their position on the disk. The LOS velocities, $v^{\rm LOS}(r,\phi)$, along the ellipses are read out according to their circular angle, $\phi$, as shown in Fig.~\ref{fig:sinfit} for distances \mbox{$r\!=\!R_{\rm S}\,+\,5$} (upper panel) and  \mbox{$r\!=\!R_{\rm S}\,+\,10$} (lower panel) \mytextbf{from the spot center}. The sunspot radius is automatically determined by the minimal deviation \mytextbf{of the ellipse} to the penumbral contour for every \mytextbf{720s intensity} map and averaged for the \mytextbf{3h period}. A manual fine adjustment was added for irregularities. 
The \mytextbf{LOS velocities}, $v^{\rm LOS}(r,\phi)$, were fitted by \mytextbf{a} sine function:
\begin{equation}
\quad v^{\rm LOS}(r,\phi)=v^{\rm LOS}_{\parallel,0}(r) \cdot \sin\phi + v^{\rm LOS}_{\perp}(r)\label{eqsinfit}\qquad.
\end{equation}
The amplitude, $v^{\rm LOS}_{\parallel,0}$, is the LOS component of the horizontal flow velocity 
\begin{equation}
\quad v_{\parallel,0}(r)=v^{\rm LOS}_{\parallel,0}(r)/\sin\theta\quad,\label{eq:veff}
\end{equation} at the heliocentric angle, $\theta$. The axial offset, $v^{\rm LOS}_{\perp}$, is the vertical flow component.

\subsection{Moat flow}
We \mytextbf{analyze}  the moat flow in the vicinity of the sunspots for the fully calibrated velocity maps, \mbox{$v_{\rm 3h}$} (listed  in Table~\ref{apptab}).

\subsubsection{Moat flow velocity}
The moat flow is \mytextbf{an outwardly directed flow, which is predominantly horizontal. } We present the flow field analysis for the velocity map, \mbox{$v_{\rm 3h}$}, of AR11084 (Figs.~\ref{fig:dopcont}c--d; Table~\ref{apptab}: No. 1b). The sunspot in the FOV center with an average spot radius of $R_{\rm S}\!=\!31\,{\rm px}\!\approx\!11.2\,{\rm Mm}$ at $\theta=51^{\circ}$ exhibits a distinct moat flow surrounding the sunspot with \mytextbf{LOS velocities} up to $-1000\,{\rm m/s}$ for the \mytextbf{blueshifted} \mytextbf{side} facing the disk center and $+1000\,{\rm m/s}$ for the \mytextbf{redshifted} \mytextbf{limb side} indicated by an arrow. The outer vicinity of the sunspot features \mytextbf{supergranular} flow cells. 

\begin{figure}[htbp]
\includegraphics*[trim = 0mm 0mm 0mm 0mm, clip,height=5.6cm]{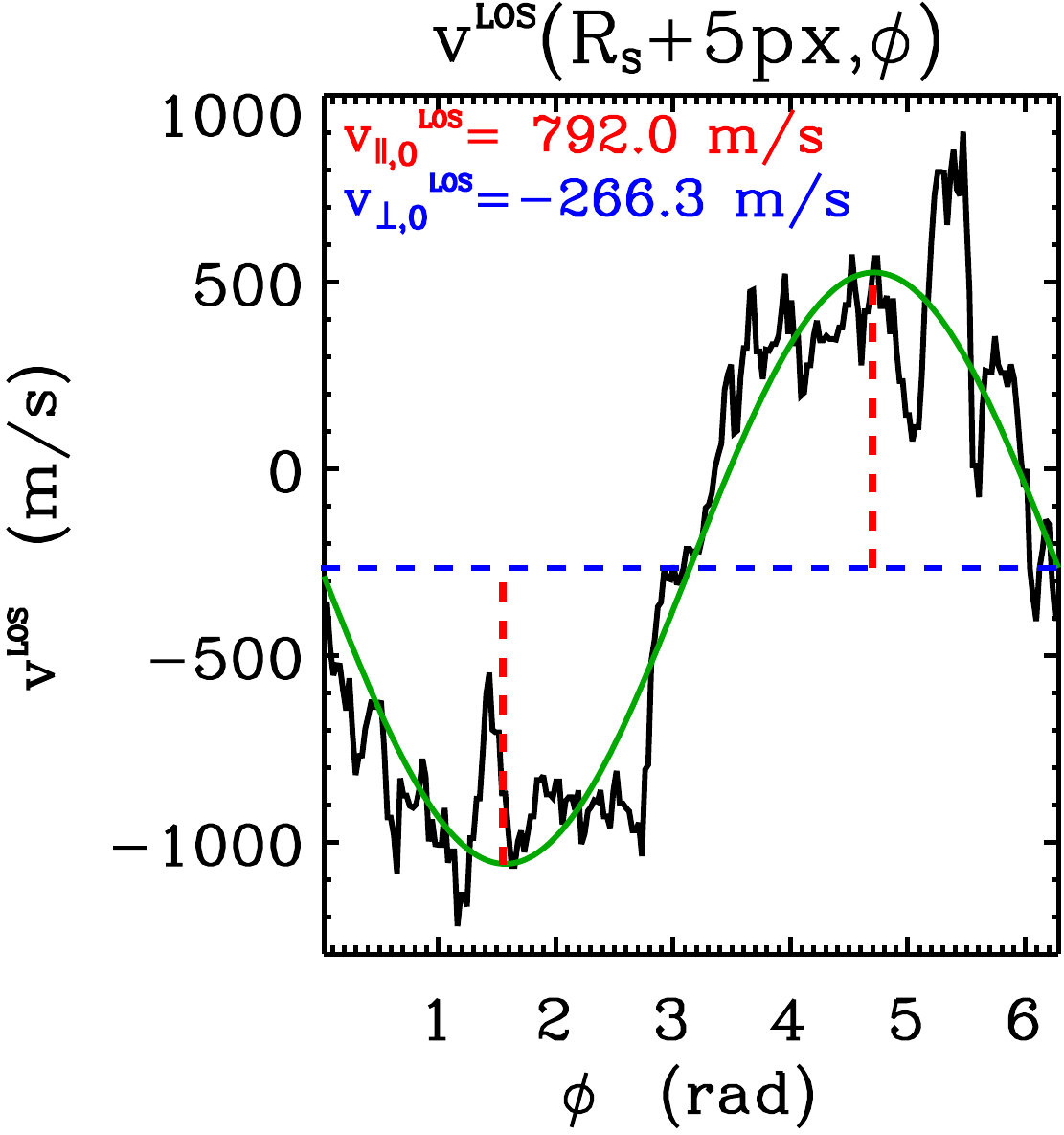}
\includegraphics*[trim = 34.0mm 0mm 0mm 0mm, clip,height=5.6cm]{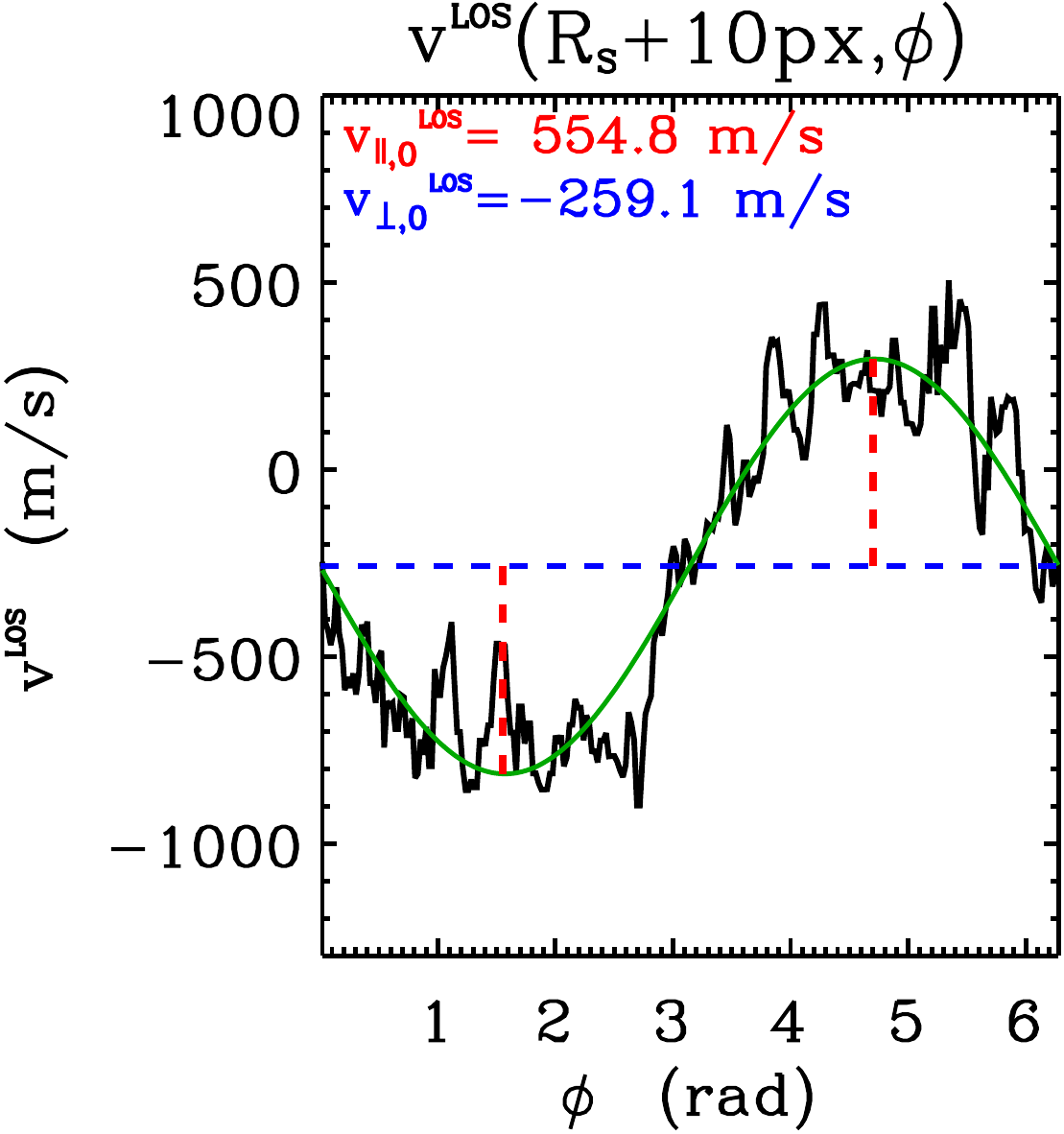}
\caption{\label{fig:sinfit}--- Angular LOS velocities, $v^{\rm LOS}(r,\phi)$, in m/s along the elliptical projection for $\theta\!=\!51^{\circ}$ of circles with radius \mbox{$r=R_{\rm S}+5{\rm px}$} (left panel) and \mbox{$r=R_{\rm S}+10{\rm px}$} (right panel) fitted by a sine curve (green, solid line) according to Eq.~\ref{eqsinfit}\mytextbf{. T}he amplitude, $v^{\rm LOS}_{\parallel,0}(r)$, and offset, $v^{\rm LOS}_{\perp}(r)$ \mytextbf{are marked by red and blue dashed lines}.}
\end{figure}

\begin{figure}[htbp]
\includegraphics*[trim = 0mm -4mm 0mm 0mm, clip,width=8cm]{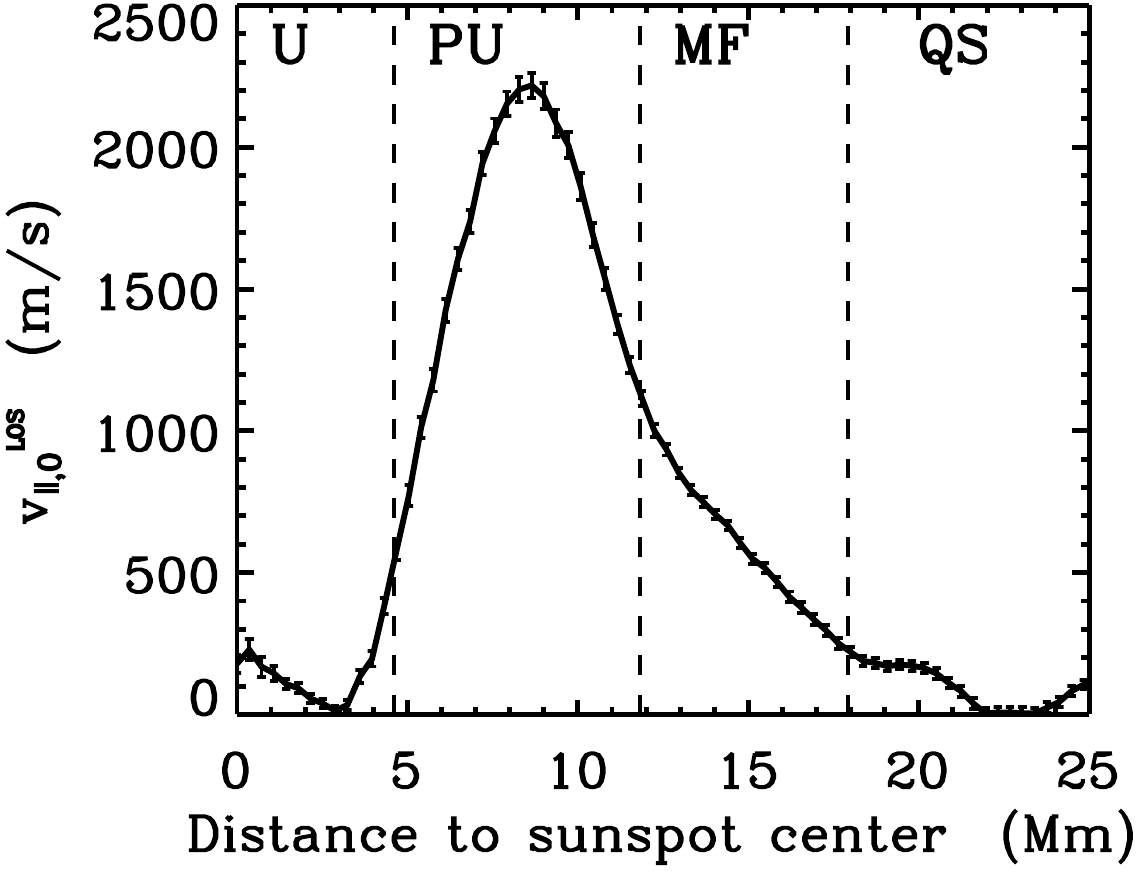}
\includegraphics[width=7.8cm]{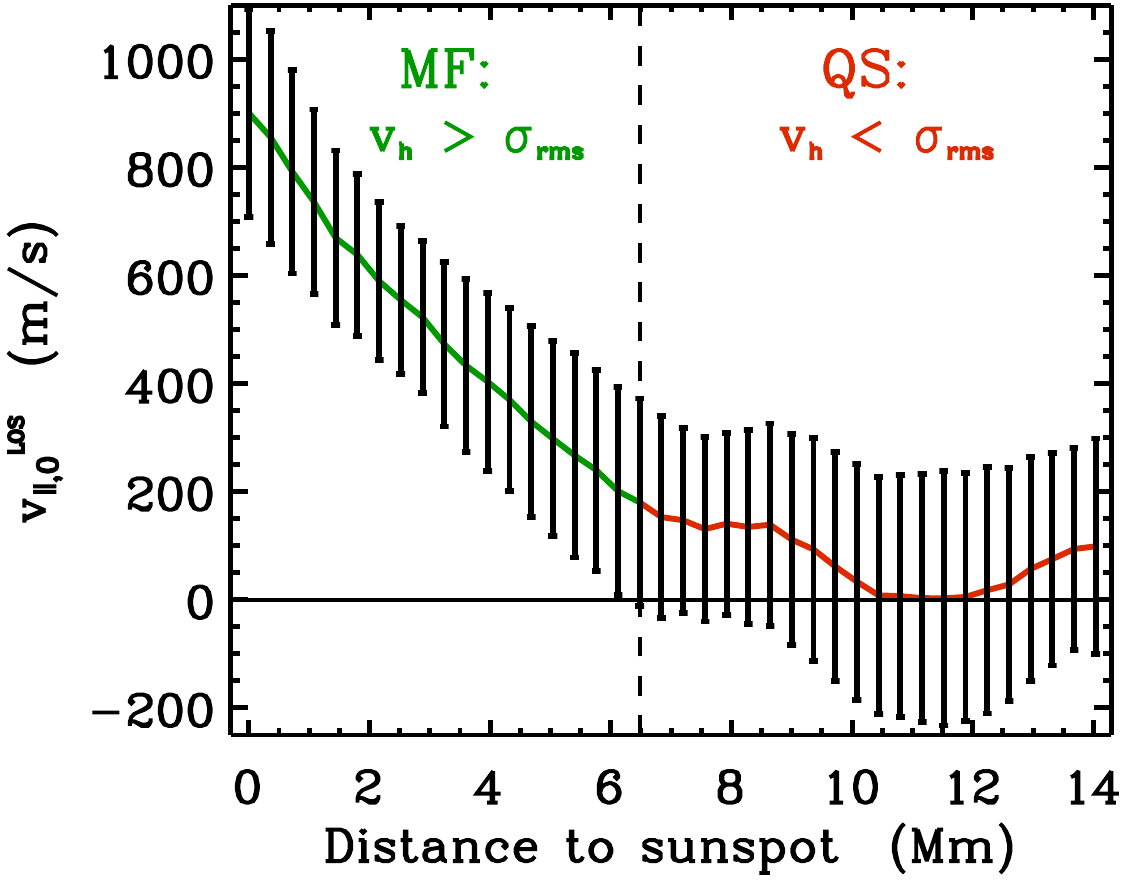}
\caption{\label{fig:vh}--- Upper panel: Horizontal flow velocities, $v_{\parallel,0}(r)$, in m/s for the distance, $r$, from the spot center in Mm\mytextbf{.  Error bars represent the standard deviation. Dashed vertical lines mark the boundaries between umbra (U), penumbra (PU), and moat flow region (MF) and the quiet Sun (QS)}. --- Lower panel: Horizontal \mytextbf{LOS velocity}, $v^{\rm LOS}_{\parallel,0}(r)$, in m/s of the MF (green curve) and the quiet Sun (red curve) against the distance from the spot in pixels with \mytextbf{$\sigma_{\rm rms}$ (see Eq.~\ref{eq:RMS}) as error bars}.}
\end{figure}

Comparing the velocity components of the MF in a distance of 5\,px and 10\,px from the sunspot (see Fig.~\ref{fig:sinfit})
\begin{eqnarray*}
&&v^{\rm LOS}_{\parallel,0}(R_{\rm S}+5{\rm px})=-792\,\pm\,17\,{\rm m/s}\\
&&v^{\rm LOS}_{\parallel,0}(R_{\rm S}+10{\rm px})=-555\pm12\,{\rm m/s}\\
&&v^{\rm LOS}_{\perp}(R_{\rm S}+5{\rm px})=-266\pm12\,{\rm m/s}\\
&&v^{\rm LOS}_{\perp}(R_{\rm S}+10{\rm px})=-259\pm9\,{\rm m/s}\quad,
\end{eqnarray*}
the most prominent result is the decrease in $v^{\rm LOS}_{\parallel,0}$ by more than 230\,m/s within these 2\,Mm. The upper panel of Fig.~\ref{fig:vh} shows the continuous run of the horizontal flow velocities, $v_{\parallel,0}(r)$, at $\theta=51^\circ$ \mytextbf{(see Eq.~\ref{eq:veff})}, from the spot center to a distance of 25\,Mm. The MF has a strong monotone decrease with increasing distance from $1150\,{\rm m/s}$ just outside the penumbra to 600\,m/s after additional 3\,Mm and 250\,m/s after 6\,Mm. 

The vertical flow component of MF and EF depends on the choice of the rest frame. Our rest frame is determined by the convective \mytextbf{blueshift} of \mytextbf{nonmagnetic} convection. Since the \mytextbf{MF} region is not \mytextbf{completely} free of magnetic fields, it may change the convective \mytextbf{blueshift} by an unknown amount. Therefore, no accurate analysis of the vertical MF component, $v^{\rm LOS}_{\perp}(r)$, can be given. Also, the inclination of the MF could not be reliably determined owing to the dependence on the vertical flow component \citep{schliche+schmidt2000}. However, on the assumption that no effect is caused by magnetic fields, we can subtract the convective blueshift \mbox{$v_{\rm b}(r)\!\approx\!-400\,{\rm m/s}$} from $v^{\rm LOS}_{\perp}(r)$ and \mytextbf{obtain} absolute vertical velocities of $\tilde{v}^{\rm LOS}_{\perp}(r)\!=\!0\ldots 200\,{\rm m/s}$. According to that, the slight \mytextbf{redshift} \mytextbf{indicates} a downward-directed MF. In comparison with horizontal MF velocities up to $v^{\rm LOS}_{\parallel,0}\!=\!1000\,m/s$, the vertical velocity is \mytextbf{an order of} magnitude less, and therefore it has \mytextbf{a} minor impact on the absolute flow velocity. 

\subsubsection{Moat flow extension}
To \mytextbf{detect} the size of the \mytextbf{MF region}, we define the criteria \mbox{$v^{\rm LOS}_{\parallel,0}\le \sigma_{\rm rms}$} to indicate the end of the detectable MF, whereas the root mean square, $\sigma_{\rm rms}$, is calculated for each distance, $r$, to the spot by 
\begin{equation} 
\sigma_{\rm rms}(r)=\sqrt{\frac{1}{n} \cdot \sum_{i=1}^n \Big(v^{\rm LOS}(r,\phi_{i})-f(\phi_{i})\Big)^2}\qquad. \label{eq:RMS}
\end{equation}
\mytextbf{In the lower panel of Fig.~\ref{fig:vh}, we display $\sigma_{\rm rms}(r)$ as error bars to the horizontal LOS velocities.} When $\sigma_{\rm rms}(r)$, i.e. the velocity fluctuation around the sine fit, $f(\phi_{i})$, with circular angles, $\phi_{i}$ (see Fig.~\ref{fig:sinfit}), exceeds the fitted amplitude,  \mbox{$v^{\rm LOS}_{\parallel,0}$}\mytextbf{, we define this distance as the end of the MF region}. For the case under consideration, we \mytextbf{obtain} an \mytextbf{MF}, which extends 6.5\,Mm into the vicinity of the sunspot. This study of the MF extension highly depends on \mytextbf{$\sigma_{\rm rms}$}, which is about $180{\rm\,m/s}$ for the \mytextbf{3h average} and which becomes smaller for longer periods. But, these longer periods cannot guarantee the stability of the observed components. We therefore believe that our method delivers a good estimate for the MF extension.

\subsection{Evershed flow}\label{sec:obs:ef}
\mytextbf{Also} the well-known penumbral Evershed flow is a predominantly horizontal flow\mytextbf{. In our analysis we assume it as axially symmetrical. In Fig.~\ref{fig:vh} (upper panel), we display the radial dependence of the horizontal flow component of AR11084 (see Fig.~\ref{fig:dopcont}c) based on \mbox{$v_{\rm 3h}$}.} The inner penumbral boundary is \mytextbf{determined as} the average extension of the umbral darkening in \mytextbf{all 720s intensity} maps. We observe EF velocities, $v_{\parallel,0}(r)$, increasing from the inner penumbra to a maximum value of \mbox{2210\,m/s} at a distance \mbox{$r\!=\!8.5\,{\rm Mm}\!\approx\!0.76\!\cdot\!R_{\rm S}$} followed by a decrease to \mytextbf{about \mbox{1150\,m/s} at} the outer penumbra. \mytextbf{Based on our analysis method, some penumbral filaments will reach into the defined MF region or end earlier. Therefore, the velocities at the transition region have impacts on both flows, and therefore the transition is smooth as seen in Fig.~\ref{fig:vh}.} But, following the run of the curve, a \mytextbf{kink} between the radially decreasing EF and MF velocities is evident at the outer penumbral boundary \mytextbf{. The mere existence of the kink is a hint at the} separation of \mytextbf{both} flow regions .

\section{Results and discussion}\label{sec:res}
\paragraph{Overview:}In this section, we compare the analysis results (see Sect.~\ref{sec:obs}) for all 51 \mytextbf{velocity maps \mbox{$v_{\rm 3h}$}} listed in Table~\ref{apptab}\mytextbf{. We} discuss the interactions between the sunspots and their moat flows. The weekly evolution of the observed components is studied for a sample of 20 sunspots, whereas the monthly evolution is \mytextbf{analyzed} for three sunspots. We compare our findings with recent studies and \mytextbf{analyze}  the relation of MF properties with sunspot properties. Our discussion \mytextbf{is focused} on whether the \mytextbf{EF and MF} are related and on what we can learn about the physical origin of the MF.

\begin{figure}[htbp]
\centering
\includegraphics*[trim = 0mm -5mm 0mm 0mm, clip,width=7.5cm]{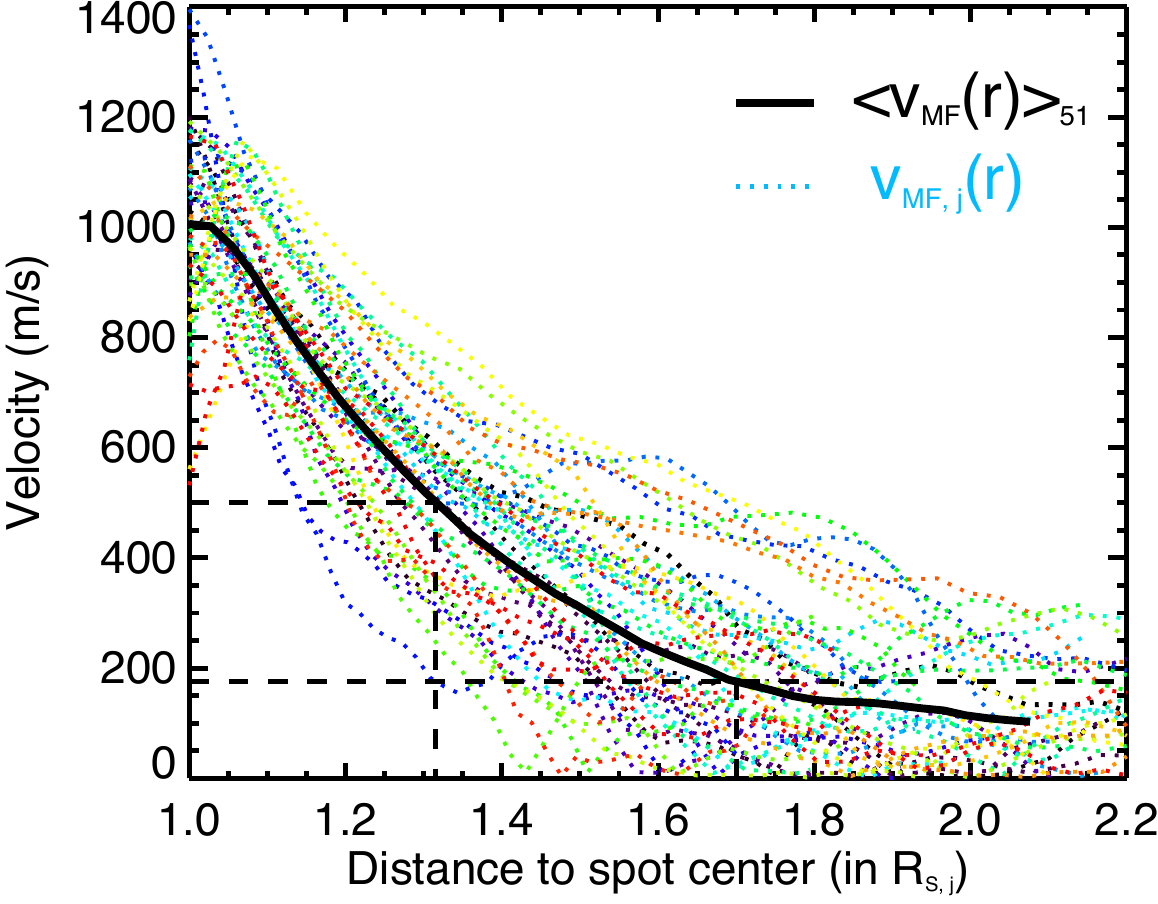}
\includegraphics*[width=7.5cm]{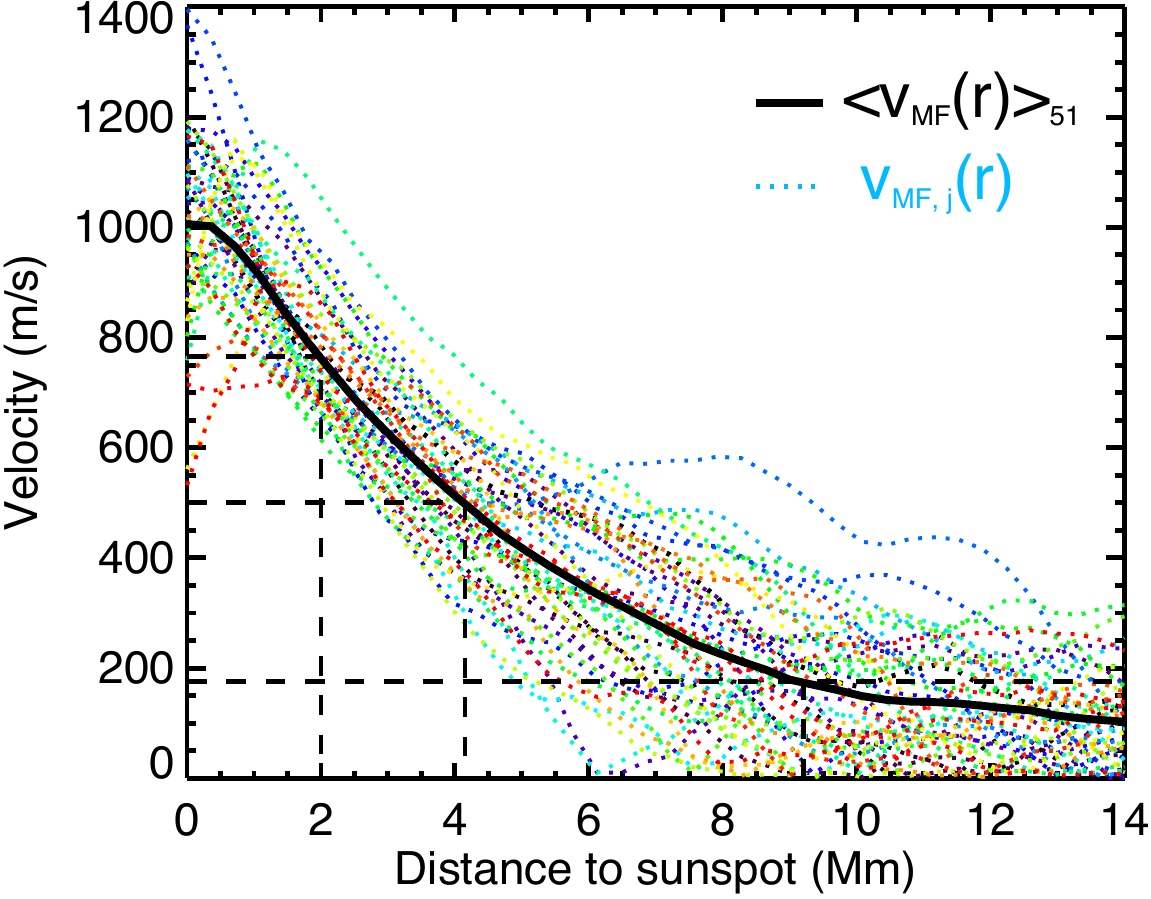}
\caption{\label{fig:vhcompare}--- Horizontal MF velocities, $v_{\rm MF}(r)$, in m/s of 51 observed \mytextbf{3h velocity maps} against the distance \mytextbf{from} the spot center in spot ratio (upper panel) and the distance to the sunspot in Mm (lower panel). The calculated average (black bold curve) is \mytextbf{drawn}.} 
\end{figure}

\subsection{Horizontal moat flow velocity: $v_{\rm MF}(r)$}
\mytextbf{In the appendix, Figs.~\ref{appfig1} and \ref{appfig2}, we display the radial dependence of the horizontal flow component for all 51 sunspot maps. Since this is the first time that the EF and MF are \mytextbf{analyzed}  coherently for more than only a few spots, it is very remarkable to find that all our spots behave very similarly. The radial dependence is qualitatively the same for all spots. At first glance, only the sunspot size and the maximum of the EF velocity vary. As described in Sec.~\ref{sec:obs:ef} the kink between the velocity decrease of the EF in the penumbra (PU) and the MF at the spot boundary is recognizable for all sunspots in the radial dependences displayed in Figs.~\ref{appfig1} and \ref{appfig2}. In the next step, we examine the flow properties of all spots in detail and compare them.} 

The moat flow was identified for all selected sunspots  as a predominantly horizontal, axially symmetrical flow starting just outside of the penumbra with velocities depending on the distance, $r$, to the sunspot. Figure~\ref{fig:vhcompare} displays the strong decay of the MF velocities, \mbox{$v_{\rm MF}(\tilde{r})\!=\!v_{\parallel,0}(\tilde{r})$} for all 51 \mytextbf{analyzed}  MF regions based on \mbox{$v_{\rm 3h}$}. The upper panel shows the decrease with respect to the distance from the spot center in the radius ratio, $\tilde{r}\!=\!\frac{R_{\rm S}+r}{R_{\rm S}}$. The lower panel plots the MF velocities against the distance, $\tilde{r}\!=\!r\!=\!0,\ldots,14\,{\rm Mm}$, from the penumbral boundary. The monotone decrease from maximum velocities of \mbox{$v_{\rm MF}(r\!=\!0)\!=\!800\ldots1200\,{\rm m/s}$} just outside the sunspot is largely similar for all sunspots. The average run, $\langle v_{\rm MF}(\tilde{r})\rangle_{\rm 51spots}$, starts with $1000\,{\rm m/s}$ just beyond the penumbral boundary at $r\!=\!0\,{\rm Mm}$ and decreases to approx. $500\,{\rm m/s}$ within $r\!=\!4\,{\rm Mm}$ or $\tilde{r}\!=\!1.3\!\cdot\!R_{\rm S}$. The average velocity falls below the typical rms values, $\sigma_{\rm rms}\!\approx\,180{\rm m/s}$, at $r\!=\!9.2\,{\rm Mm}$ or $\tilde{r}\!=\!1.7\!\cdot\!R_{\rm S}$. Consequently, the MF velocity highly depends on the location within the MF region. We compare both panels of Fig.~\ref{fig:vhcompare} and notice that the runs according to the radius ratio in the left hand panel are spread wider than according to the sheer distance from the spot. Consequently, the MF appears to be independent of the spot size. 

This result is in line with recent studies describing the MF velocity on the surface ranging between $v_{\rm MF}\!=\!500\ldots1000\,{\rm m/s}$ \citep{balthasar+etal1996,rimmele1997,dominguez+etal2007}, whereas the continuous velocity decrease with increasing distance from the spot boundary has not been measured before.

\subsection{Maximum Evershed flow velocity: $v_{\rm EF}$}
\begin{figure*}[htdp]
\begin{center}
\begin{tabular}{|c|c|c|c|}
\hline
\multicolumn{1}{|l|}{\large{a)}}&\multicolumn{1}{l|}{\large{b)}}&\multicolumn{1}{l|}{\large{c)}}&\multicolumn{1}{l|}{\large{d)}}\\
\includegraphics[height=4.35cm]{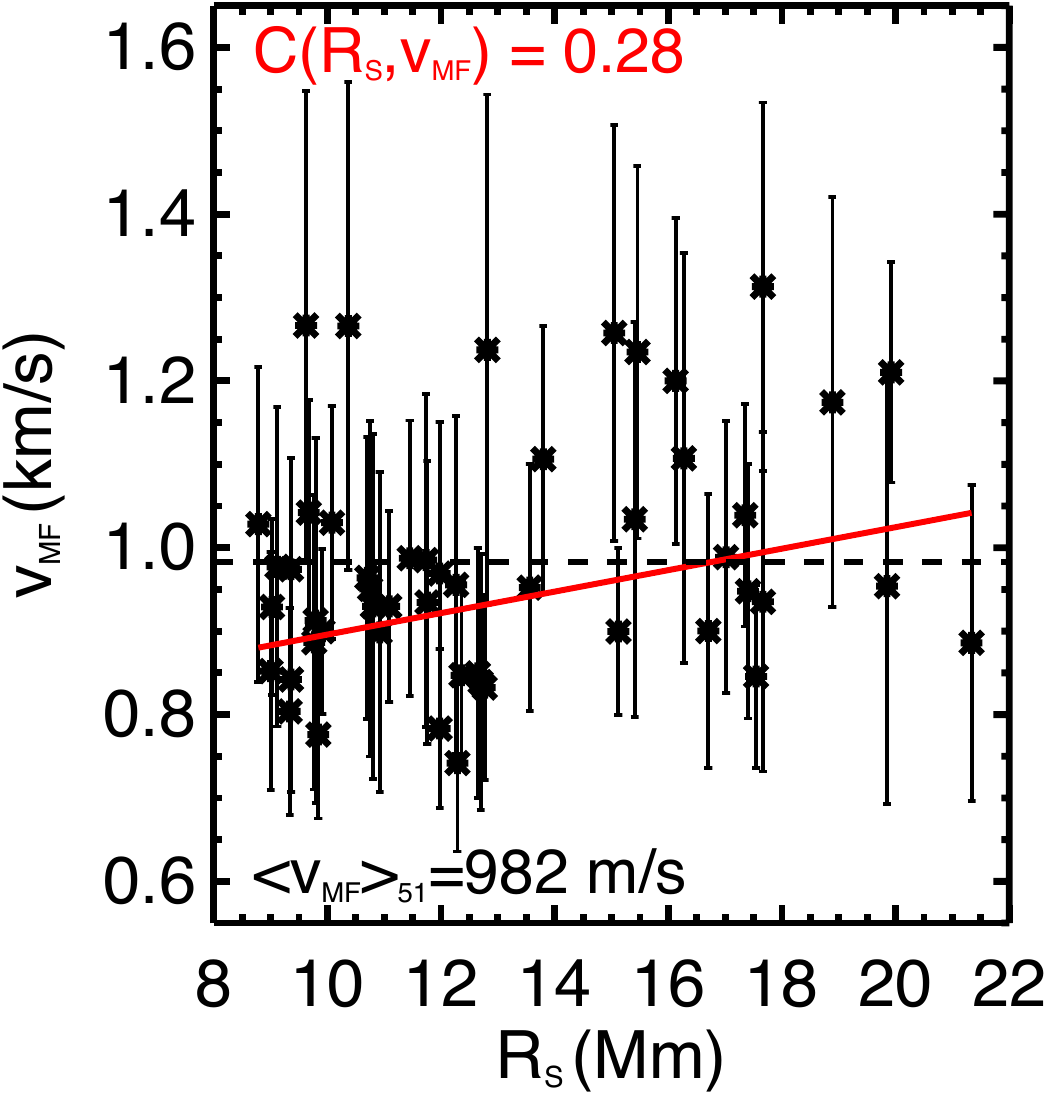}
&\includegraphics[height=4.35cm]{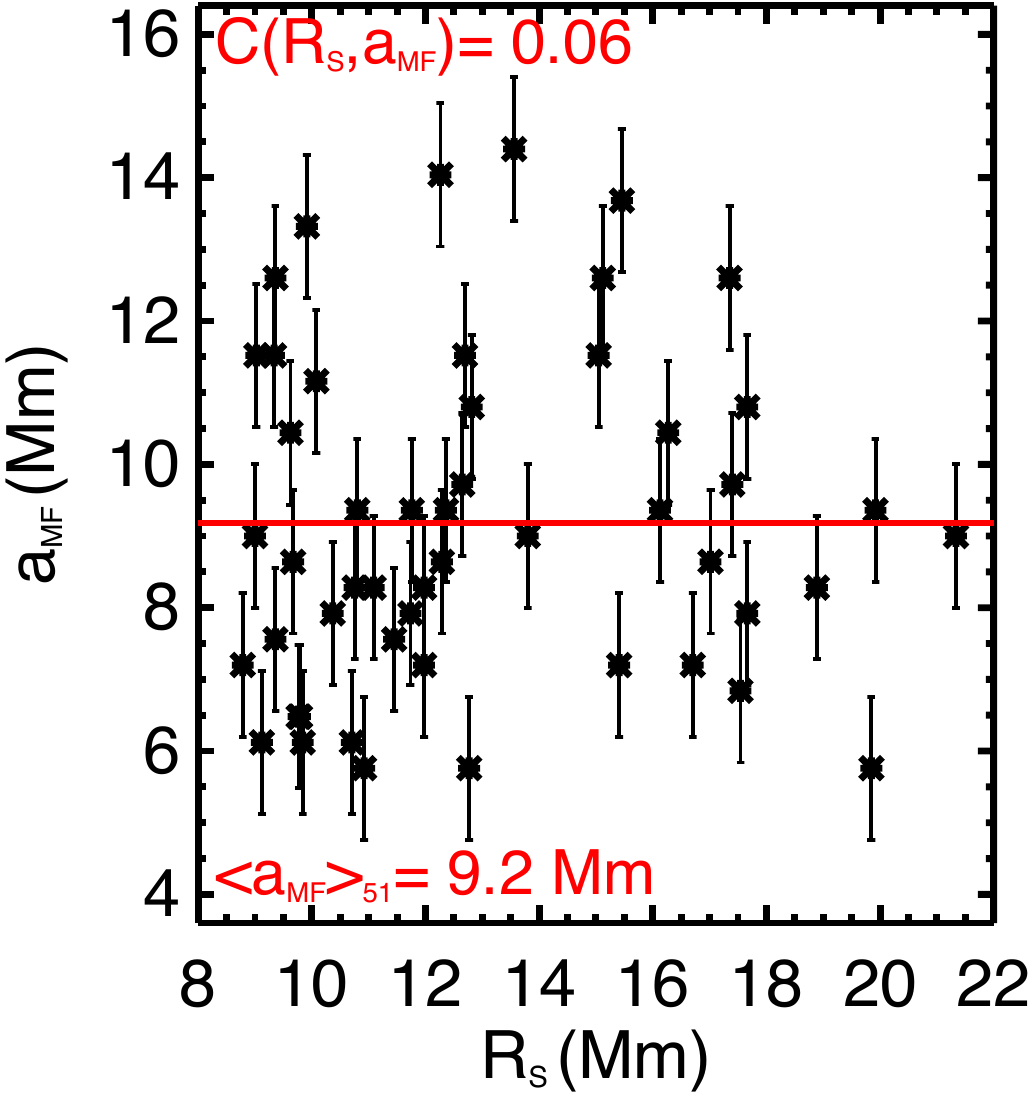}
&\includegraphics[height=4.35cm]{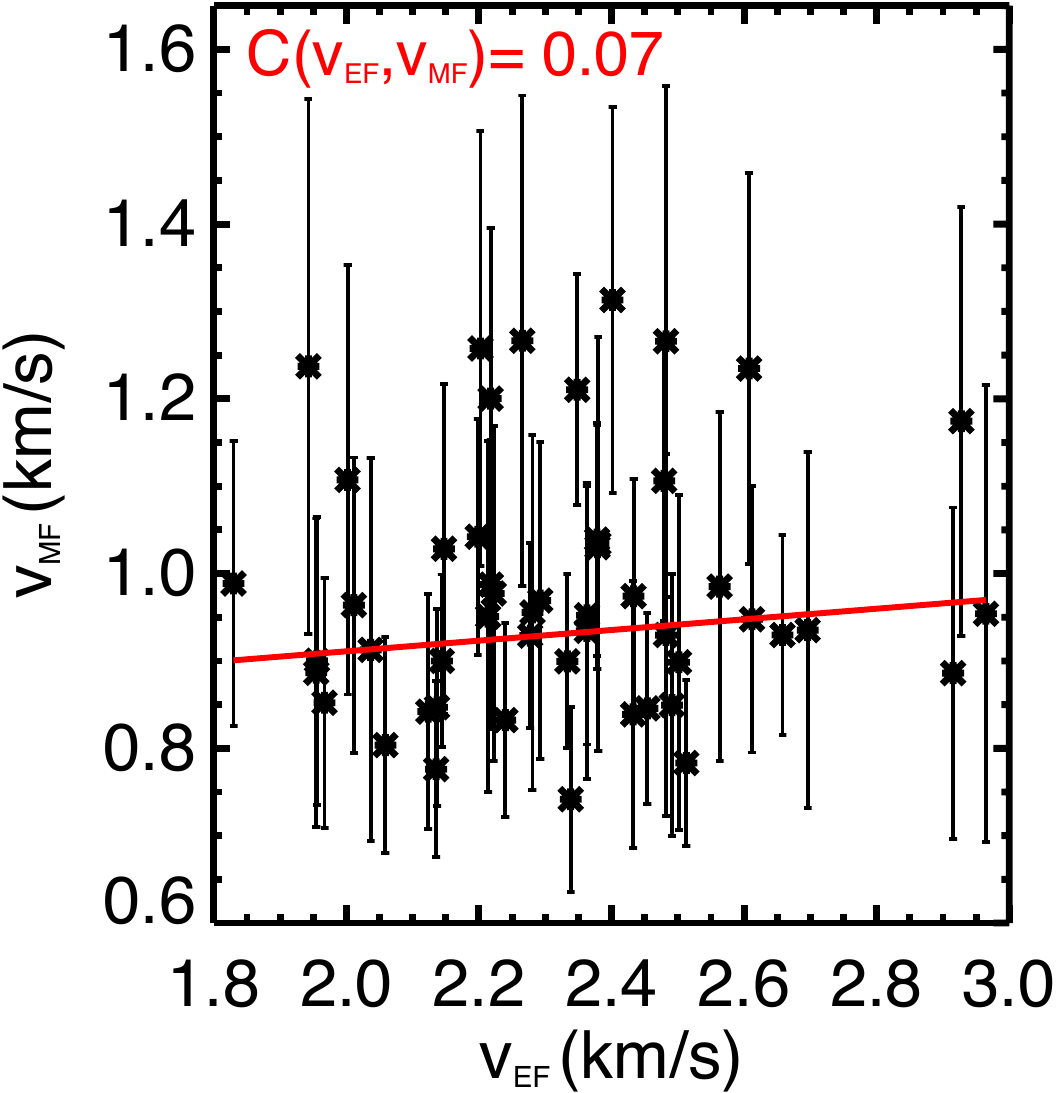}
&\includegraphics[height=4.35cm]{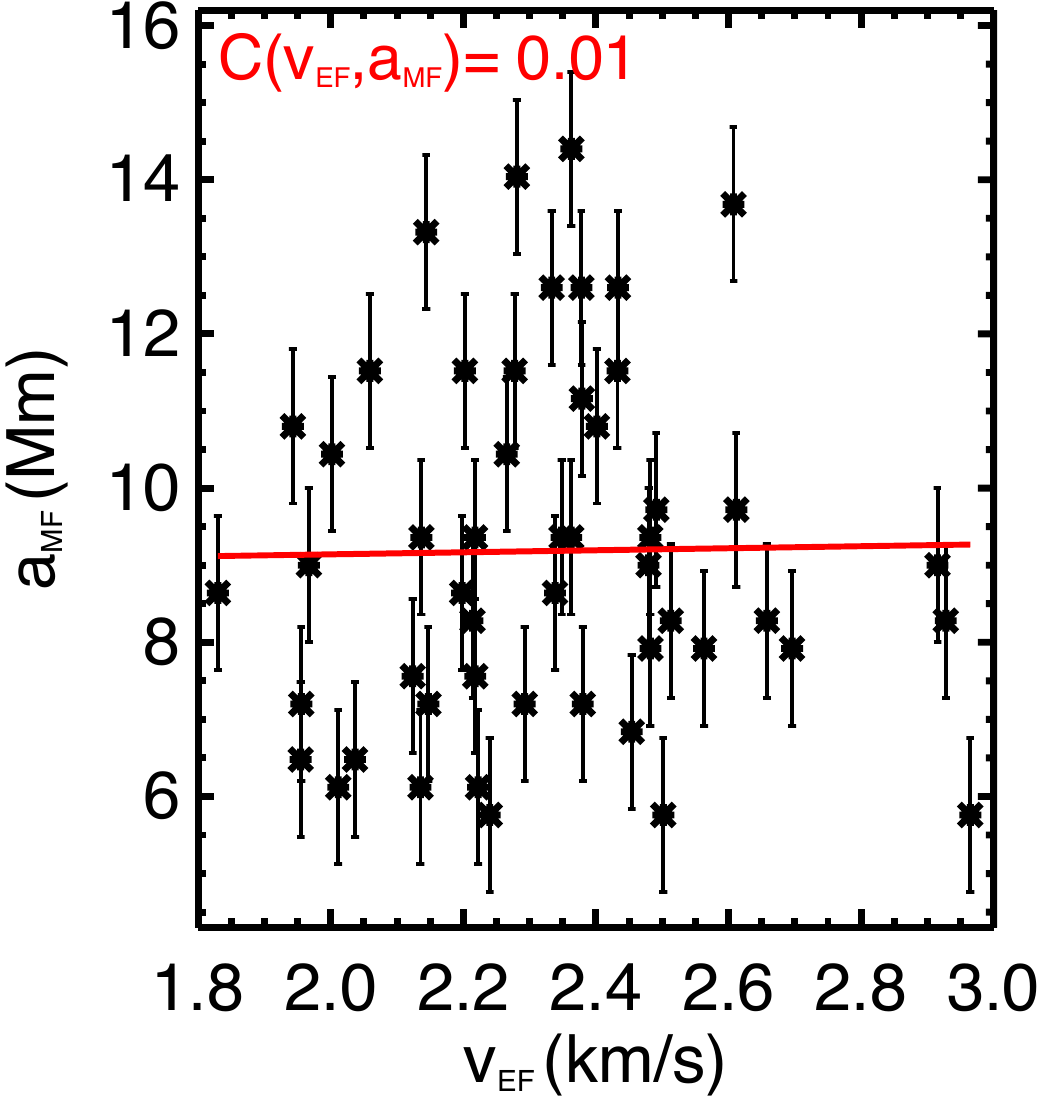}
\\
\hline
\multicolumn{1}{|l|}{\large{e)}}&\multicolumn{1}{l|}{\large{f)}}&\multicolumn{1}{l|}{\large{g)}}&\multicolumn{1}{l|}{\large{h)}}\\
\includegraphics[height=4.35cm]{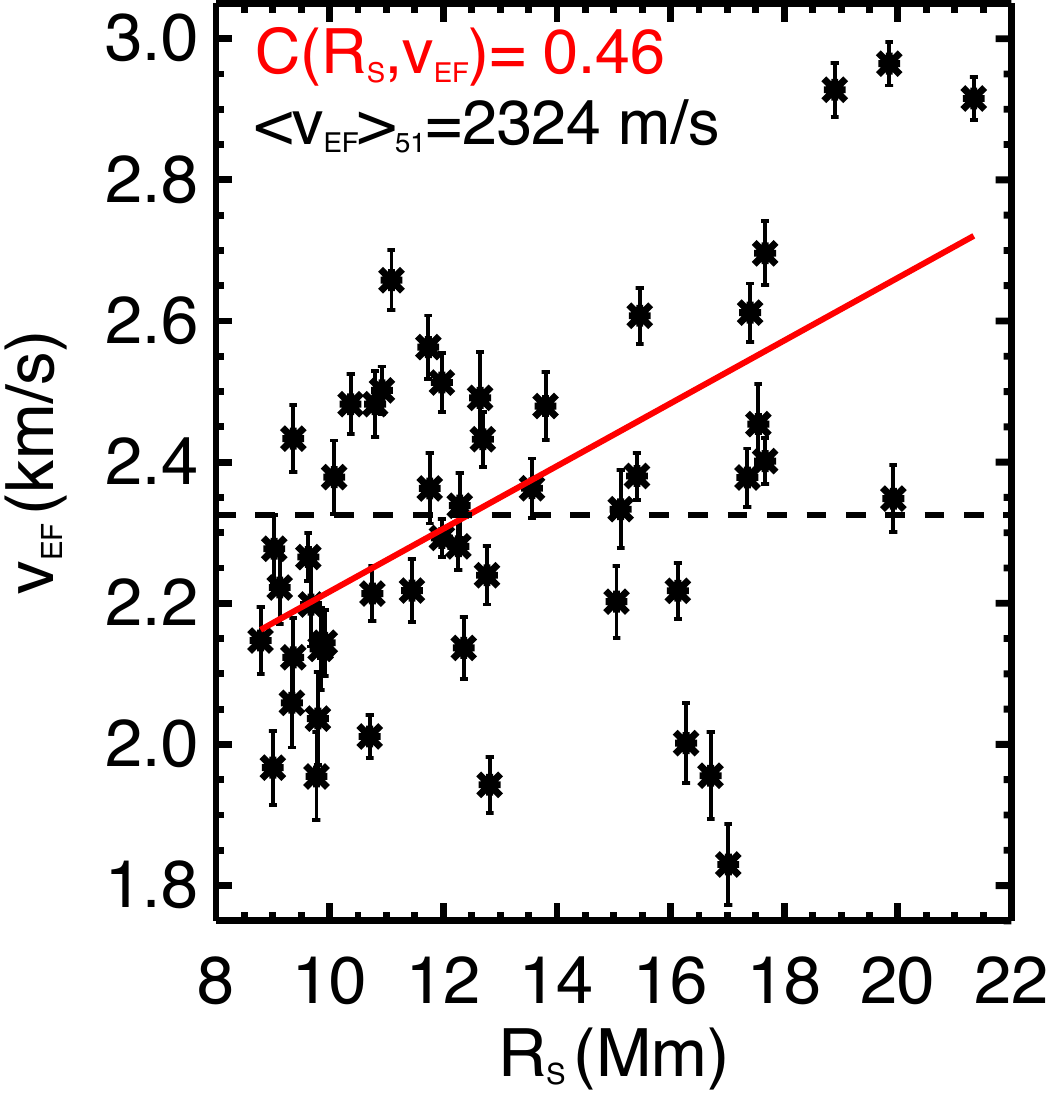}
&\includegraphics[height=4.35cm]{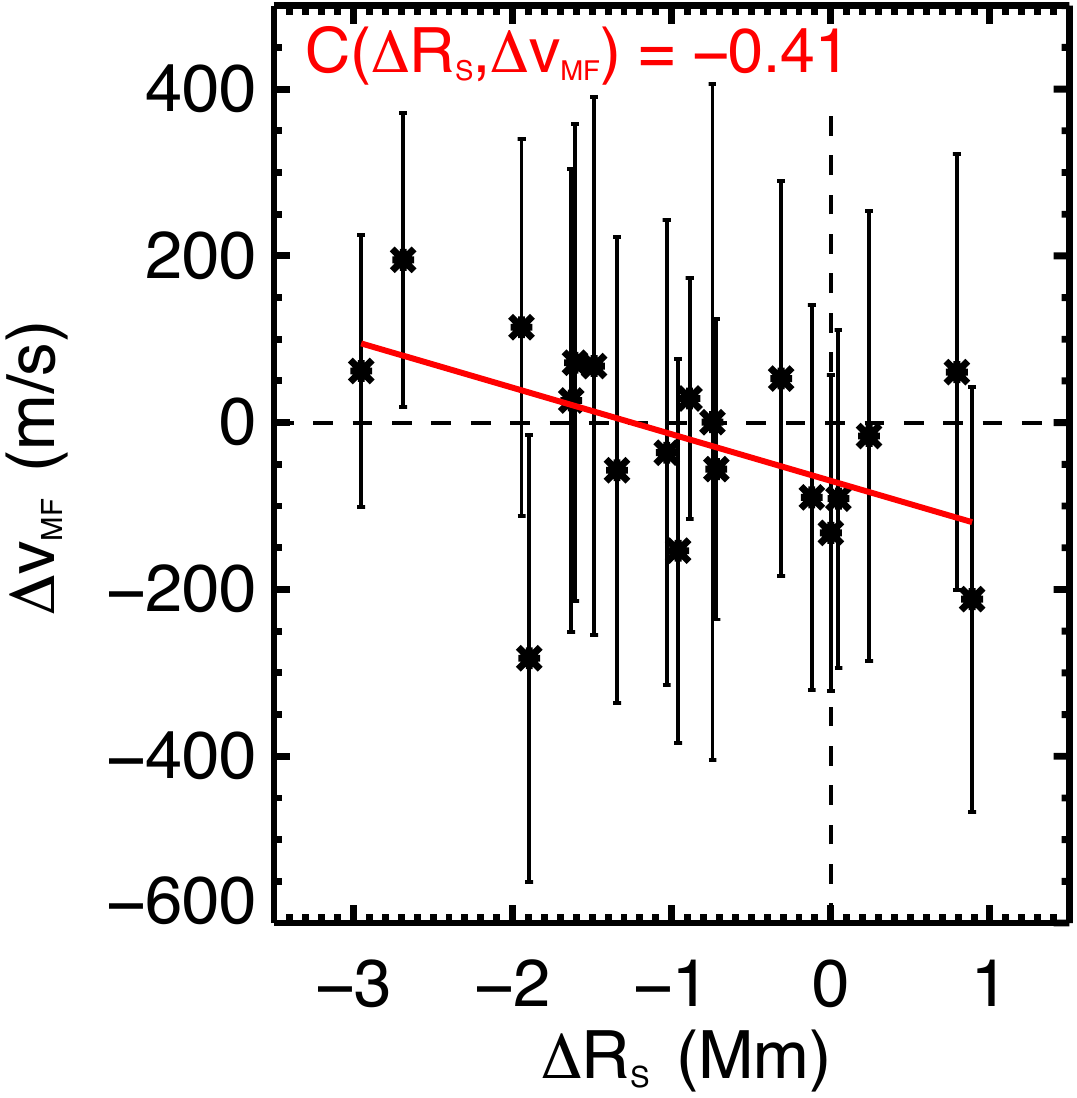}
&\includegraphics[height=4.35cm]{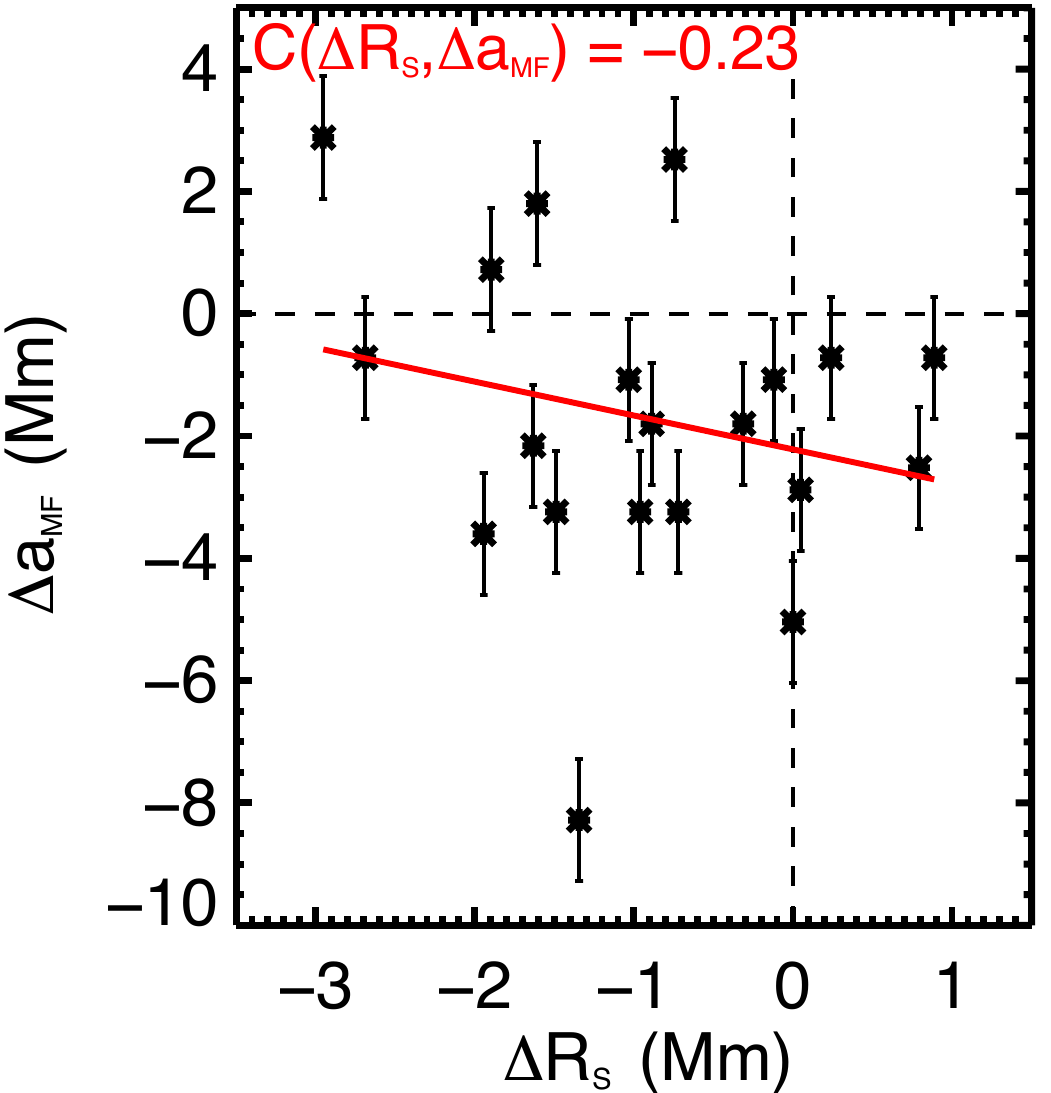}
&\includegraphics[height=4.35cm]{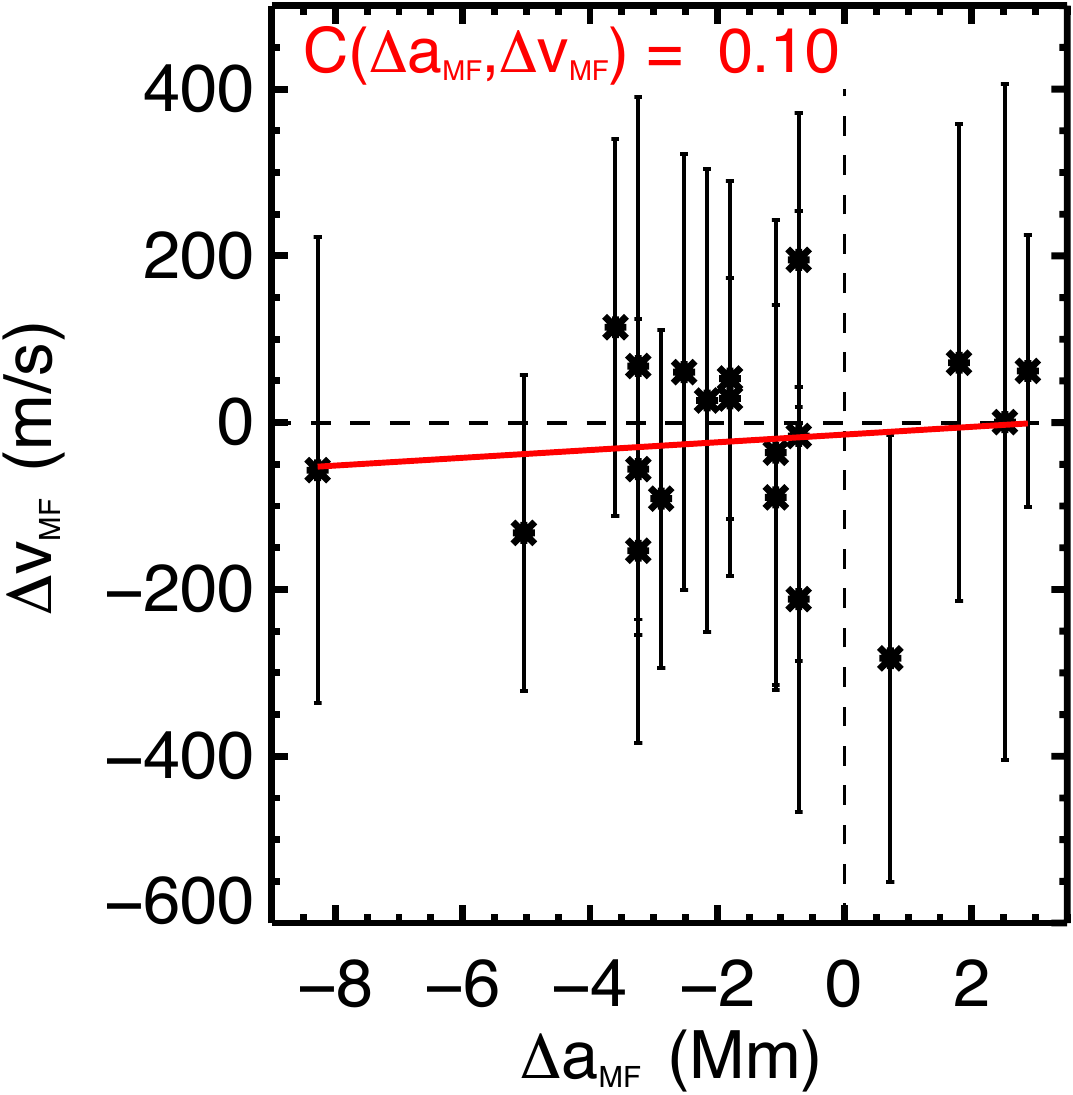}
\\
\hline
\end{tabular}
\end{center}
\caption{\label{fig:relations}--- Correlations of sunspot and \mytextbf{MF} properties \mytextbf{for all 51 sunspot \mytextbf{maps (a\,--\,e)},  \mbox{$v_{\rm 3h}$},} and their evolution within one week for 20 sunspots \mytextbf{(f\,--\,h)} indicated \mytextbf{by linear regressions (red lines) and the coefficients}, ${\cal C}(\cdot, \cdot)$. 
\newline-- \textbf{a)} MF velocities, $v_{\rm MF}$, in m/s and \textbf{b)} MF extensions, $a_{\rm MF}$, in Mm against the sunspot radii, $R_{\rm S}$, in Mm with the average values, $\langle v_{\rm MF}\rangle_{51}\!=\!982\,{\rm m/s}$ and $\langle a_{\rm MF}\rangle_{51}\!=\!9.2\,{\rm Mm}$. -- \textbf{c)} MF velocities, $v_{\rm MF}$, in m/s and \textbf{d)} MF extensions, $a_{\rm MF}$, in Mm against the maximum EF velocities, $v_{\rm EF}$, in m/s. -- \textbf{e)} Maximum EF velocities, $v_{\rm EF}$, in m/s against $R_{\rm S}$ in Mm with the average value, $\langle v_{\rm EF}\rangle_{51}\!=\!2324\,{\rm m/s}$. -- \textbf{f)} Weekly changes, $\Delta v_{\rm MF}$, in m/s  and \textbf{g)} $\Delta a_{\rm MF}$ in Mm of the moat flow against $\Delta R_{\rm S}$ in Mm. -- \textbf{h)} Weekly changes \mytextbf{of MF properties,} $\Delta v_{\rm MF}$, in m/s against $\Delta a_{\rm MF}$ in Mm.}
\end{figure*}

The maximum EF velocity, \mbox{$v_{\rm EF}\!=\!max\,[v_{\parallel,0}\,(R_{\rm U}\le r\le R_{\rm S})]$}, for all observed penumbrae is in the range of \mbox{$v_{\rm EF}\!=\!1830\ldots3000\,{\rm m/s}$} (\mytextbf{Fig.~\ref{fig:relations}e}) with an average of 2325\,m/s. Since we measure azimuthally averaged flow speeds of \mytextbf{3h averages}, our maximum EF velocities are slightly lower than the maximum velocities of up to $3\ldots4\,\rm{ km/s}$ found in recent studies \citep{schliche+schmidt2000,shine1994,rouppe2002,bellotrubio2003a}. The location, $r_{\rm EF}$, of the extreme \mytextbf{EF} velocity can by found at  \mbox{$0.65\!\cdot\!R_{\rm S}\le r_{\rm EF}\le 0.87\!\cdot\!R_{\rm S}$}, with an average ratio of \mbox{$r_{\rm EF}\!=\!0.78\!\cdot\!R_{\rm S}$} which is not as far in the outer spot boundary as \mbox{$r_{\rm EF}\!=\!0.8\ldots0.9\!\cdot\!R_{\rm S}$} \citep{tritschler,franzthesis}. The velocity run within the penumbra is largely similar for all sunspots with some differences in the skewness \mytextbf{as seen in Figs.~\ref{appfig1} and \ref{appfig2}}. 

\subsection{Correlations}\label{cor:vmfrspot}
In the following section, we \mytextbf{analyze the} correlations between $v_{\rm MF}$, $a_{\rm MF}$, $v_{\rm EF}$, and $R_{\rm S}$ based on the \mytextbf{3h average sample.} 

\subsubsection{Correlation of spot size and MF velocity}\label{cor:vmfrspot}
In order to work out the interaction of the sunspots with their moat flows, we plot the MF velocities, \mbox{$v_{\rm MF}\!=\!\langle v_{\rm MF}(r)\rangle_{r=0,\ldots,5{\rm px}}$}, against the sunspot size, $R_{\rm S}$, as it can be seen in \mytextbf{Fig.~\ref{fig:relations}a} with a \mytextbf{nonsignificant} rise of the linear regression (red \mytextbf{line}) and a correlation coefficient of ${\cal C}(v_{\rm MF}, R_{\rm S})\!=\!0.28$. Hence, we find that the MF velocity is in average around 1000\,m/s and \mytextbf{does not depend} on the spot size. Since we measure close to the outer sunspot boundary the indicated slight increase  may be due to a stronger EF in bigger sunspots (see Sec.~\ref{cor:vefrspot} and \ref{cor:vefvmf}). 

\subsubsection{Correlation of spot size and MF extension}\label{cor:amfrspot}
According to Eq.~\ref{eq:RMS} the MF extensions, $a_{\rm MF}$, were determined for all sunspots with \mytextbf{a} boundary criteria of \mbox{$\sigma_{\rm rms}\!\approx\!180\,\rm{m/s}$} and \mytextbf{range} between \mbox{$a_{\rm MF}=5\ldots15\,\rm{ Mm}$} for the spot sizes \mytextbf{of} \mbox{$R_{\rm S}=8.6\ldots21.2\,\rm{ Mm}$}, cf. \mytextbf{Fig.~\ref{fig:relations}b}. Since the linear regression and the correlation coefficient \mytextbf{of} ${\cal C}(a_{\rm MF}, R_{\rm S})\!=\!0.06$ indicate uncorrelated components, we average the \mytextbf{MF extensions} to \mbox{$\langle a_{\rm MF}\rangle_{51\,spots}=9.2\,\rm{ Mm}$}. Comparing our results to \mytextbf{other} studies \citep[e.g.,][]{brickhouse+labonte1988} the MF extension lies \mytextbf{below} the mentioned \mbox{$a_{\rm MF}=10\ldots20\,\rm{ Mm}$} for small sunspots and far below the double spot radius for larger spots. In our studies, the MF extension is independent of the spot size and therefore is in the order of the sunspot size only for small but not for bigger spots. \mytextbf{This finding goes in line with studies by \citet{sobotka+roudier2007}. In summary it can be stated that  we find four indications that the MF is not correlated to the spot size: (i) the kink of $v_{\parallel,0}(r)$ at the outer spot boundary (e.g. Fig.~\ref{fig:vh}: upper panel), (ii) the smaller spread of all 51 MFs by merely displaying them with their radial distance from the spot in Fig.~\ref{fig:vhcompare}, (iii) the trend in Fig.~\ref{fig:relations}a and (iv) the trend in Fig.~\ref{fig:relations}b.}

\subsubsection{Correlation of spot size and EF velocity}\label{cor:vefrspot}
To yield the correlation of the sunspot properties, i.~e.~ the maximum horizontal EF velocity between $v_{\rm EF}\!=\!1830\ldots3000\,{\rm m/s}$ and the sunspot radius, we plot both components against each other in \mytextbf{Fig.~\ref{fig:relations}e} and fit a linear regression \mytextbf{by considering all values. We \mytextbf{obtain}}  a slightly positive trend for bigger sunspots harboring higher \mytextbf{maximum} EF velocities.  \mytextbf{Following the regression}, sunspots with \mytextbf{$R_{\rm S}=10\,{\rm Mm}$ would have maximum} EF velocities \mytextbf{of 2200\,m/s whereas sunspots with the double size yield maximum EF velocities that are almost \mbox{500\,m/s}} faster. \mytextbf{Although there are some outliers between 16-17\,Mm, even} the correlation coefficient, ${\cal C}(a_{\rm MF}, R_{\rm S})\!=\!0.46$, suggests a positive link \mytextbf{of the EF velocity} with the spot size. 

\subsubsection{Correlation of Evershed flow and moat flow}\label{cor:vefvmf}
As we pointed out in Sects.~\ref{cor:vmfrspot} and \ref{cor:amfrspot} the moat flow seems to be independent of the sunspot radius. In the following, we examine the largely unknown link between the penumbral EF and the adjacent MF. It is therefore important to find out whether higher EF velocities also cause higher MF velocities or a bigger MF extension, which would argue for a partial drive of the MF by the EF. Due to \mytextbf{Figs.~\ref{fig:relations}c--d} which plot the MF velocities, $v_{\rm MF},$ and MF extensions, $a_{\rm MF}$, against the EF velocities, $v_{\rm EF}$, with correlation coefficients of ${\cal C}(v_{\rm MF}, v_{\rm EF})\!=\!0.07$ and ${\cal C}(a_{\rm MF}, v_{\rm EF})\!=\!0.01$, we draw the conclusion that the EF has no impact on the MF\mytextbf{. This means that there is a greater decrease in speed for spots with} higher maximum EF velocities \mytextbf{toward} the outer penumbral boundary  (see Figs.~\ref{appfig1} and \ref{appfig2}). \mytextbf{Although the existence of the MF is directly coupled to the existence of a penumbra and its EF, the observed properties indicate the MF and the EF are} independent sunspot flows. 

\subsection{Weekly evolution}
To study the temporal evolution of the \mytextbf{analyzed}  components, i.~e.~ $R_{\rm S}$, $v_{\rm MF}$, $a_{\rm MF}$, and $v_{\rm EF}$, we have tracked 20 sunspots (see Table~\ref{apptab}) from the eastern to the western \mytextbf{limb sides} and performed the analysis of \mbox{$v_{\rm 3h}$} for a second time at $\theta\!\approx\!50^\circ$. According to the latitude and the differential rotations, the observing dates range between 6 to 8 days. 

The weekly changes are in the range of \mbox{$\Delta R_{\rm S}\!=\!-3\ldots1\,{\rm Mm}$}, \mbox{$\Delta v_{\rm EF}\!=\!\pm500\,{\rm m/s}$}, \mbox{$\Delta v_{\rm MF}\!=\!\pm200\,{\rm m/s}$} and \mbox{$\Delta a_{\rm MF}\!=\!-8\ldots3\,{\rm Mm}$}. \mytextbf{Figures~\ref{fig:relations}f--g} plot the changes, $\Delta v_{\rm MF}$ and $\Delta a_{\rm MF}$, of the MF velocities and extensions against the change of the spot radius, $\Delta R_{\rm S}$, with linear regressions according to the correlation coefficients, ${\cal C}(\Delta v_{\rm MF}, \Delta R_{\rm S})\!=\!-0.41$ and ${\cal C}(\Delta a_{\rm MF}, \Delta R_{\rm S})\!=\!-0.23$, indicating slightly negative trends. Following this, the MF velocity increases for strongly decaying sunspots and tends to decrease slightly for small $\Delta R_{\rm S}$. The MF extension, $a_{\rm MF}$, decreases for the majority of the sunspots with some exceptions for strongly decaying sunspots, but yields no significant correlation with the sunspot evolution. \mytextbf{Figure~\ref{fig:relations}h} plots the change in the MF velocities, $\Delta v_{\rm MF}$, against the changing MF extension, $\Delta a_{\rm MF}$, with no significant correlation, \mbox{${\cal C}(\Delta v_{\rm MF}, \Delta a_{\rm MF})\!=\!0.10$}. Because they are \mbox{${\cal C}(\Delta v_{\rm EF}, \Delta R_{\rm S})\!=\!-0.06$}, \mbox{${\cal C}(\Delta v_{\rm MF}, \Delta v_{\rm EF})\!=\!0.23$}, \mbox{${\cal C}(\Delta a_{\rm MF}, \Delta v_{\rm EF})\!=\!-0.18$}, all other correlations are \mytextbf{insignificant} and yield no further impact. According to these results, we would draw the conclusions, that a strong sunspot decay leads to an additional drive of the moat flow by accelerating its velocity and sporadically expanding its outreach, whereas it has no impact on the EF velocity.

\subsection{Monthly evolution}
Tracking three long-lasting sunspots across the far side of the Sun, they reappear at the eastern \mytextbf{front side} \mytextbf{(Table~\ref{apptab}: No. 4$\,\rightarrow$\,6, 7\,$\rightarrow$\,9, 11\,$\rightarrow$\,13; Figs.~\ref{fig:dopcont}e--h)}\mytextbf{. This} allows us to study the long-term evolution for one month in order to crosscheck the results of the weekly evolution. The changes in sunspot size, $\rm{\Delta R_{\rm S}}$, the MF velocity, $\rm{\Delta v_{\rm MF}}$, the MF extension, $\rm{\Delta a_{\rm MF}}$, and the maximum EF velocity, $\rm{\Delta v_{\rm EF}}$, as listed in Table~\ref{verlauf} are based on the mean difference between the first and second appearances.

\begin{table}[htbp]
\caption{Monthly evolution \mytextbf{for three long-lasting sunspots}.}
\centering
\begin{tabular}{c c c c c}
\hline
\hline
Spot-No.&$\rm{\Delta R_{\rm S}}$&$\rm{\Delta a_{\rm MF}}$&$\rm{\Delta v_{\rm MF}}$&$\rm{\Delta v_{\rm EF}}$\\
\mytextbf{(Table~\ref{apptab})}&\mytextbf{(Mm)}&\mytextbf{(Mm)}&\mytextbf{(m/s)}&\mytextbf{(m/s)}\\
\hline
$4 \to 6$&$-2.8$&$-3.1$&$+77$&$-65$\\
$7 \to 9$&$-6.1$&$+6.6$&$+173$&$-455$\\
$11 \to 13$&$-5.2$&$-5.4$&$+64$&$-115$\\
\hline
\end{tabular}
\label{verlauf}
\end{table}%

Coupled to the strong decay in sunspot sizes, the MF velocity significantly increases, whereas the EF velocity shows a decrease of several hundred m/s.  These trends are in line with the correlations for the weekly evolution . The size of the \mytextbf{MF} region shows no unique trend, but the huge widening $\rm{\Delta a_{\rm MF}}\!=\!+6.6\,{\rm Mm}$ of Spot-No.\,7 (in Table~\ref{verlauf}) by more than 6\,Mm could underline an additional outflow of   plasma over the moat cell due to a strong sunspot decay. 

\section{Conclusion}\label{sec:concl}
\subsection{\mytextbf{Conclusions on} the moat flow}\label{sec:conclusions:mf}
\mytextbf{\citet{meyer1979} suggested that a sunspot is embedded in a supergranular cell. If one puts MF cells in a context with supergranules, one should realize that they have a diameter between 30 and 60\,Mm, i.e., are up to twice as large as supergranules. Also, the MF velocities are more than twice as high as supergranular velocities \citep[e.g.,][]{brickhouse+labonte1988}. As a result, the MF needs a driving mechanism that is distinct from the driving mechanism of normal supergranules. \citet{nye1988} modeled the MF being driven by a surplus gas pressure beneath the penumbra, arguing that the lack of radiative cooling beneath the penumbra generates a surplus of heat, hence gas pressure. In these models the MF velocity depends on the depth of the penumbra, but not on the size of a spot. At that time this posed a problem, as it was commonly believed that the MF extension scales with the spot radius. Later, \citet{sobotka+roudier2007} provided evidence that there is no correlation between the MF and the spot radius. Our investigation based on Doppler shift measurements rather than on local correlation tracking, independently confirms these later findings. The evidence therefore supports the moat flow model in which the driving forces are due to surplus gas pressure beneath sunspots.}

\mytextbf{A moat flow that is driven from beneath the penumbra would naturally push away the granules from the spot as seen with local correlation techniques. This implies that the MF is present in the deep photosphere. In contrast, the Evershed flow or the fraction of it that extends from the penumbra outwards into the moat is present in the magnetic canopy that surrounds the sunspot in the mid and upper photosphere \citep{rezaei+al2006}. In the immediate surroundings of the sunspot two types of flows exist: (1) the (magnetic) EF that partially continues in the magnetic canopy, which ascends outwards from the mid-photosphere at the spot boundary; (2) the (largely nonmagnetic) MF in the deep photosphere and beneath.}

\mytextbf{We consider the moving magnetic features (MMFs) to be distinct from the moat flow. They migrate into or are advected by the MF. MMFs are associated with inclined magnetic field lines (relative to horizontal) that reach up into the higher atmospheric layers. These MMF field lines possibly originate in the sunspot and either witness the decay of sunspots (uni-polar MMFs) or are due to some waves that propagate outwards (bipolar MMFs), \citep[see e.g.,][]{dalda+pillet2005, dalda+rubio2008a, dalda+rubio2008b, schliche2002}. Because of their vertical configuration, outwardly migrating MMFs can also be detected in higher atmospheric layers as reported by \citet{sobotka+roudier2007}.}

\subsection{\mytextbf{Summary}}\label{sec:conclusions:sum}
\mytextbf{We calibrated HMI velocity maps such that systematic absolute errors are below
$\rm{150\,m/s}$ on an absolute scale }. The synoptic CLV of the convective blueshift was measured in good accordance with \mytextbf{previous findings and} the theoretical synthesis curve. \mytextbf{We find a} maximum velocity of the solar rotation of 1992\,m/s, \mytextbf{which agrees} with previous measurements and reveals no significant impact by stray light. The instrumental artifacts were identified in line with \mytextbf{HMI calibration} studies \citep{centeno+etal2011}. \mytextbf{These artifacts make the meridional flow amplitude too small to be measured.}

\mytextbf{We \mytextbf{analyzed} 3h time averages of sunspot flow velocities. We constructed 51 velocity maps of 31 sunspots. The flow in and around circular sunspots with a fully developed penumbra was \mytextbf{analyzed}  using azimuthal averages, thereby assuming axial symmetry. In both, the MF and EF, the horizontal velocity component was dominant. The vertical flow components of MF and EF were determined to be small. Since the exact amount of convective blueshift is unknown, we do not address the sign of vertical flows.}

\mytextbf{The radial dependence of the velocity fields was observed to be similar for all 51 sunspot maps.} The analysis of the EF yielded results that are \mytextbf{consistent} with recent studies. A higher EF velocity was detected for bigger sunspots\mytextbf{, and the EF velocity scales with the spot size.} The MF is \mytextbf{a} convective outflow with \mytextbf{radially} decreasing velocity. The MF velocity turned out to be \mytextbf{similar in size} and \mytextbf{independently of} the spot size. \mytextbf{Also, t}he MF extension is \mytextbf{not correlated with any} sunspot property. \mytextbf{As MF and EF properties turn out to be uncorrelated, we inferred that the driving mechanisms of the two flows should be distinct. Therefore, we favor the MF model of \citet{nye1988}: The MF is driven by surplus gas pressure beneath the penumbra.}

\mytextbf{We find a tendency toward increasing MF velocity for strongly decaying sunspots on time scales of one week (spot rotates from the east to west limbs) and one month (spot reappears on the east limb). It is beyond the scope of this paper to put this into the context of a model for decaying sunspots.}

\begin{acknowledgements} The HMI instrument of NASA's SDO mission is operated by the HMI/AIA Joint Science Operations Center (JSOC) at Stanford University. The data was provided by the JSOC webpage. We want to thank Hans-G\"unter Ludwig for the spectral line synthesis, Rebecca Centeno and Sebastien Couvidat for their correspondence on HMI questions, and Juan Manuel Borrero for his fruitful comments on the manuscript.
\end{acknowledgements}


\begin{thebibliography}{55}
\expandafter\ifx\csname natexlab\endcsname\relax\def\natexlab#1{#1}\fi

\bibitem[{{Balthasar}(1985)}]{balthasar1985}
{Balthasar}, H. 1985, \solphys, 99, 31

\bibitem[{{Balthasar} {et~al.}(1996){Balthasar}, {Schleicher}, {Bendlin}, \&
  {Volkmer}}]{balthasar+etal1996}
{Balthasar}, H., {Schleicher}, H., {Bendlin}, C., \& {Volkmer}, R.
  1996, \aap, 315, 603

\bibitem[{{Balthasar} \& {Muglach}(2010)}] {balthasar+muglach}
{Balthasar}, H., {Muglach}, K. 2010, \aap, 511, 67

\bibitem[{{Beeck} {et~al.}(2012)}]{beeck}
{Beeck}, B., {Collet}, R., {Steffen}, M., {Asplund}, M.,	{Cameron}, R.~H., {Freytag}, B., {Hayek}, W., {Ludwig}, H.-G., {Sch{\"u}ssler}, M. 2012, \aap, 539, 121

\bibitem[{{Brickhouse} \& {Labonte}(1988)}]{brickhouse+labonte1988}
{Brickhouse}, N.S., {Labonte}, B.J. 1988, \solphys, 115, 43

\bibitem[{{Bellot Rubio} {et~al.}(2003a)}]{bellotrubio2003a}
{Bellot Rubio}, L.R., {Balthasar}, H., {Collados}, M., {Schlichenmaier}, R. 2003a, \aap, 403, 47

\bibitem[{{Cabrera Solana} {et~al.}(2006)}]{cabrera2006}
{Cabrera Solana}, D., {Bellot Rubio}, L. R., {Beck}, C., {del Toro Iniesta}, J. C. 2006, The
Astrophysical Journal Letters, 649, L41

\bibitem[{{Cabrera Solana} {et~al.}(2007)}]{cabrera2007}
{Cabrera} Solana, D., {Bellot Rubio}, L. R., {Beck}, C., {del Toro Iniesta}, J. C.: 2007,
\aap, 475, 1067

\bibitem[{{Centeno} {et~al.}(2011)}]{centeno+etal2011}
{Centeno}, R., {Tomczyk}, S., {Borrero}, J.M., {et~al.} 2011, Proceedings of the Solar Polarisation Workshop 6, 437, 147

\bibitem[{{Cruz Rodr\'iguez} {et~al.}(2011)}]{cruzrodriguez2011}
{Cruz Rodr\'iguez}, J. de La, {Kiselman}, D., {Carlsson}, M. 2011, \aap, 528, 113

\bibitem[{{Delbouille} {et~al.}(1990)}]{liege}
{Delbouille}, L., {Roland}, G., {Neven}, L. 1990, Atlas photometrique DU spectre solaire, Universit\'e de Li\`ege

\bibitem[{{Evershed}(1909)}]{evershed1909}
{Evershed}, J. 1909, Monthly Notices Royal Astron. Soc. 69, 454

\bibitem[{{Featherstone} {et~al.}(2011)}]{featherstone+etal2011}
{Featherstone}, N.A., {Hindman}, B.W., {Thompson}, M.J. 2011, J. Phys.: Conf. Ser., 271, 012002

\bibitem[{{Franz} \& {Schlichenmaier}(2009)}]{franz2009}
{Franz}, M., {Schlichenmaier}, R. 2009, \aap, 508, 1453

\bibitem[{{Franz}(2011)}]{franzthesis}
{Franz}, M. 2011, Spectropolarimetry of Sunspot Penumbrae, Cuvillier G\"ottingen

\bibitem[{{Freytag} {et~al.}(2012)}]{freytag}
{Freytag}, B., {Steffen}, M., {Ludwig}, H.-G., {et~al.}, {Wedemeyer-B{\"o}hm}, S., {Schaffenberger}, W., {Steiner}, O. 2012, J. Comp. Phys., 231, 919

\bibitem[{{Gizon} {et~al.}(2000)}]{gizon+etal2000}
{Gizon}, L., {Duvall}, T.L. Jr., {Larsen}, R.M. 2000, J. Astrophys. Astron., 21, 339

\bibitem[{{Gizon} {et~al.}(2009)}]{gizon+etal2009}
{Gizon}, L., {Schunker}, H., {Baldner}, C.S., {et~al.} 2009, Space Sci. Rev., 144, 249

\bibitem[{{Harvey} \& {Harvey}(1973)}]{harvey+harvey1973}
{Harvey}, K., {Harvey}, J. 1973, \solphys, 28, 61

\bibitem[{{Hathaway} \& {Rightmire}(2010)}]{HathRight2010}
{Hathaway}, D., {Rightmire}, L. 2010, Science, 327, 1350

\bibitem[{{Howe} {et~al.}(2011)}]{howe+etal2011}
{Howe}, R., {Jain}, K., {Hill}, F., {et~al.} 2011, J. Phys.: Conf. Ser., 271, 012060

\bibitem[{{Komm} {et~al.}(2011)}]{komm2011}
{Komm}, R., {Howe}, R., {Hill}, F., {et~al.} 2011, J. Phys.: Conf. Ser., 271, 012077

\bibitem[{{Maltby}(1964)}]{maltby1964}
{Maltby}, P. 1964, Astrophys. Norvegica 8, 205

\bibitem[{{Meyer} {et~al.}(1979)}]{meyer1979}
{Meyer}, F., {Schmidt}, H.U.,{Weiss}, N.O. 1979, \aap, 76, 35

\bibitem[{{Michard}(1951)}]{michard1951}
{Michard}, R. 1951, Annales dÕAstrophysique 14, 101

\bibitem[{{Nye} {et~al.}(1988)}]{nye1988}
{Nye}, A., {Bruning}, D., {LaBonte}, B.J., 1988, \solphys, 115, 251

\bibitem[{{Pardon} {et~al.}(1979)}]{pardon1979}
{Pardon}, L., {Worden}, S.P., {Schneeberger}, T.J. 1979, \solphys, 63, 247

\bibitem[{{Rezaei} {et~al.}(2006)}]{rezaei+al2006}
{Rezaei}, R., {Schlichenmaier}, R., {Beck}, C., {Bellot Rubio}, L. R. 2006, \aap, 454, 975

\bibitem[{{Rimmele}(1994)}]{rimmele1994}
{Rimmele}, T. R. 1994, \aap, 290, 972

\bibitem[{{Rimmele}(1997)}]{rimmele1997}
{Rimmele}, T. R. 1997, \apj, 490, 458

\bibitem[{{Rouppe van der Voort}(2002)}]{rouppe2002}
{Rouppe van der Voort}, L.H.M. 2002, \aap, 389, 1020

\bibitem[{{Sainz Dalda} \& {Mart\'inez Pillet}(2005)}]{dalda+pillet2005}
{Sainz Dalda}, A., {Mart\'inez Pillet}, V. 2005, \apj, 632, 1176

\bibitem[{{Sainz Dalda} \& {Bellot Rubio}(2008a)}]{dalda+rubio2008a}
{Sainz Dalda}, A., {Bellot Rubio}, L.R. 2008a, \aap, 481, L21

\bibitem[{{Sainz Dalda} \& {Bellot Rubio}(2008b)}]{dalda+rubio2008b}
{Sainz Dalda}, A., {Bellot Rubio}, L.R. 2008b, \aap, 481, L21

\bibitem[{{Schlichenmaier} \& {Schmidt}(1999)}]{schliche1999}
{Schlichenmaier}, R., {Schmidt}, W. 1999, \aap, 349, L37

\bibitem[{{Schlichenmaier}(2002)}]{schliche2002}
{Schlichenmaier}, R. 2002, Astron. Nachr., 323, 303

\bibitem[{{Schlichenmaier} \& {Collados}(2002)}]{schliche+collados2002}
{Schlichenmaier}, R., {Collados}, M. 2002, \aap, 381, 668

\bibitem[{{Schlichenmaier} \& {Schmidt}(2000)}]{schliche+schmidt2000}
{Schlichenmaier}, R., {Schmidt}, W. 2000, \aap, 358, 1122

\bibitem[{{Schlichenmaier} {et~al.}(2004)}]{schliche2004}
{Schlichenmaier}, R., {Bellot Rubio}, L.R., {Tritschler}, A. 2004, \aap, 415, 731

\bibitem[{{Schlichenmaier} {et~al.}(2010)}]{schlichenmaier+etal2010}
{Schlichenmaier}, R., {Rezaei}, R., {Bello Gonz\'alez}, N., {Waldmann}, T. A. 2010, \aap, 512, L1

\bibitem[{{Sheeley}(1969)}]{sheeley1969}
{Sheeley}, N.R. Jr. 1969, \solphys, 9, 347

\bibitem[{{Sheeley}(1972)}]{sheeley1972}
{Sheeley}, N.R. Jr. 1972, \solphys, 25, 98

\bibitem[{{Shine} {et~al.}(1994)}]{shine1994}
{Shine}, R. A., {Title}, A. M., {Tarbell}, T. D., {Smith}, K., {Frank}, Z. A., and {Scharmer}, G.
1994, \apj, 430, 413

\bibitem[{{Snodgrass}(1984)}]{snodgrass1984}
{Snodgrass}, H.B. 1984, \solphys, 94, 13

\bibitem[{{Sobotka} {et~al.}(1999)}]{sobotka1999}
{Sobotka}, M., {V\'azquez}, M., {Bonet}, J.A., Hanslmeier, A., Hirzberger, J. 1999, \apj, 511, 436

\bibitem[{{Sobotka} \& {Roudier}(2007)}]{sobotka+roudier2007}
{Sobotka}, M., {Roudier}, T. 2007, \aap, 472, 277

\bibitem[{{Solanki}(2003)}]{solanki2003}
{Solanki}, S.K. 2003, \aap, Rev. 11, 153

\bibitem[{{Stix}(2002)}]{stix2002}
{Stix}, M. 2002, The Sun , Springer Berlin

\bibitem[{{Sun} {et~al.}(1997)}]{sun1997}
{Sun}, M.T., {Chou}, D.Y., {Lin}, C.H. and the TON Team 1997, \solphys, 176, 59

\bibitem[{{Tritschler} {et~al.}(2004)}]{tritschler}
{Tritschler}, A., {Schlichenmaier}, R., {Bellot Rubio}, L.R., {KAOS Team}, {Berkefeld}, T., {Schelenz}, T. 2004, \aap, 415, 717

\bibitem[{{Vargas Dom\'inguez} {et~al.}(2007)}]{dominguez+etal2007}
{Vargas Dom\'inguez}, S., {Bonet}, J.A., {Mart\'inez Pillet}, V., {Katsukawa}, Y., {Kitakoshi}, Y., {Rouppe van der Voort}, L. 2007, \apjl, 660, L165

\bibitem[{{Verma} {et~al.}(2012)}]{verma+etal2012}
{Verma}, M., {Balthasar}, H., {Deng}, N., {et~al.} 2012, \aap, 538, 109

\bibitem[{{Vrabec}(1971)}]{vrabec1971}
{Vrabec}, D. 1971, In: Howard, R. (ed.) Solar Magnetic Fields, IAU Sympos. 43, 329

\bibitem[{{Vrabec}(1974)}]{vrabec1974}
{Vrabec}, D. 1974, In: Athay, R.G. (ed.) Chromospheric Fine Structure, IAU Sympos. 56, Reidel, Dordrecht, 201

\bibitem[{{Wiehr} \& {Degenhardt}(1992)}]{wiehr1992}
{Wiehr}, E., {Degenhardt}, D. 1992, \aap, 259, 313


\end{thebibliography}


\clearpage
\begin{appendix}
\begin{table*}[htdp]
\section{Analyzed sunspots}
\caption{\mytextbf{List of the 31 numbered, observed sunspots. The corresponding NOAA number of the active region (AR) is given. The number, date, and starting time of 3h observation, as well as the average spot location, the spot radius, $R_{\rm S}$, (in Mm), and the heliocentric angle, $\theta$ (in $^\circ$), are listed. The spots with} No. 4+6, 7+9, and 11+13 describe the same recurrent sunspots, respectively.}\label{apptab}
\begin{center}
\begin{tabular}{l l l l l l l l}
\hline
No.&AR&Obs.-No.&Date&Time&Location&$R_{\rm S}$\,[Mm]&$\theta$ [$^\circ$]\\
\hline\hline
1&11084&a&2010-06-29&00:00&S19 E45&11.5&$49$\\
&&b&2010-07-06&00:00&S19 W47&11.2&$51$\\
2&11092&a&2010-08-06&20:00&N13 W44&15.1&$45$\\
3&11101&a&2010-08-27&00:00&N13 E45&11.5&$46$\\
&&b&2010-09-03&08:00&N12 W48&11.9&$49$\\
4 ($\rightarrow$\,6)&11108&a&2010-09-25&00:00&S30 W36&17.3&$49$\\
&&b&2010-09-25&10:00&S30 W40&17.3&$52$\\
5&11113&a&2010-10-16&00:00&N17 E48&12.6&$49$\\
&&b&2010-10-23&10:00&N16 W48&10.4&$48$\\
6 ($\leftarrow$\,4)&11115&a&2010-10-17&18:00&S28 E40&15.1&$52$\\
&&b&2010-10-24&04:00&S29 W40&13.7&$52$\\
7 ($\rightarrow$\,9)&11131&a&2010-12-05&00:00&N30 E39&21.2&$49$\\
&&b&2010-12-11&08:00&N30 W38&19.8&$47$\\
8&11133&a&2010-12-14&04:00&N14 W49&10.1&$51$\\
9 ($\leftarrow$\,7)&11140&a&2011-01-02&20:00&N32 E40&15.1&$51$\\
&&b&2011-01-08&20:00&N33 W35&13.3&$47$\\
10&11147&a&2011-01-18&16:00&N25 E41&12.6&$45$\\
&&b&2011-01-24&12:00&N24 W43&11.9&$49$\\
11 ($\rightarrow$\,13)&11195&a&2011-04-28&10:00&S17 W44&15.1&$48$\\
12&11203&a&2011-05-01&12:00&N18 E45&12.2&$47$\\
&&b&2011-05-08&09:00&N17 W47&10.8&$50$\\
13 ($\leftarrow$\,11)&11216&a&2011-05-18&12:00&S15 E47&10.4&$48$\\
&&b&2011-05-25&14:00&S15 W46&9.0&$47$\\
14&11251&a&2011-07-13&15:00&N16 E49&12.2&$49$\\
&&b&2011-07-20&20:00&N17 W46&9.4&$46$\\
15&11260&a&2011-08-02&18:00&N19 W46&17.6&$50$\\
16&11277&a&2011-08-27&00:00&N17 E47&9.0&$51$\\
&&b&2011-09-03&09:00&N19 W46&9.7&$50$\\
17&11287&a&2011-09-06&00:00&S31 E37&12.2&$49$\\
&&b&2011-09-11&18:00&S29 W39&9.0&$52$\\
18&11305&a&2011-10-04&20:00&N11 W48&14.8&$50$\\
19&11312&a&2011-10-07&00:00&N22 E48&17.6&$49$\\
&&b&2011-10-14&03:00&N23 W43&17.3&$45$\\
20&11314&a&2011-10-19&00:00&N26 W42&16.6&$46$\\
21&11317&a&2011-10-14&00:00&S27 E42&9.4&$51$\\
&&b&2011-10-20&06:00&S27 W41&9.4&$51$\\
22&11338&a&2011-11-10&17:00&S12 W48&12.6&$53$\\
23&11340&a&2011-11-07&00:00&S09 E49&10.1&$50$\\
&&b&2011-11-14&10:00&S08 W49&9.4&$50$\\
24&11342&a&2011-11-15&00:00&N17 W47&12.6&$48$\\
25&11343&a&2011-11-16&00:12&N28 W39&8.6&$44$\\
26&11355&a&2011-11-27&21:00&N14 W45&9.7&$47$\\
27&11356&a&2011-11-27&21:00&N16 W23&10.8&$30$\\
28&11366&a&2011-12-05&00:00&N17 E51&9.7&$53$\\
&&a&2011-12-12&08:00&N18 W47&9.0&$48$\\
29&11384&a&2011-12-23&00:00&N12 E36&19.8&$37$\\
&&b&2011-12-26&00:00&N13 W01&18.7&$14$\\
30&11388&a&2011-12-30&00:00&S23 E42&10.8&$45$\\
&&b&2012-01-05&00:00&S24 W35&9.7&$40$\\
31&11389&a&2011-12-31&00:00&S23 E43&15.8&$45$\\
&&b&2012-01-07&00:00&S21 W45&16.9&$50$\\
\hline
\end{tabular}
\end{center}
\end{table*}%

\clearpage

\begin{figure*}[htdp]
\caption{Horizontal flow velocities, $v_{\parallel,0}(r)$, in km/s \mytextbf{from the sunspot center ($r=0\,\rm{Mm}$) to a distance of 36Mm in the quiet Sun(QS) for spots No. 1--17 (see Table~\ref{apptab}). The boundaries of the umbra (U), the penumbra (PU), and the end of the moat flow region (MF) are indicated as vertical dashed lines.}}\label{appfig1}
\begin{center}
\begin{tabular}{|c|c|c|c|c|}
\hline
\includegraphics[trim = 42mm 140mm 104.5mm 57mm, clip,height=3.7cm]{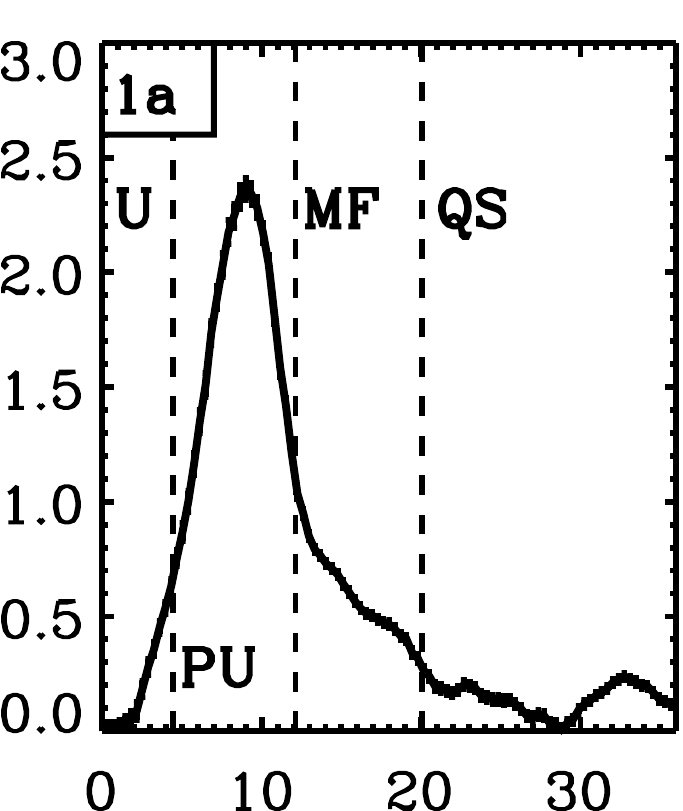}
&\includegraphics[trim = 42mm 140mm 104.5mm 57mm, clip,height=3.7cm]{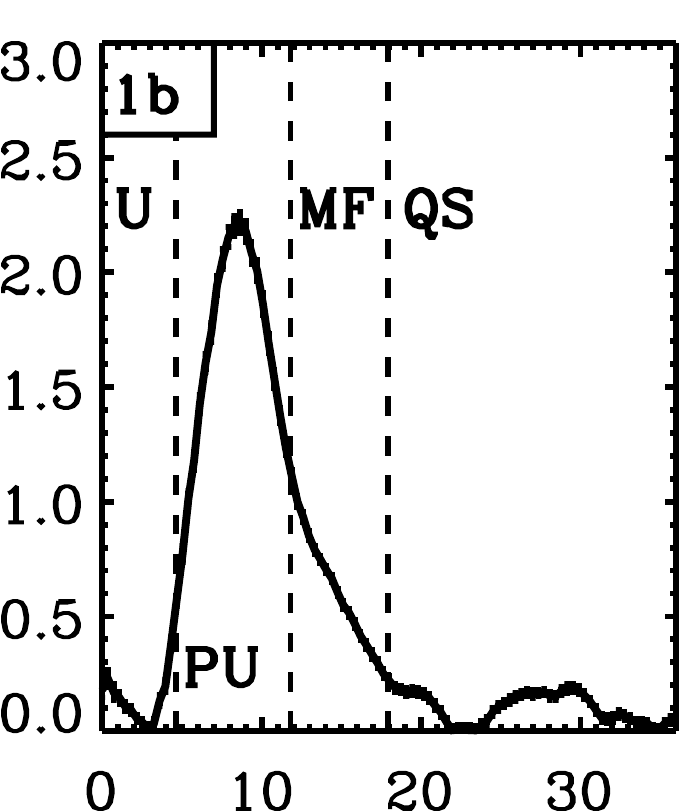}
&\includegraphics[trim = 42mm 140mm 104.5mm 57mm, clip,height=3.7cm]{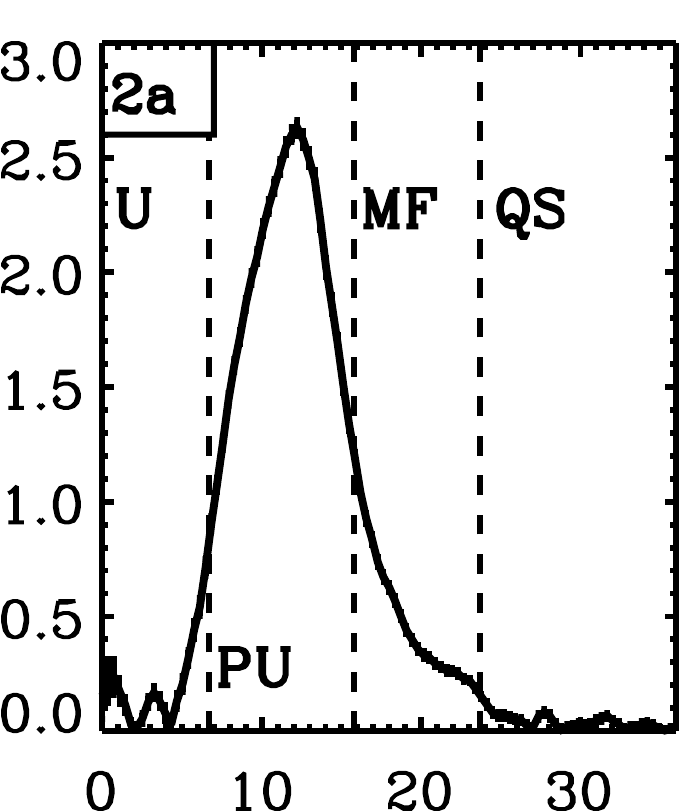}
&\includegraphics[trim = 42mm 140mm 104.5mm 57mm, clip,height=3.7cm]{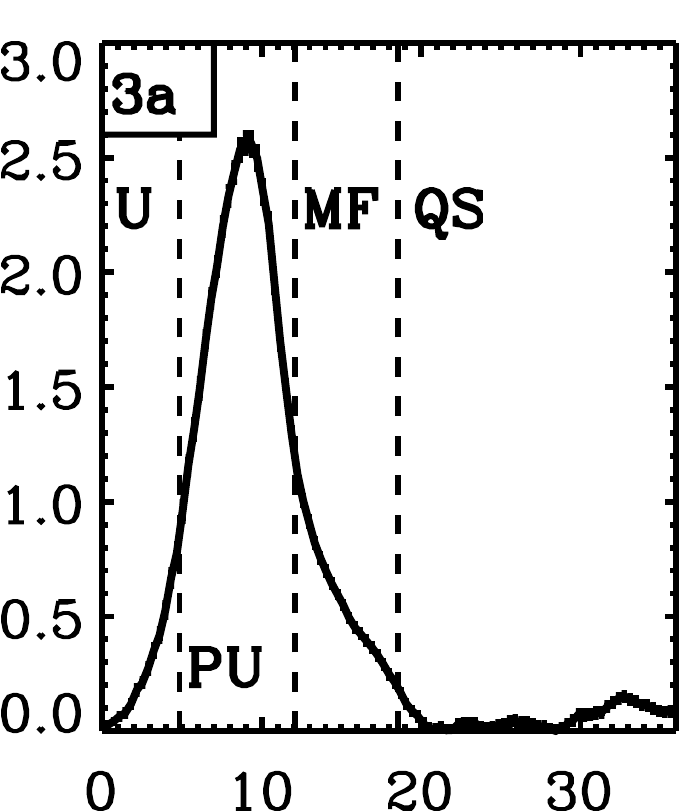}
&\includegraphics[trim = 42mm 140mm 104.5mm 57mm, clip,height=3.7cm]{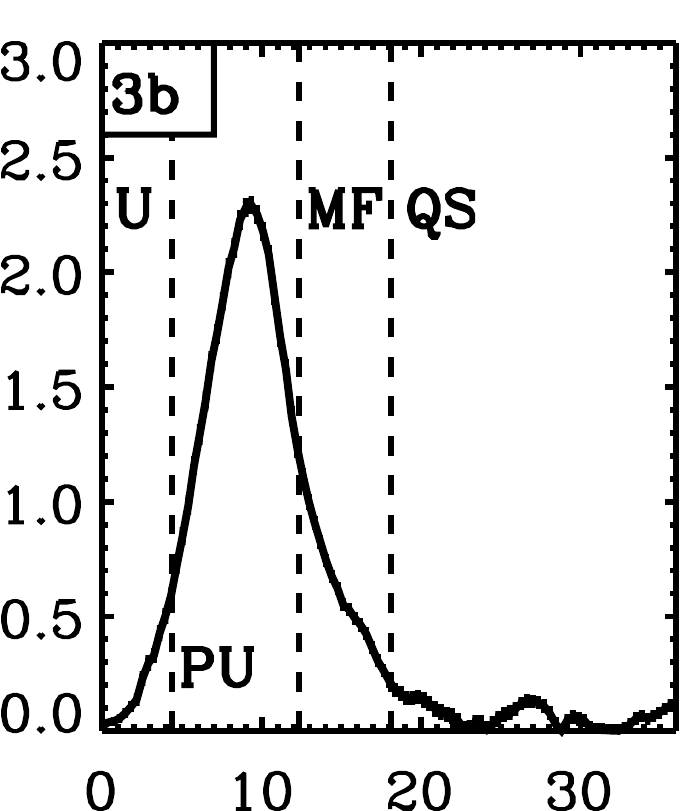}
\\
\hline
\includegraphics[trim = 42mm 140mm 104.5mm 57mm, clip,height=3.7cm]{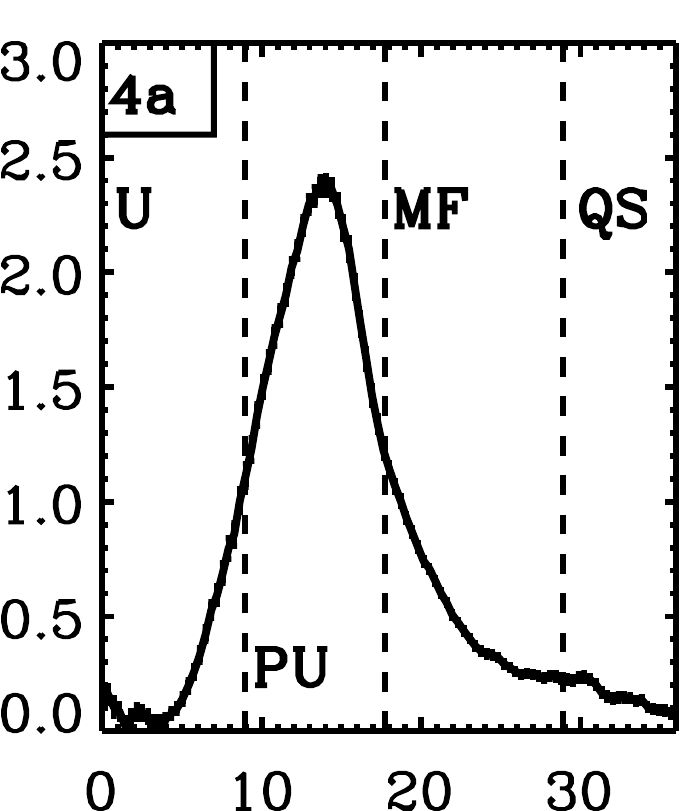}
&\includegraphics[trim = 42mm 140mm 104.5mm 57mm, clip,height=3.7cm]{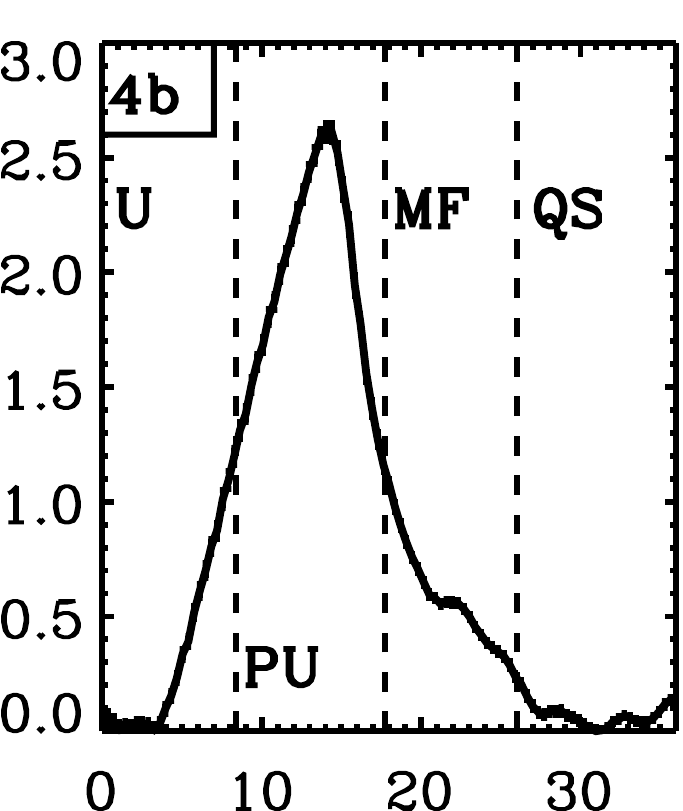}
&\includegraphics[trim = 42mm 140mm 104.5mm 57mm, clip,height=3.7cm]{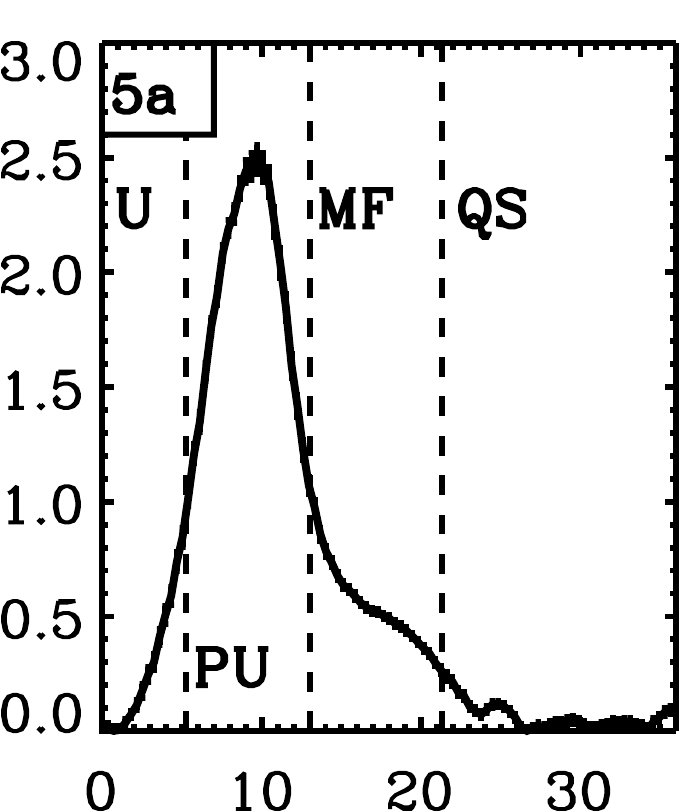}
&\includegraphics[trim = 42mm 140mm 104.5mm 57mm, clip,height=3.7cm]{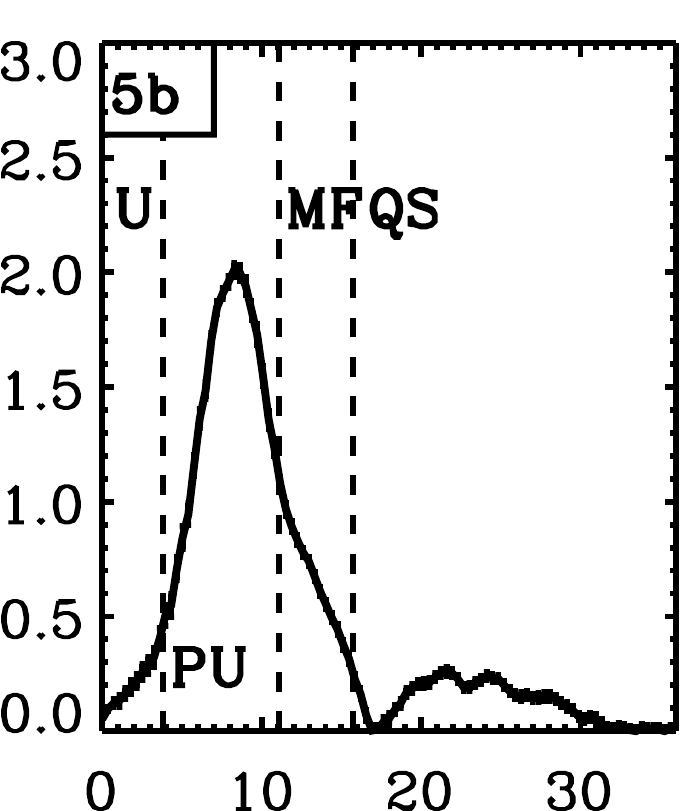}
&\includegraphics[trim = 42mm 140mm 104.5mm 57mm, clip,height=3.7cm]{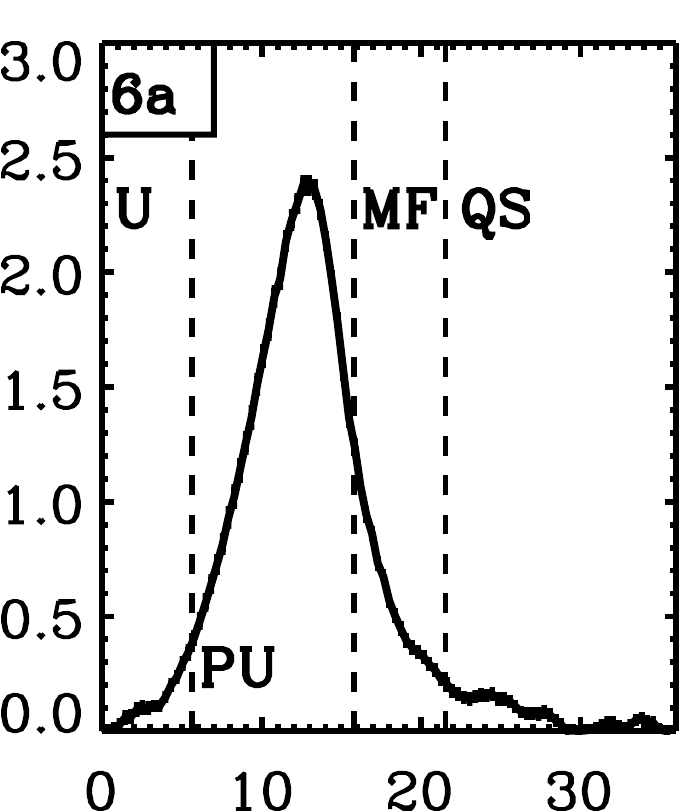}
\\
\hline
\includegraphics[trim = 42mm 140mm 104.5mm 57mm, clip,height=3.7cm]{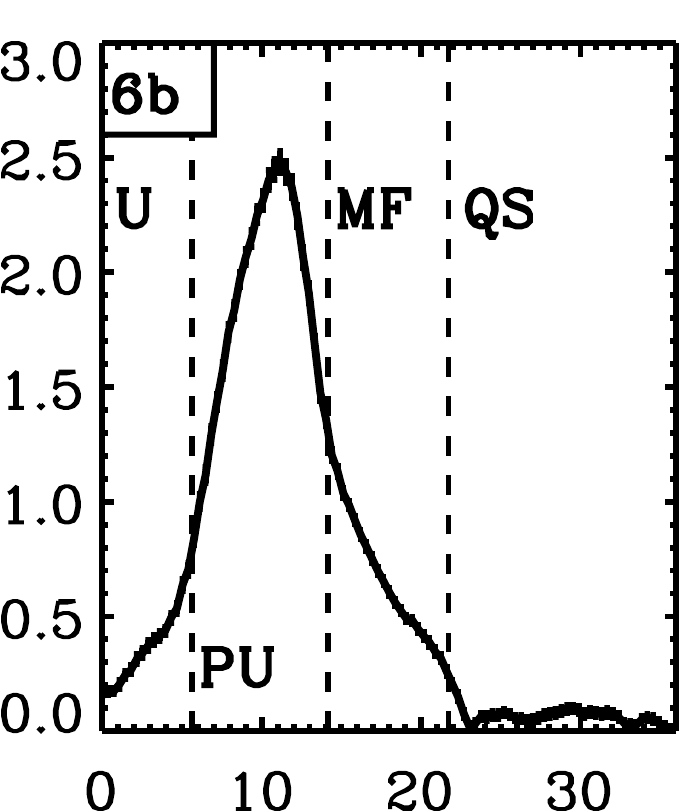}
&\includegraphics[trim = 42mm 140mm 104.5mm 57mm, clip,height=3.7cm]{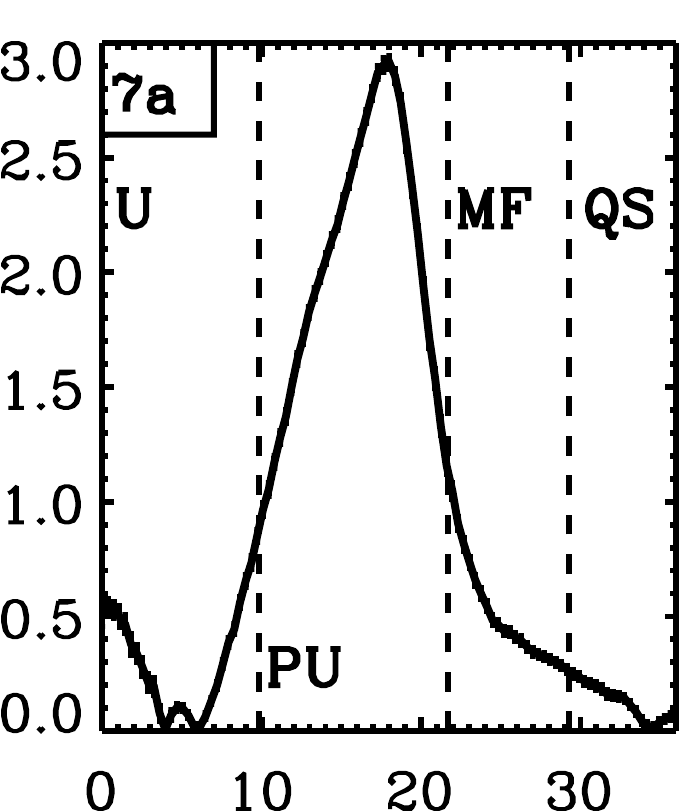}
&\includegraphics[trim = 42mm 140mm 104.5mm 57mm, clip,height=3.7cm]{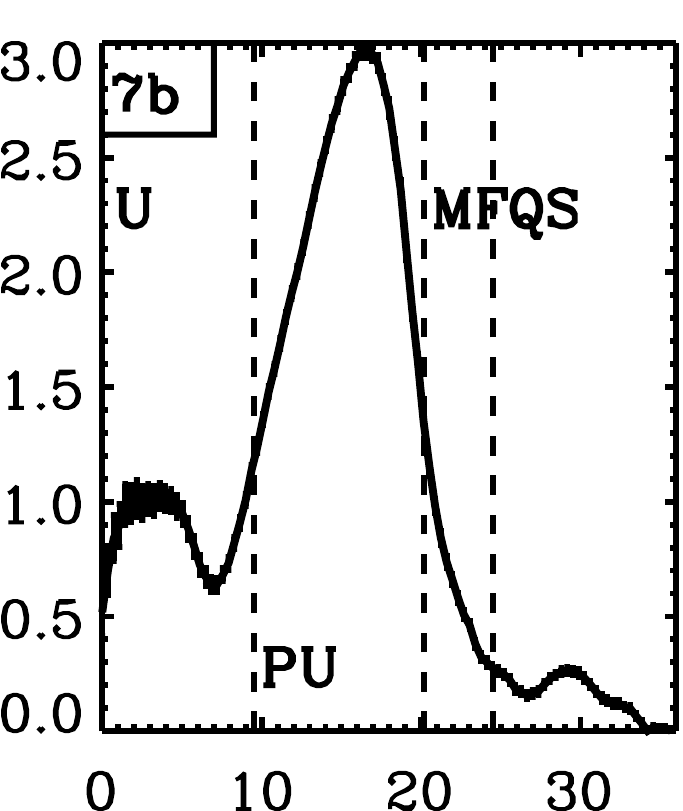}
&\includegraphics[trim = 42mm 140mm 104.5mm 57mm, clip,height=3.7cm]{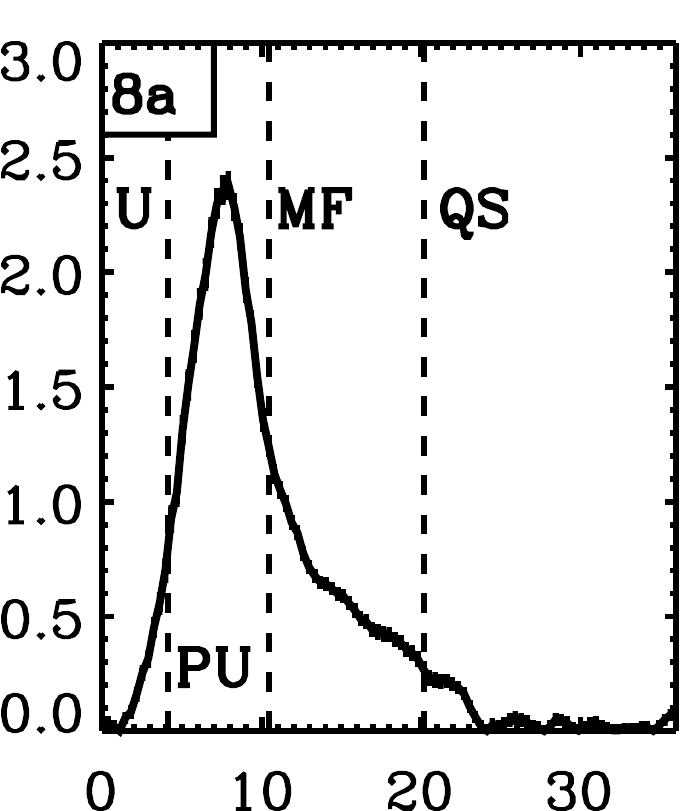}
&\includegraphics[trim = 42mm 140mm 104.5mm 57mm, clip,height=3.7cm]{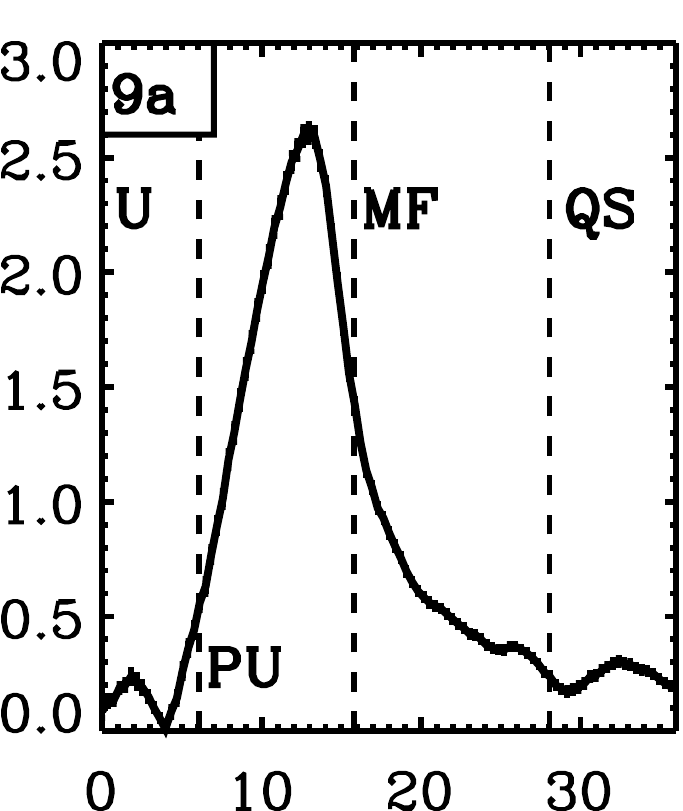}
\\
\hline
\includegraphics[trim = 42mm 140mm 104.5mm 57mm, clip,height=3.7cm]{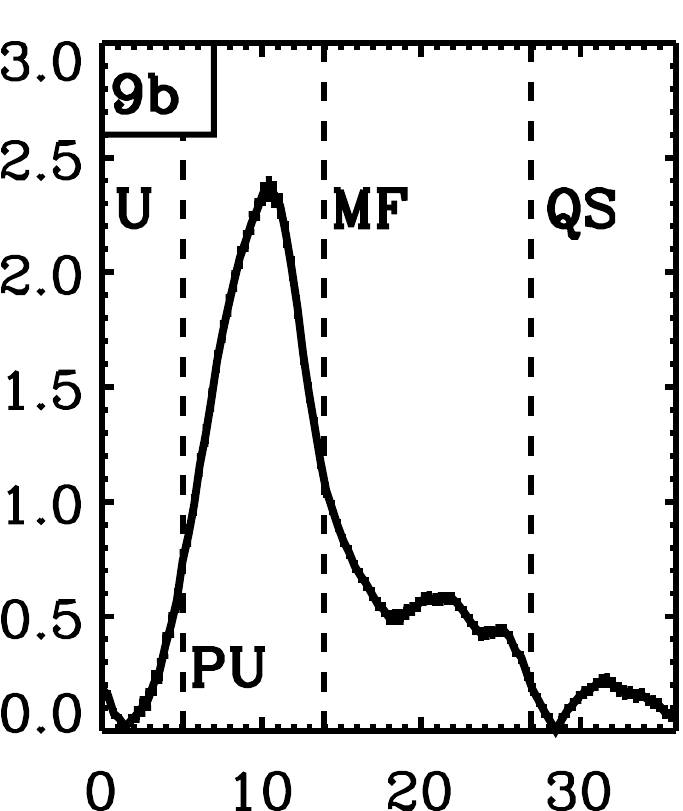}
&\includegraphics[trim = 42mm 140mm 104.5mm 57mm, clip,height=3.7cm]{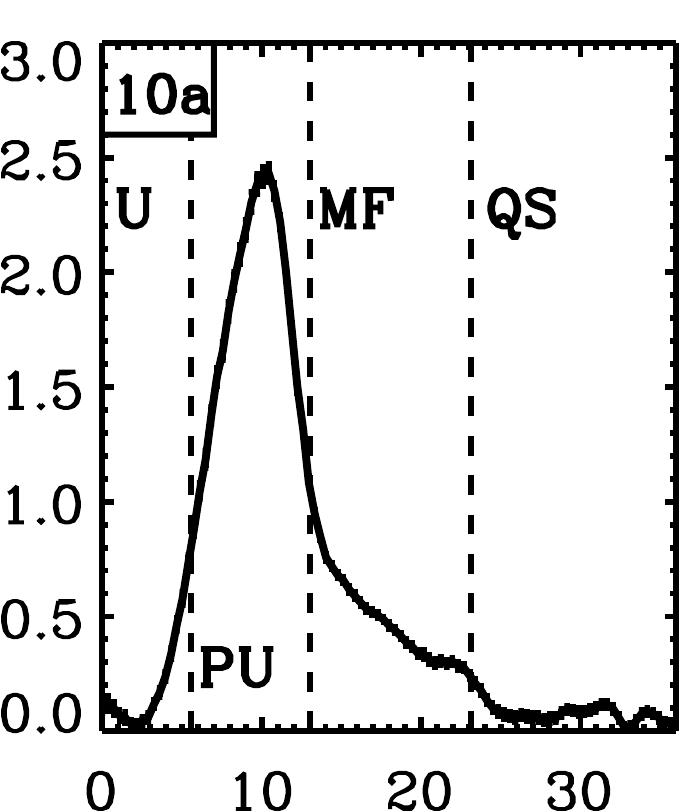}
&\includegraphics[trim = 42mm 140mm 104.5mm 57mm, clip,height=3.7cm]{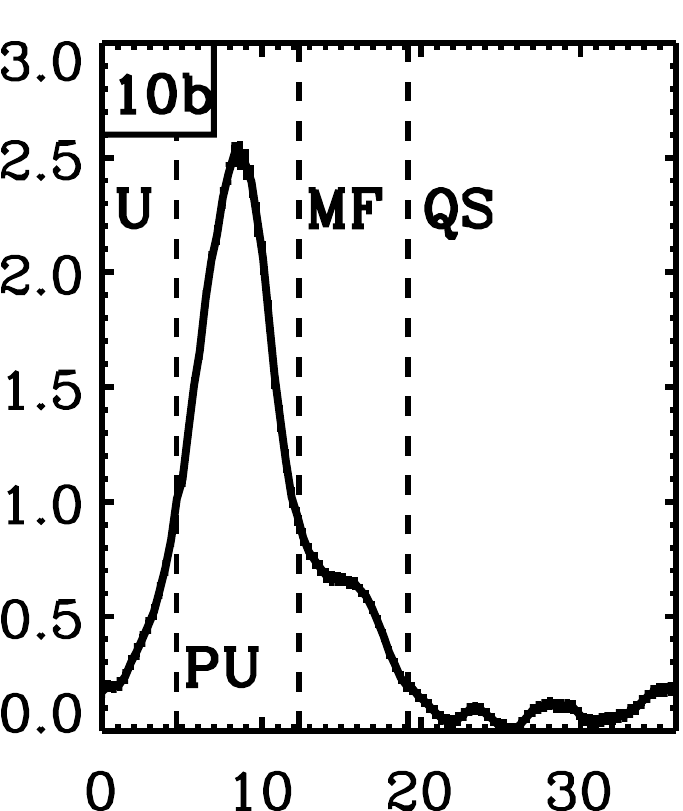}
&\includegraphics[trim = 42mm 140mm 104.5mm 57mm, clip,height=3.7cm]{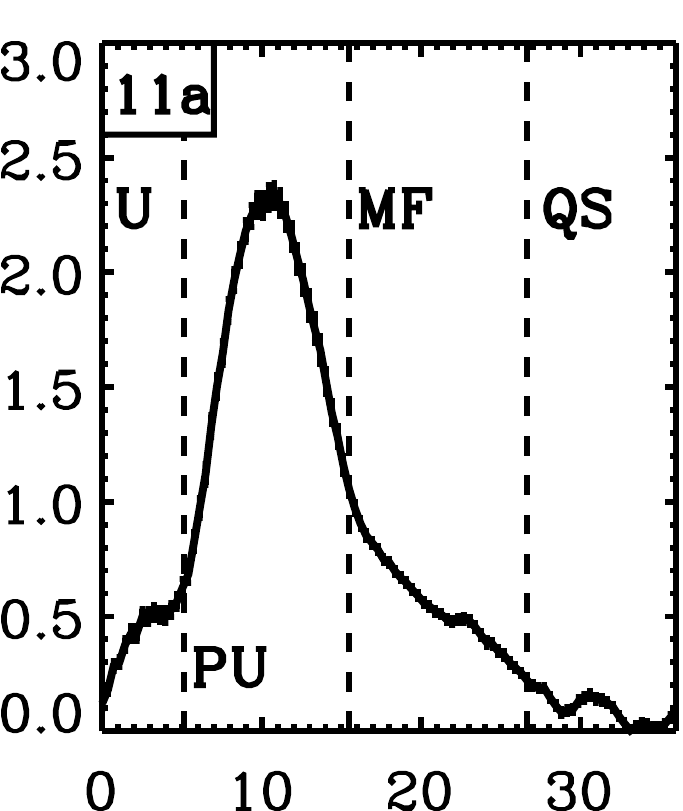}
&\includegraphics[trim = 42mm 140mm 104.5mm 57mm, clip,height=3.7cm]{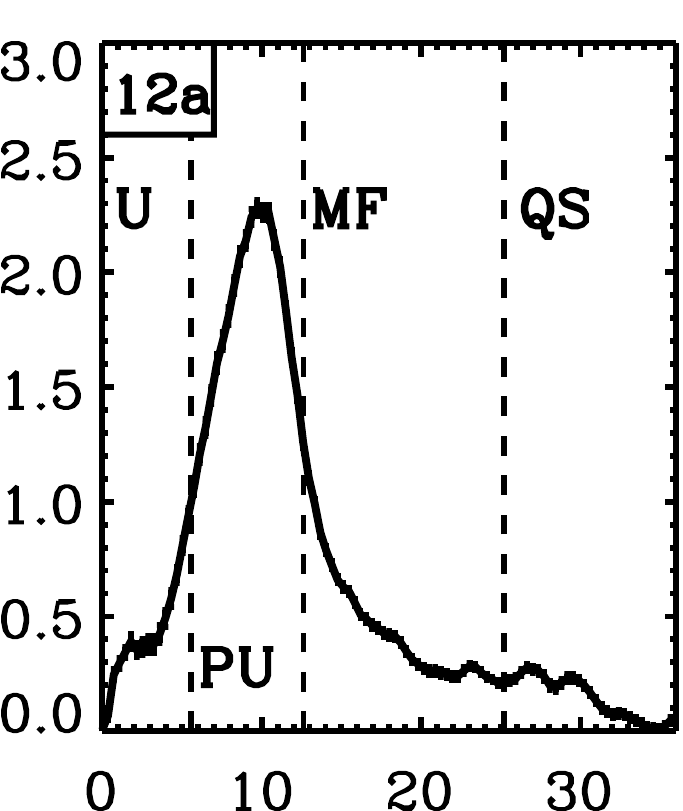}
\\
\hline
\includegraphics[trim = 42mm 140mm 104.5mm 57mm, clip,height=3.7cm]{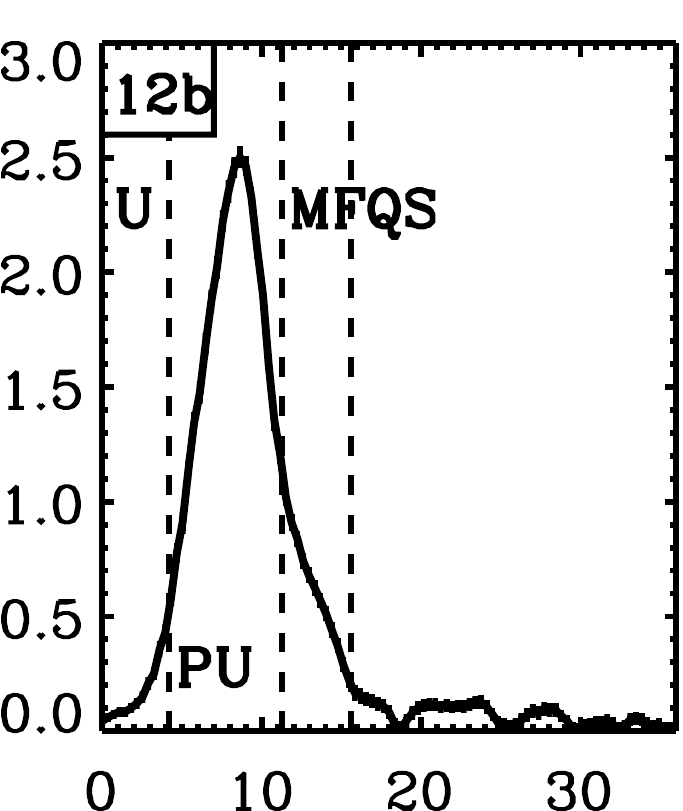}
&\includegraphics[trim = 42mm 140mm 104.5mm 57mm, clip,height=3.7cm]{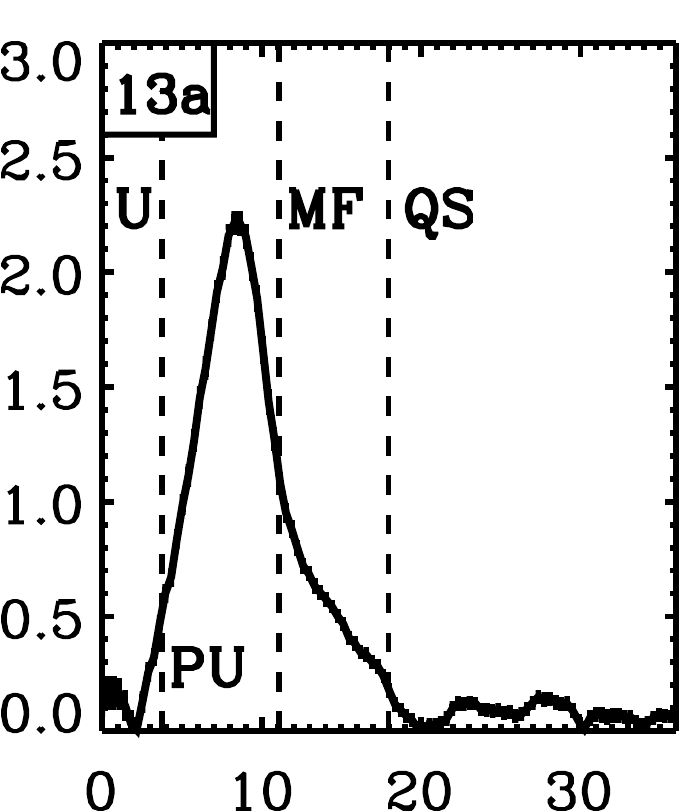}
&\includegraphics[trim = 42mm 140mm 104.5mm 57mm, clip,height=3.7cm]{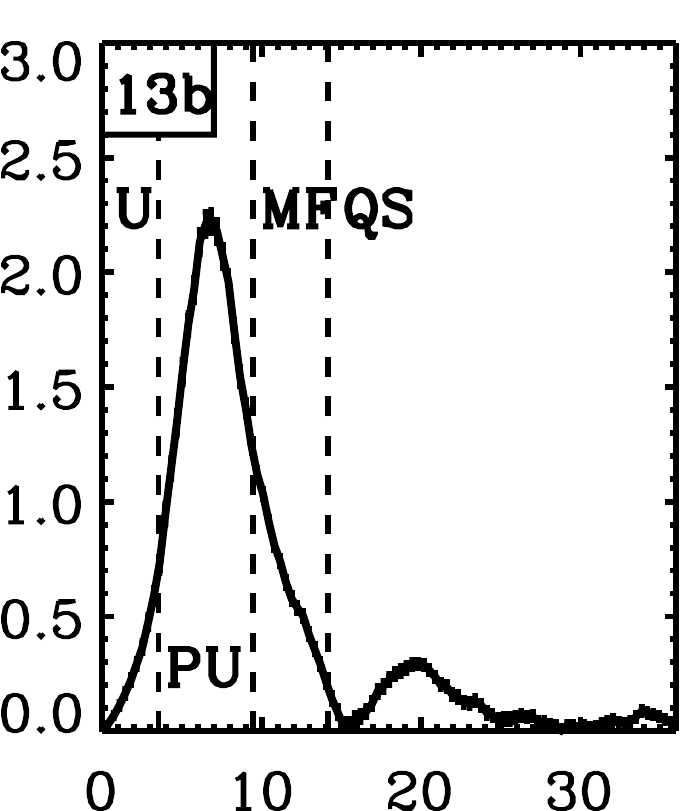}
&\includegraphics[trim = 42mm 140mm 104.5mm 57mm, clip,height=3.7cm]{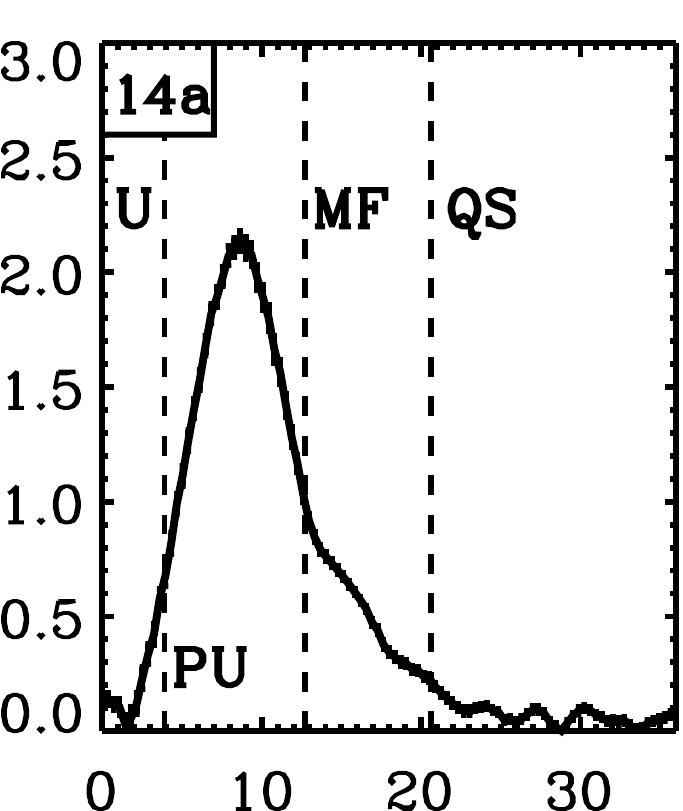}
&\includegraphics[trim = 42mm 140mm 104.5mm 57mm, clip,height=3.7cm]{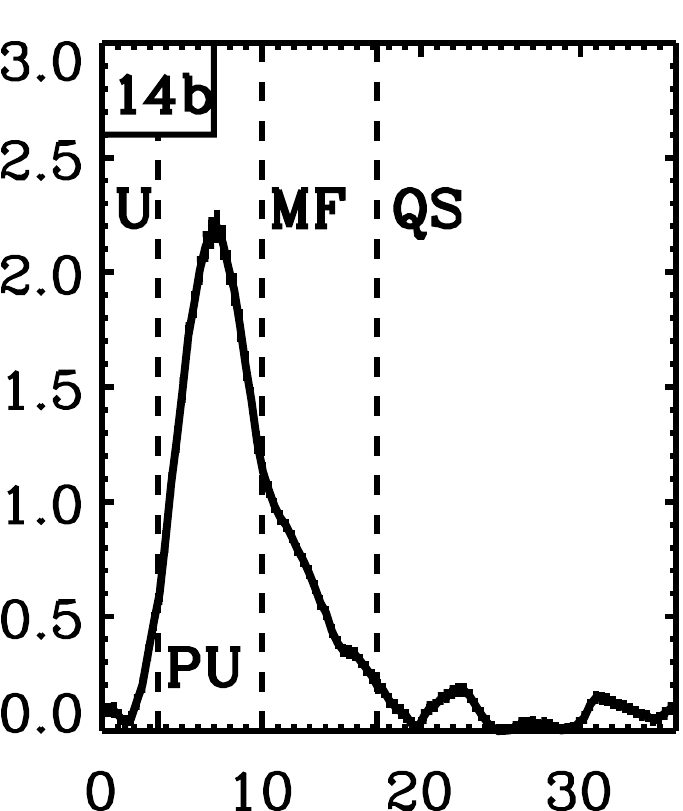}
\\
\hline
\includegraphics[trim = 42mm 140mm 104.5mm 57mm, clip,height=3.7cm]{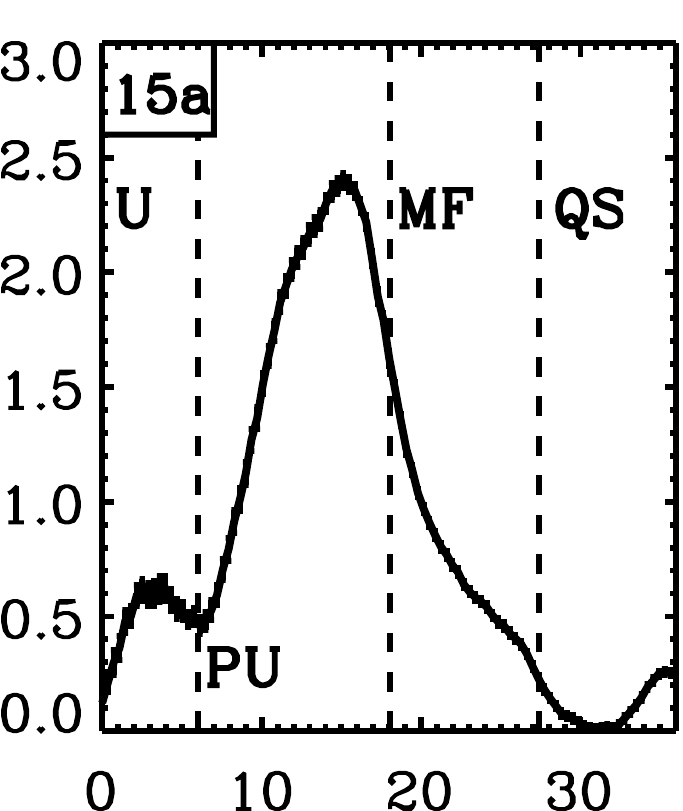}
&\includegraphics[trim = 42mm 140mm 104.5mm 57mm, clip,height=3.7cm]{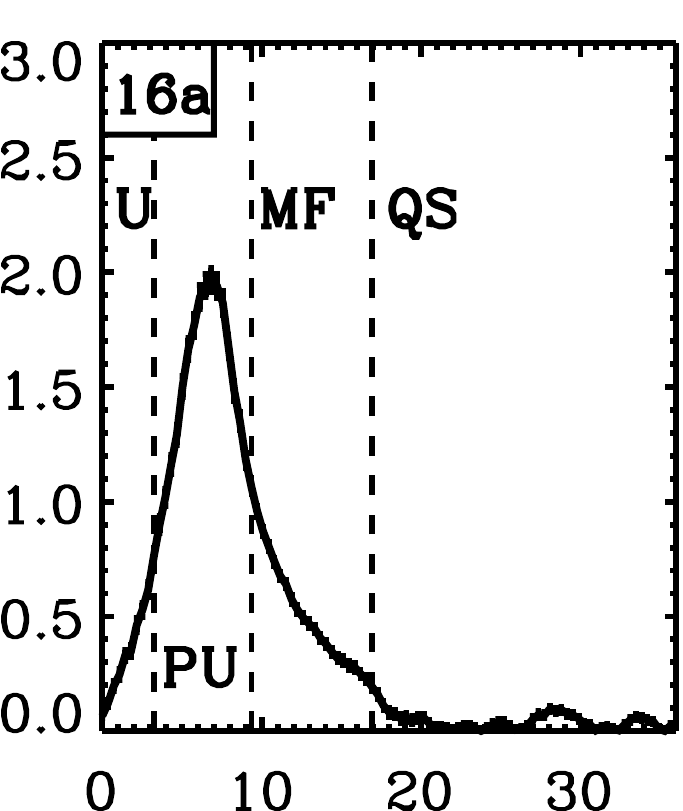}
&\includegraphics[trim = 42mm 140mm 104.5mm 57mm, clip,height=3.7cm]{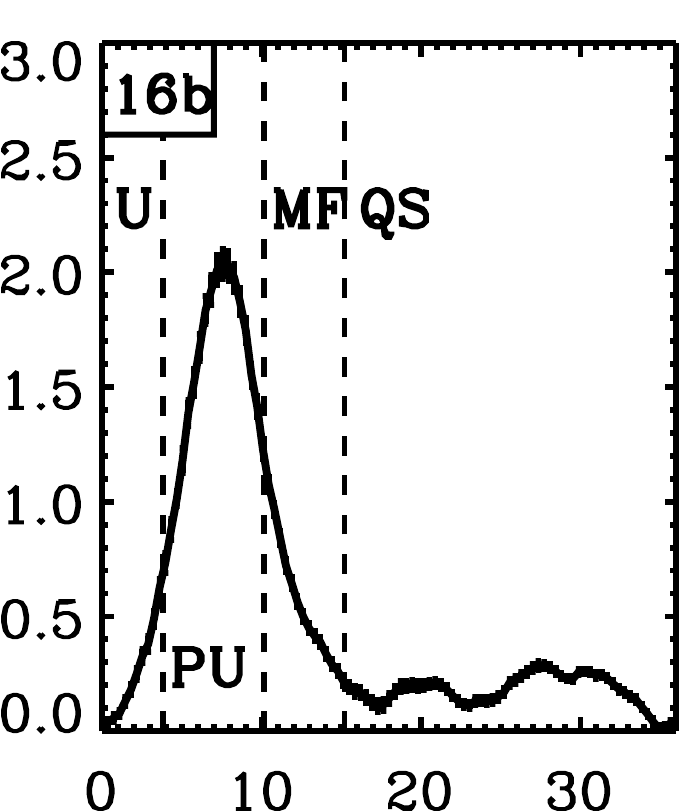}
&\includegraphics[trim = 42mm 140mm 104.5mm 57mm, clip,height=3.7cm]{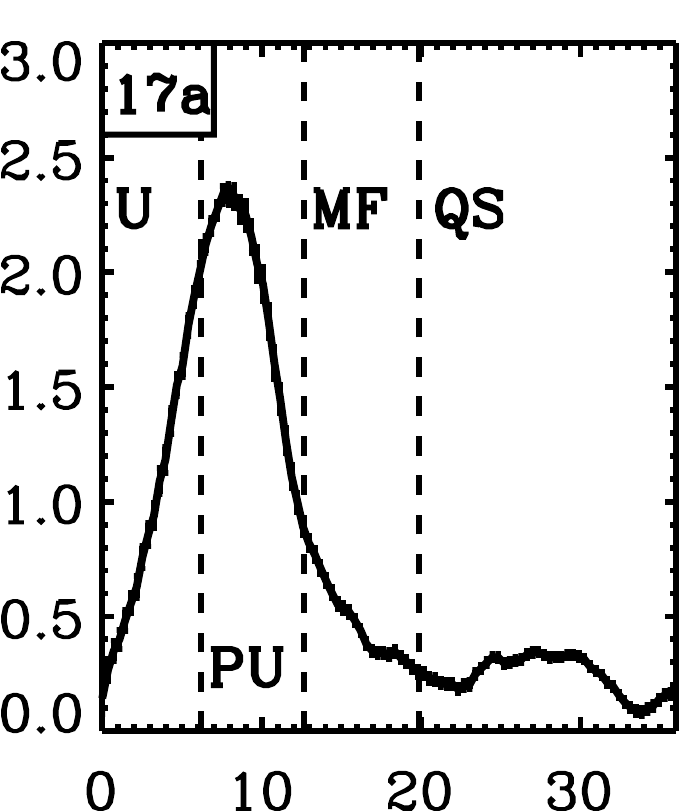}
&\includegraphics[trim = 42mm 140mm 104.5mm 57mm, clip,height=3.7cm]{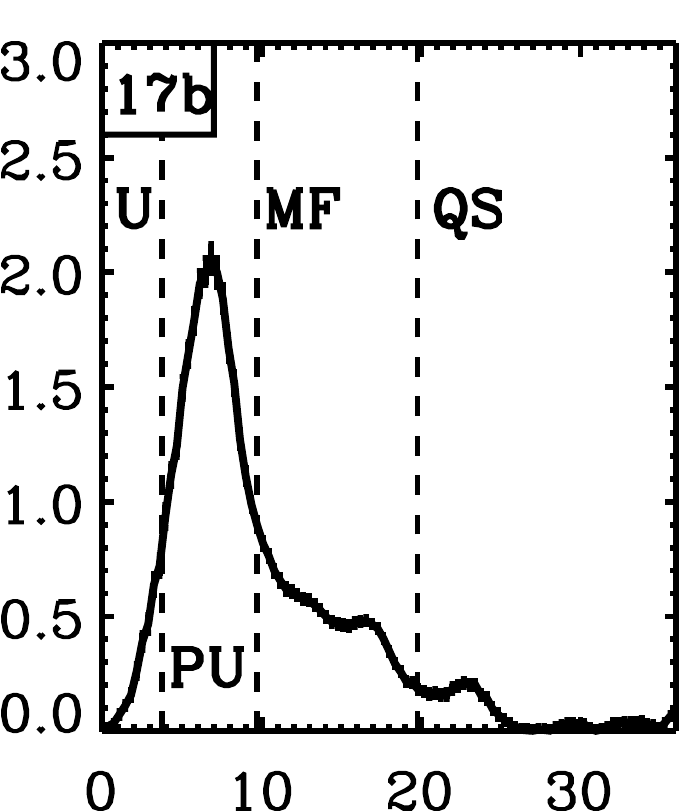}
\\
\hline
\end{tabular}
\end{center}
\end{figure*}%

\begin{figure*}[htdp]
\caption{Horizontal flow velocities, $v_{\parallel,0}(r)$, in km/s \mytextbf{from the sunspot center ($r=0\,\rm{Mm}$) to a distance of 36Mm in the quiet Sun(QS) for spots No. 18--31 (see Table~\ref{apptab}). The boundaries of the umbra (U), the penumbra (PU), and the end of the moat flow region (MF) are indicated as vertical dashed lines.}}\label{appfig2}
\begin{center}
\begin{tabular}{|c|c|c|c|c|}
\hline
\includegraphics[trim = 42mm 140mm 104.5mm 57mm, clip,height=3.7cm]{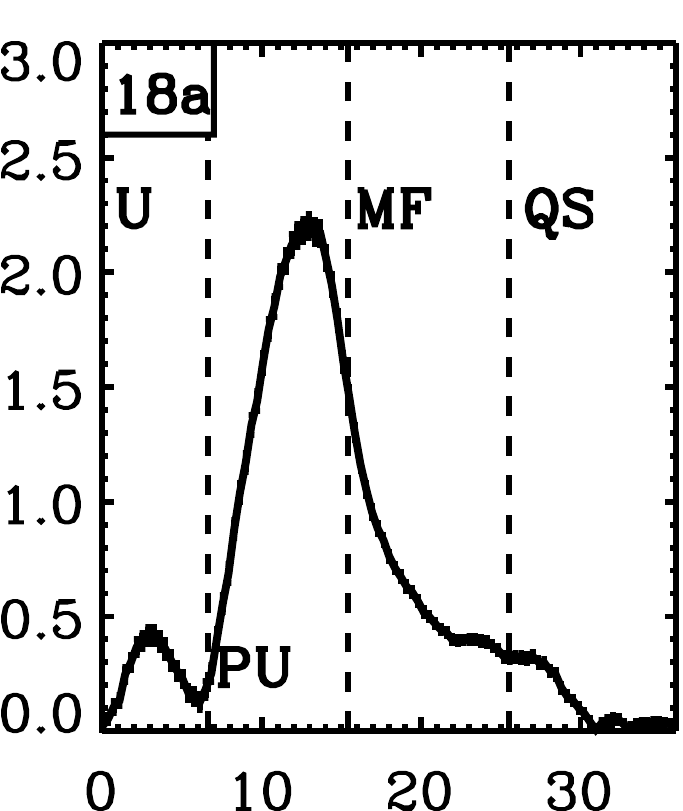}
&\includegraphics[trim = 42mm 140mm 104.5mm 57mm, clip,height=3.7cm]{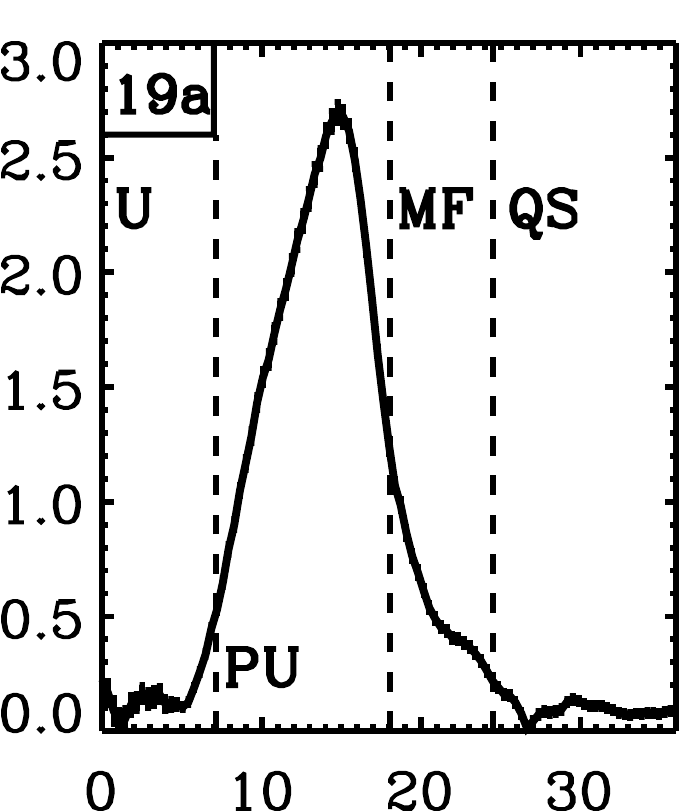}
&\includegraphics[trim = 42mm 140mm 104.5mm 57mm, clip,height=3.7cm]{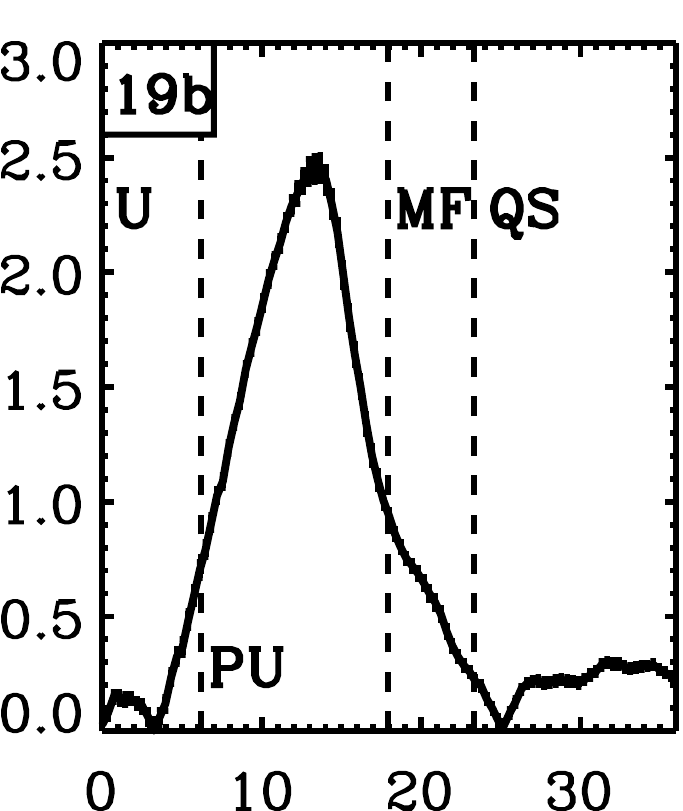}
&\includegraphics[trim = 42mm 140mm 104.5mm 57mm, clip,height=3.7cm]{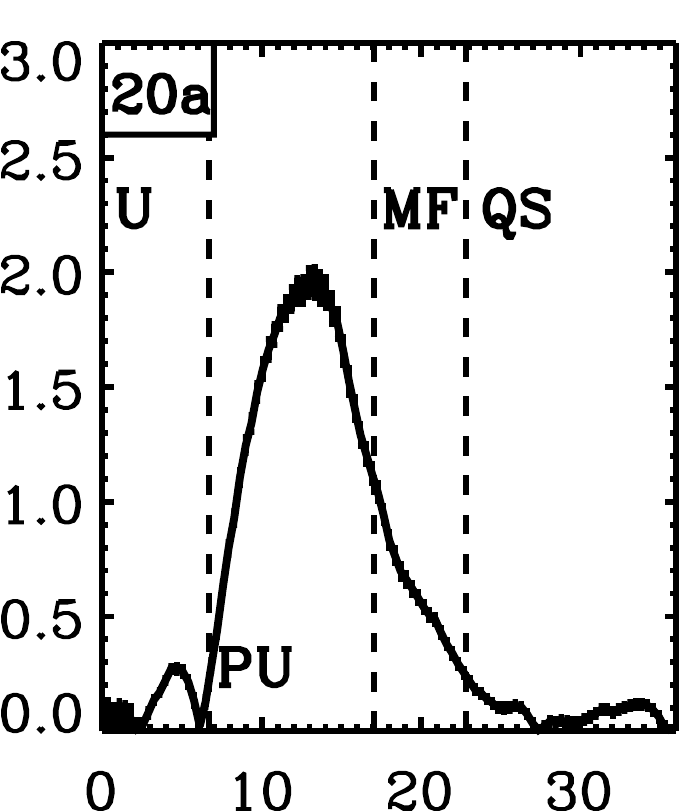}
&\includegraphics[trim = 42mm 140mm 104.5mm 57mm, clip,height=3.7cm]{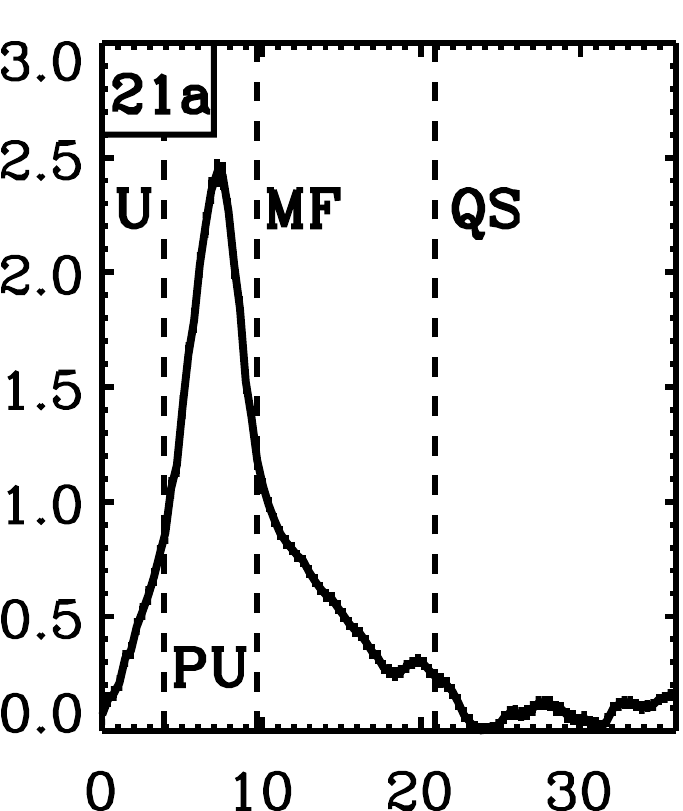}
\\
\hline
\includegraphics[trim = 42mm 140mm 104.5mm 57mm, clip,height=3.7cm]{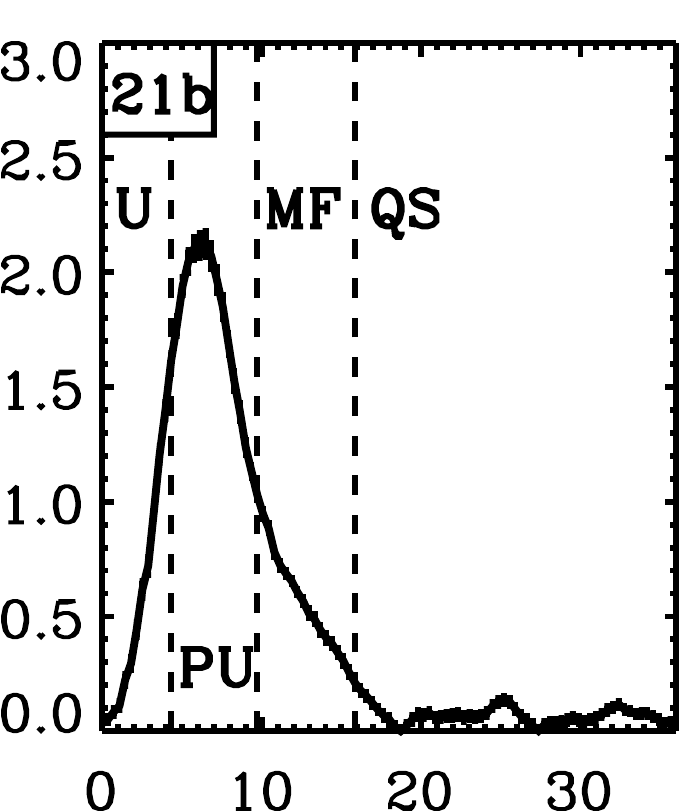}
&\includegraphics[trim = 42mm 140mm 104.5mm 57mm, clip,height=3.7cm]{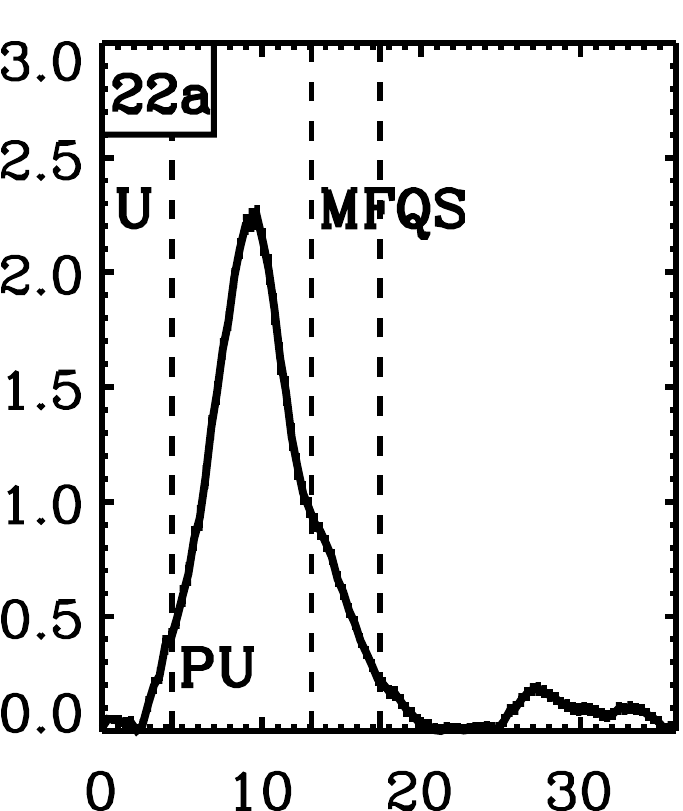}
&\includegraphics[trim = 42mm 140mm 104.5mm 57mm, clip,height=3.7cm]{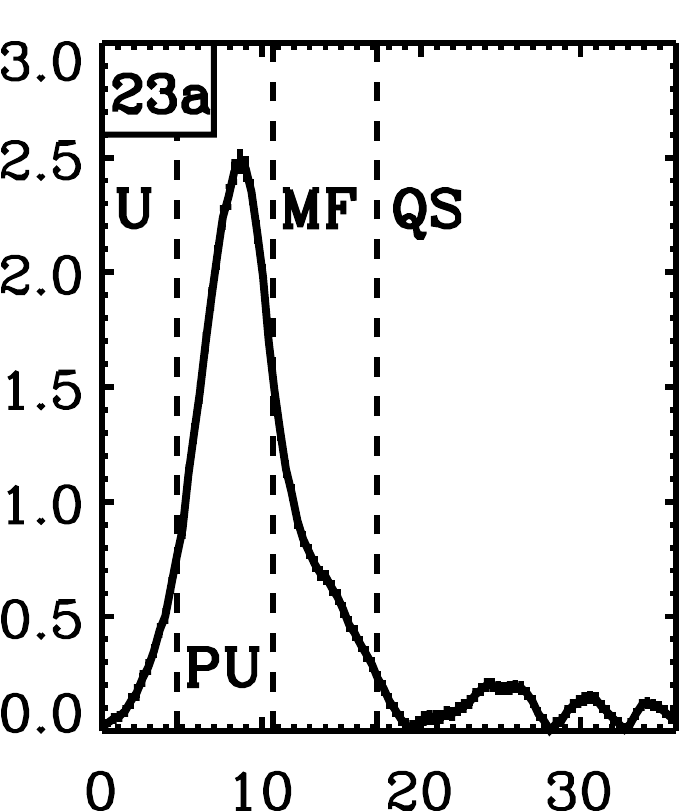}
&\includegraphics[trim = 42mm 140mm 104.5mm 57mm, clip,height=3.7cm]{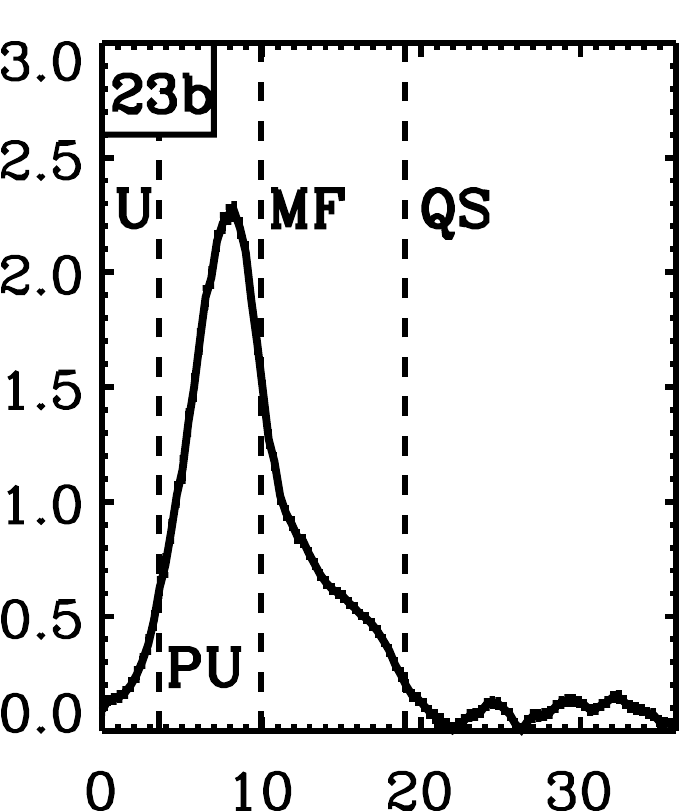}
&\includegraphics[trim = 42mm 140mm 104.5mm 57mm, clip,height=3.7cm]{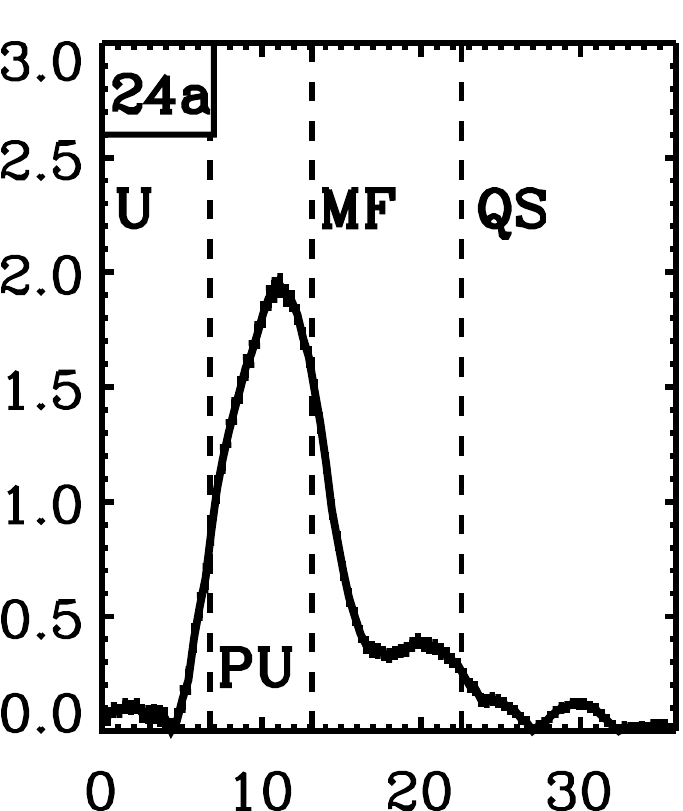}
\\
\hline
\includegraphics[trim = 42mm 140mm 104.5mm 57mm, clip,height=3.7cm]{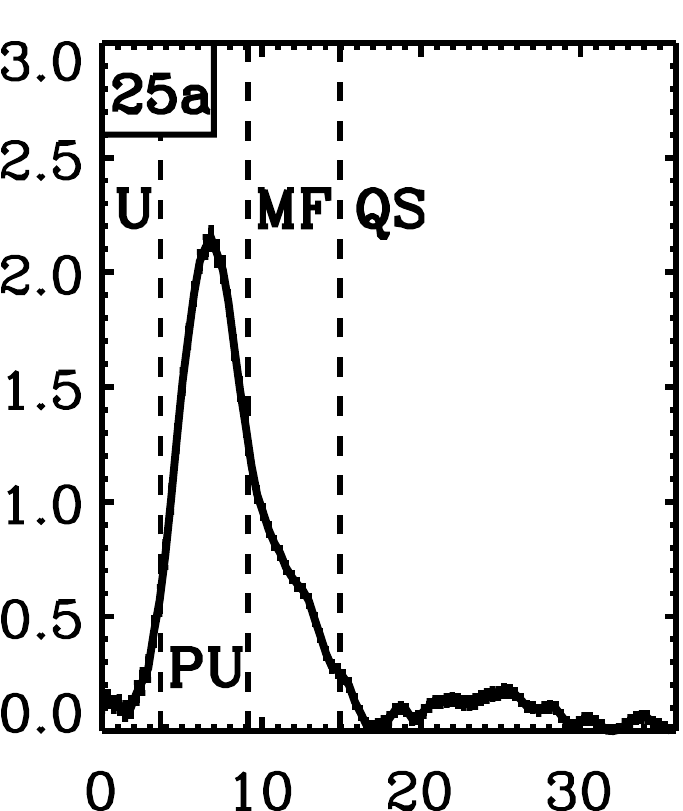}
&\includegraphics[trim = 42mm 140mm 104.5mm 57mm, clip,height=3.7cm]{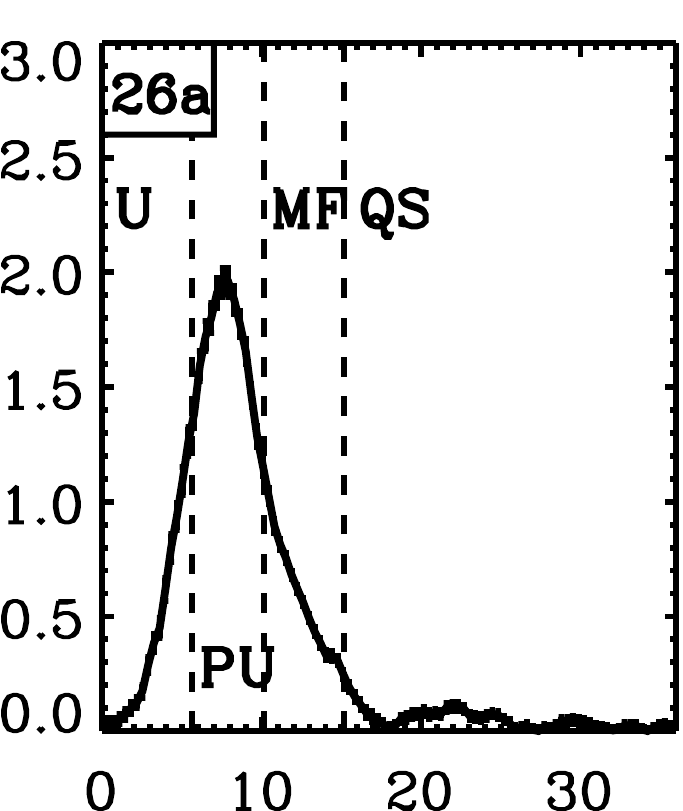}
&\includegraphics[trim = 42mm 140mm 104.5mm 57mm, clip,height=3.7cm]{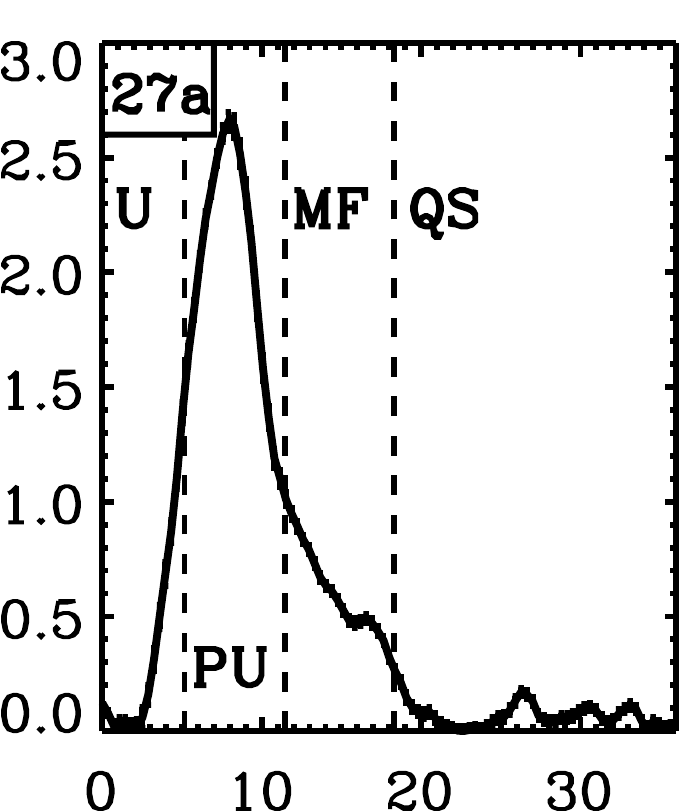}
&\includegraphics[trim = 42mm 140mm 104.5mm 57mm, clip,height=3.7cm]{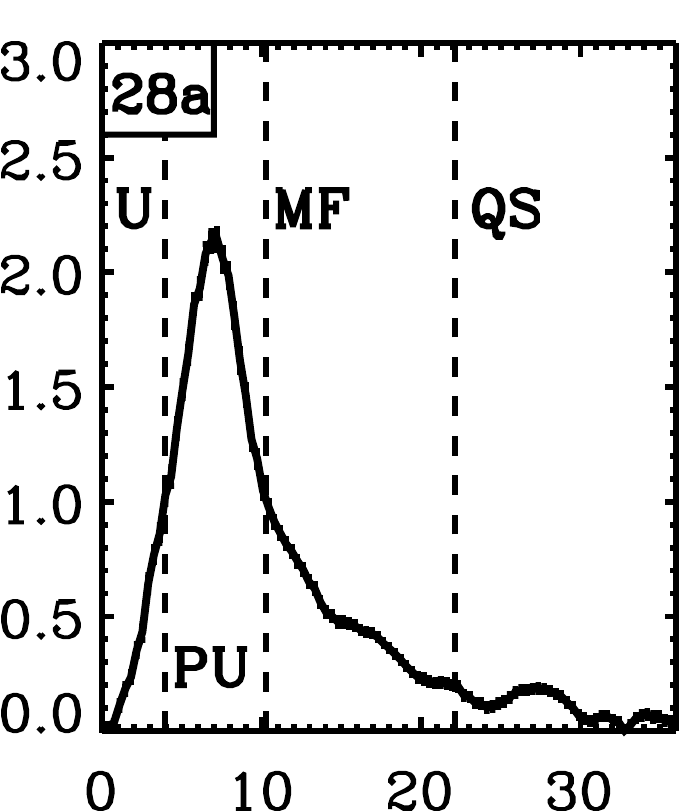}
&\includegraphics[trim = 42mm 140mm 104.5mm 57mm, clip,height=3.7cm]{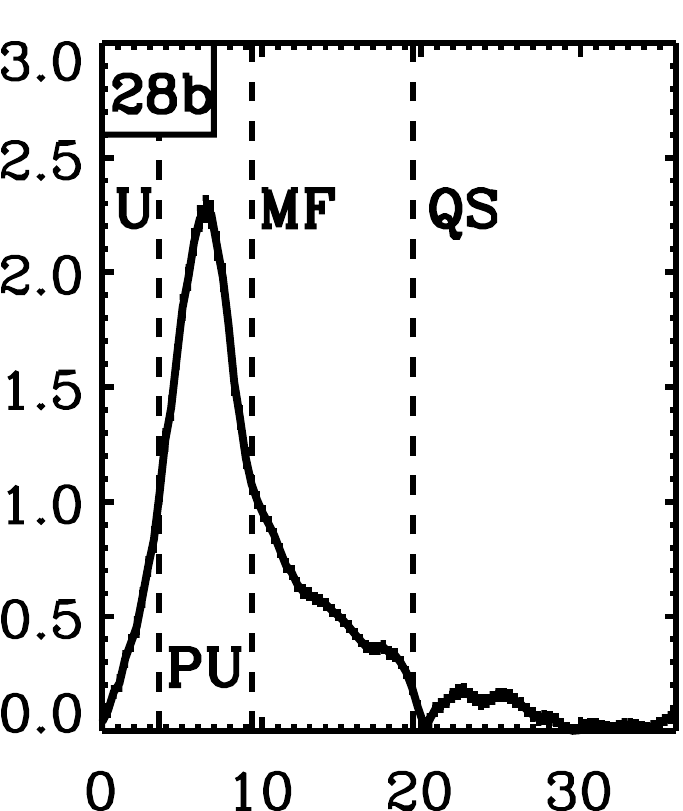}
\\
\hline
\includegraphics[trim = 42mm 140mm 104.5mm 57mm, clip,height=3.7cm]{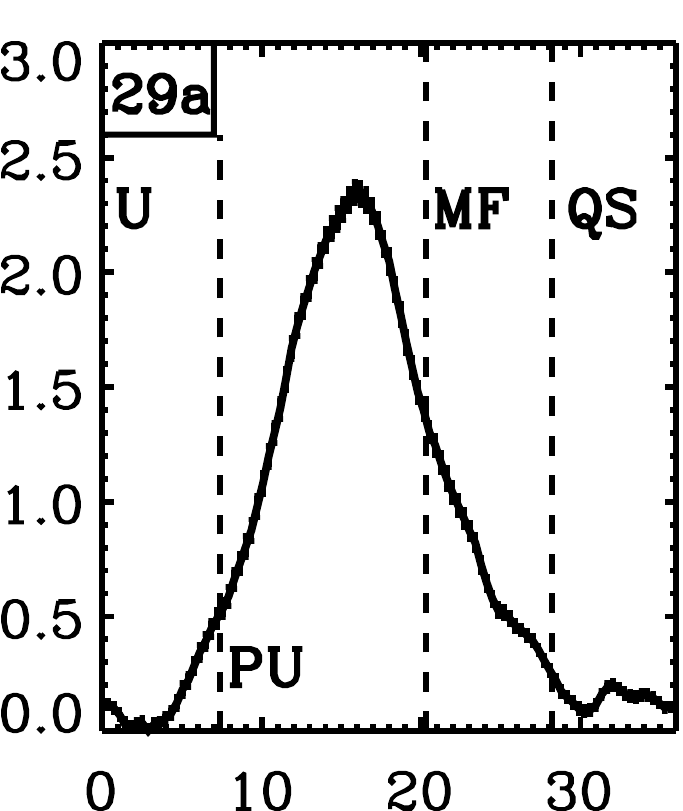}
&\includegraphics[trim = 42mm 140mm 104.5mm 57mm, clip,height=3.7cm]{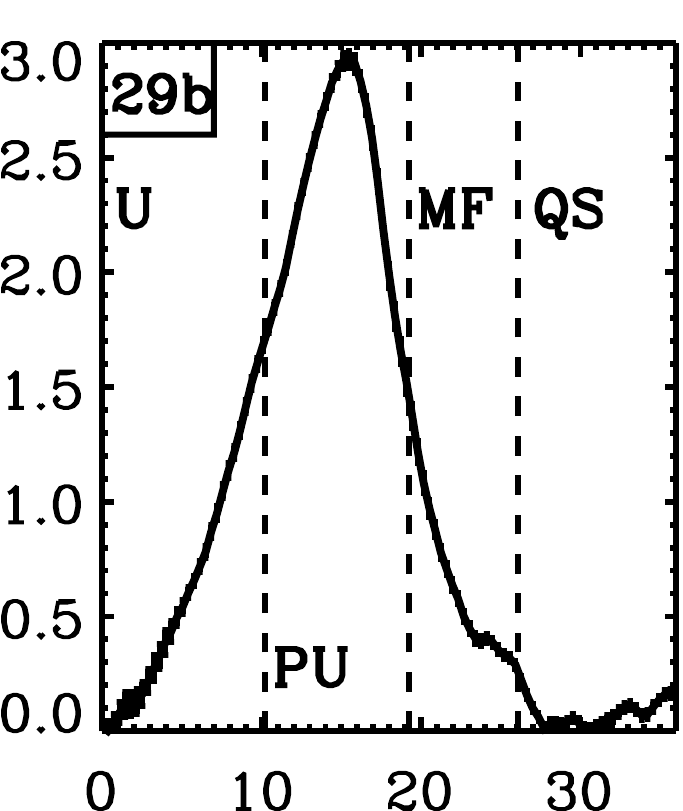}
&\includegraphics[trim = 42mm 140mm 104.5mm 57mm, clip,height=3.7cm]{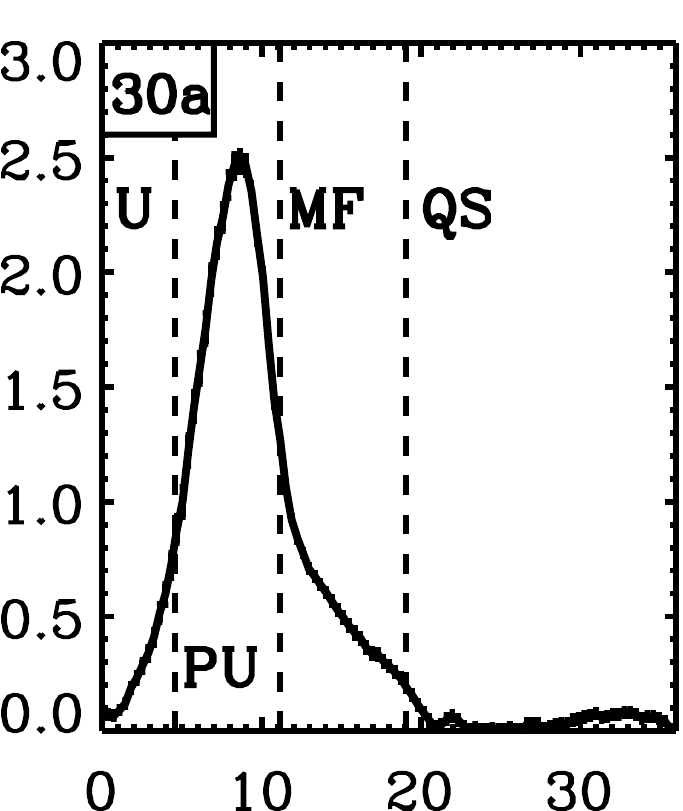}
&\includegraphics[trim = 42mm 140mm 104.5mm 57mm, clip,height=3.7cm]{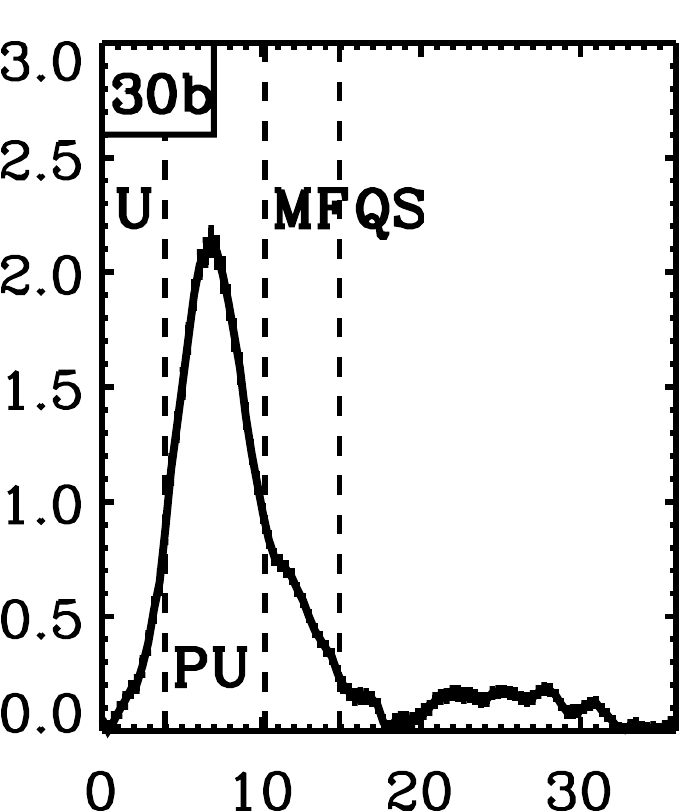}
&\includegraphics[trim = 42mm 140mm 104.5mm 57mm, clip,height=3.7cm]{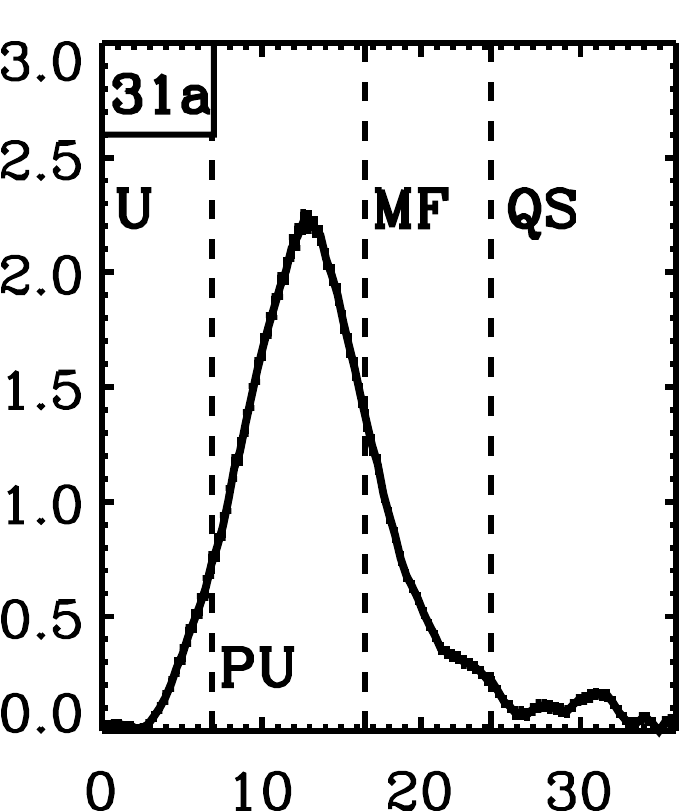}
\\
\hline
\includegraphics[trim = 42mm 140mm 104.5mm 57mm, clip,height=3.7cm]{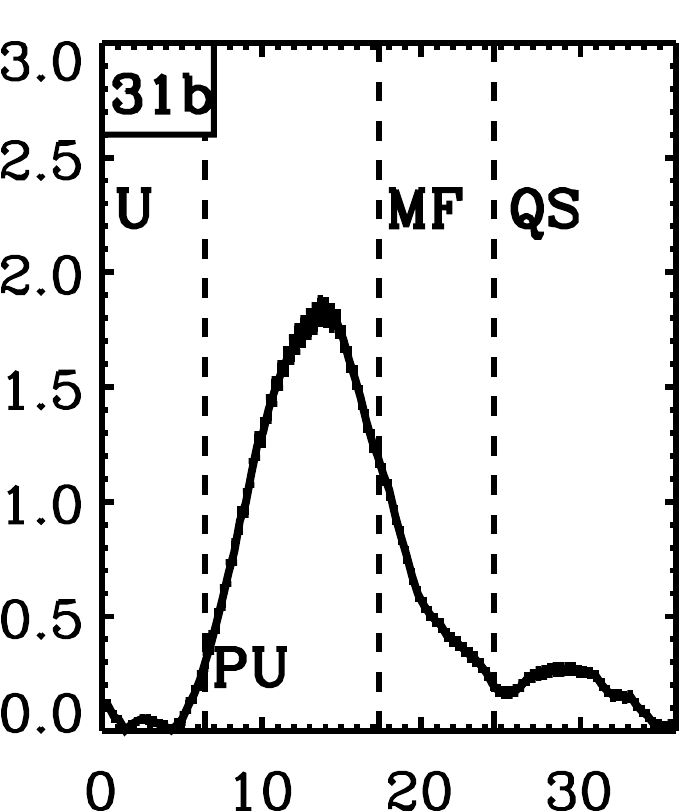}
&&&&\\
\hline
\end{tabular}
\end{center}
\end{figure*}%

\end{appendix}

\end{document}